\documentclass[pra,aps,amssymb,amsmath,amsmath,showpacs,reprint,twocolumn]{revtex4-1}
\usepackage{color}
\usepackage{graphicx}
\usepackage{epstopdf}
\usepackage{natbib}
\usepackage{hyperref}
\usepackage{dcolumn}
\usepackage{bm}
\usepackage{dcolumn}
\usepackage{multirow}

\newcolumntype{d}[1]{D{.}{.}{#1}}

\newcommand{\Fkt}[1]{\,\mathsf {#1}}
\def\openone{\leavevmode\hbox{\small1\kern-3.3pt\normalsize1}}

\ifx\Tr\renewcommand{\Tr}{\Fkt{Tr}} 
\else\newcommand{\Tr}{\Fkt{Tr}}
\fi

\usepackage{booktabs}
\usepackage{graphicx}
\usepackage{amsmath}
\usepackage{indentfirst}
\begin{document}
\title{First-principles calculation of the frequency-dependent dipole polarizability of argon}

\author{\sc Micha\l\ Lesiuk}
\email{e-mail: m.lesiuk@uw.edu.pl}
\author{\sc Bogumi\l\ Jeziorski}
\email{e-mail: jeziorsk@chem.uw.edu.pl}
\affiliation{\sl Faculty of Chemistry, University of Warsaw\\
Pasteura 1, 02-093 Warsaw, Poland}
\date{\today}

\begin{abstract}
In this work we report state-of-the-art theoretical calculations of the dipole polarizability of the argon atom. Frequency dependence of the polarizability is taken into account by means of the dispersion coefficients (Cauchy coefficients) which is sufficient for experimentally relevant wavelengths below the first resonant frequency. In the proposed theoretical framework, all known physical effects including the relativistic, quantum electrodynamics, finite nuclear mass, and finite nuclear size corrections are accounted for. We obtained $\alpha_0=11.0775(19)$ for the static polarizability and $\alpha_2=27.976(15)$ and $\alpha_4=95.02(11)$ for the second and fourth dispersion coefficients, respectively. The result obtained for the static polarizability agrees (within the estimated uncertainty) with the most recent experimental data [C. Gaiser and B. Fellmuth, Phys. Rev. Lett. \textbf{120}, 123203 (2018)], but is less accurate. The dispersion coefficients determined in this work appear to be most accurate in the literature, improving by more than an order of magnitude upon previous estimates. By combining the experimentally determined value of the static polarizability with the dispersion coefficients from our calculations, the polarizability of argon can be calculated with accuracy of around $10\,$ppm for wavelengths above roughly $450\,$nm. This result is important from the point of view of quantum metrology, especially for a new pressure standard based on thermophysical properties of gaseous argon. Additionally, in this work we calculate the static magnetic susceptibility of argon which relates the refractive index of dilute argon gas with its pressure. While our results for this quantity are less accurate than in the case of the polarizability, they can provide, via Lorenz- 
Lorentz formula, the best available theoretical estimate of the refractive index of argon. 
\end{abstract}

\maketitle

\section{Introduction}
\label{sec:intro}

The electric dipole polarizability $\alpha(\omega)$ is an intrinsic microscopic property of atomic and molecular systems describing their response to an external electric field oscillating with frequency $\omega$. Focusing on dilute gases of noble atoms, the polarizability appears in the fundamental Clausius-Mossotti equation
\begin{align}
 \frac{\epsilon_r-1}{\epsilon_r+2} = \frac{4\pi}{3}\alpha\rho,
\end{align}
which relates the relative electric permittivity $\epsilon_r$   of an atomic gas with the density of the gas, $\rho$. We can further express the gas density trough the ideal gas formula $p=kT\rho$, where $k$ is the Boltzmann constant. After some rearrangements we arrive at the relation
\begin{align}
 p = \frac{3}{4\pi} \frac{\epsilon_r-1}{\epsilon_r+2} \frac{kT}{\alpha}
\end{align}
which is the basis for the new primary gas-pressure standard established in 2020~\cite{gaiser20,gaiser22}. Indeed, according to the recent revisions of the fundamental constants~\cite{mohr18,fischer19,machin19}, the Boltzmann constant $k$ has a fixed predefined value. Therefore, by measuring the temperature and electric permittivity of a gas~\cite{fellmuth14,gaiser15,guenz17,gaiser19,gaiser20}, the macroscopic pressure $p$ can be found, as long as the atomic polarizability is known. By progressive improvements to the experimental setup and accuracy of the polarizability determined from theory, the new pressure standard is competitive with the best mechanical pressure measurements, as illustrated with the recent stress test~\cite{gaiser22}.

The aforementioned pressure standard uses helium as the medium gas. This choice is justified, among other things, by high accuracy of theoretical predictions that can be obtained for this relatively simple two-electron atom~\cite{johnson96,bhatia98,pachucki00,cencek01,lach04,puchalski11,piszczu15,puchalski16,puchalski20}. However, the disadvantage of helium is its relatively small polarizability which makes this setup sensitive to impurities, requires high quality materials free of contaminants, etc. A natural way to avoid these problems is to replace helium by a heavier noble atom such as neon or argon. As both of them are significantly more polarizable than helium, the sensitivity problems are marginalized. Unfortunately, as the electronic structure of neon and argon is much more complicated, it is impossible to maintain the same accuracy of theoretical predictions. In fact, while the polarizability of helium can be calculated from first principles~\cite{puchalski20} with relative accuracy of about $10^{-7}$, which is entirely sufficient from the point of view of metrology, the same is not true for neon and argon. Two recent papers~\cite{lesiuk20,hellmann22} devoted to the theoretical calculation of the polarizability of neon were the first studies where all known relevant physical effects were systematically included. Despite significant effort and immense amount of computational time, the best theoretical estimate   still has an uncertainty about five times larger than the experiment~\cite{gaiser18} in the case of the static polarizability. However, the theoretically-derived frequency dependence of the polarizability, which is more difficult to get experimentally, is a useful supplement for the measurements~\cite{rourke21}.

In comparison with neon and, especially, helium, the best available theoretical results for argon lag behind in terms of accuracy. The most reliable theoretical data for the polarizability reported by Lupinetti and Thakkar~\cite{lupinetti05} and by Pawłowski \emph{et al.}~\cite{pawlowski05} can be estimated to have uncertainties of several parts per thousand. This is insufficient for the purposes of metrology and hence in the present work we report state-of-the-art \emph{ab initio} calculations in order to improve the current state of theory. We employ a sequence of coupled-cluster (CC) methods~\cite{bartlett07,crawford07} that converge to the exact solution of the non-relativistic clamped-nuclei Schr\"{o}dinger equation, combined with large basis set up to nonuple-zeta quality. This enables reliable extrapolation to the complete basis set limit and estimation of the residual error which is particularly important in applications to metrology. Equally importantly, in the theoretical framework we include for the first time  all known physical effects, including relativistic, quantum electrodynamics (QED), finite nuclear mass and finite nuclear size contributions.

The static polarizability of argon measured by Gaiser and Fellmuth~\cite{gaiser18} is accurate to about 2 parts per million (ppm). Within the current state of the theory, it is unreasonable to expect that a comparable accuracy can be achieved from first principles. However, this is not the goal of the present work; comparison with the experimental data for the static polarizability will be used primarily to verify that the adopted theoretical framework is adequate. We shall also focus on determination of the so-called dispersion coefficients (defined further in the text) which describe the frequency dependence of the polarizability and are much more difficult to determine experimentally. However, by combining the experimental result for the static polarizability with the frequency dependence derived from theory, high level of accuracy can be obtained for the dynamic polarizability at experimentally relevant frequencies. 

Besides the polarizability, in the present work we consider the static magnetic susceptibility of argon atom, $\chi_0$. It is defined as the second derivative of the energy (with sign reversed) with respect to the strength of the external magnetic field. The importance of the magnetic susceptibility is motivated by the Lorentz-Lorenz formula
\begin{align}
\label{eqll}
\frac{n^2-1}{n^2+2} = \frac{4\pi}{3}(\alpha+\chi_0)\rho,
\end{align}
which relates the the refractive index, $n$, of a gas with its density, $\rho$. The magnetic susceptibility of argon is several orders of magnitude smaller than the polarizability. Therefore, the value of $\chi_0$ may be determined less accurately without a significant impact on the accuracy of $n$. This allows us to neglect the frequency-dependence of the magnetic susceptibility and consider only its static value. Additionally, we neglect several minor corrections in our theoretical framework which are included in case of polarizability.

Unless explicitly stated otherwise, atomic units (a.u.) are used throughout the present work.
Following the CODATA recommendations~\cite{codata18}, we adopt the following values of the fundamental physical constants: speed of light in vacuum, $c=137.035\,999\,084 $, atomic mass unit, $1\,$Da = $1822.888\,486\,209(53)$, Bohr radius, $ a_0 = 0.529\,177\,210\,903\,$\AA{}. We consider only the most naturally abundant ($99.6\%$) stable isotope $^{40}$Ar with atomic mass $39.962\,383\,$Da. Most of the available experimental data related to the molar polarizability of argon is reported in the literature in the units of cm$^3$/mol. To express such quantities in the atomic units we use the conversion factor $1\,$cm$^3$/mol$\;=11.205\,872\,a_0^3$.

\section{Basis sets preparation}
\label{sec:basis}

The family of correlation-consistent~\cite{dunning89} Gaussian basis sets for argon, usually abbreviated as cc-pVXZ, were optimized by Dunning and collaborators~\cite{woon93,mourik00,dunning01,peterson02} up to the sextuple-zeta level of quality. Moreover, additional sets of diffuse and core-valence augmenting functions are also available in the literature. Unfortunately, the standard cc-pVXZ basis sets are not adequate for the purposes of the present work, because of considerable irregularities in the convergence pattern of the results to the complete basis set limit. While for total energies these irregularities were negligible, a significant deterioration was observed for atomic polarizabilities which are the main focus herein. It is worth pointing out that the quality of the results reported in this work depends significantly on the reliability of the extrapolation procedure used to eliminate the residual basis set incompleteness error. The presence of the aforementioned irregularities precludes a robust extrapolation and complicates the error estimation. Therefore, we have decided to optimize a new family of Gaussian basis sets for argon that match the specific requirements of this work.

In the design of the new basis sets we follow the general principles of correlation consistency introduced by Dunning~\cite{dunning89}. First, we optimized a large set of $s$- and $p$-type Gaussian functions to variationally minimize the Hartree-Fock energy of argon. Note that at this level of theory functions with angular momentum $l\ge 2$ do not contribute to the ground-state energy. The number of $s$- and $p$-type functions was increased progressively and the exponents of the Gaussian-type orbitals (GTO) were constrained to form a geometric sequence
\begin{align}
\label{temper}
 \zeta_{ln} = \alpha_l\cdot\beta_l^n\;\;\;\mbox{or}\;\;\; \log\zeta_{ln} = \log\alpha_l+n\log\beta_l
\end{align}
where $n=0,1,\ldots$, and $\alpha_l$ and $\beta_l$ are subject to the optimization. Starting with a small number of functions taken from the cc-pVDZ basis set, the size of the basis was increased by one function at a time, followed by re-optimization of the $\alpha_l$ and $\beta_l$ parameters. We finally settled for the basis set of size $34s27p$ which leads to the accuracy of about 0.3$\,\mu$H (nine significant digits in the energy) in comparison with the numerical Hartree-Fock results of Cinal~\cite{cinal20} which are assumed to be exact for the present purposes.

At some stages of the calculations we shall require an even more accurate basis for the Hartree-Fock calculations. However, extending the geometric sequence (\ref{temper}) further leads to progressive accumulation of numerical noise due to the increasing linear dependencies. It is hence difficult to use the formula (\ref{temper}) for basis sets with more than about $30-40$ functions. To circumvent this problem we used a generalization of Eq.~(\ref{temper}), namely
\begin{align}
\label{temper4}
 \log\zeta_{ln} = \log\alpha_l+n\log\beta_l+n^2\log\gamma_l+n^3\log\delta_l,
\end{align}
$\gamma_l$ and $\delta_l$ are additional variational parameters. Employing this formula, we optimized a $37s37p$ basis for Hartree-Fock calculations which is the smallest basis that reaches the accuracy of a few nH (eleven significant digits in the energy). No significant numerical issues were encountered for this basis.

The next step of the basis set optimization is the addition of polarization functions necessary to recover the electronic correlation effects. At this stage it is customary to contract the $sp$ part of the basis optimized in order to reduce the size of the basis. We follow this protocol; however, in some calculations we will use uncontracted basis sets when explicitly stated. The contraction coefficients were obtained from expansion coefficients of the Hartree-Fock orbitals within a given basis. The polarization functions were added according to the correlation consistency principle, i.e.
the double-zeta basis contains a single $d$ polarization function, triple-zeta -- two $d$ and one $f$, quadruple-zeta -- three $d$, two $f$, and one $g$, etc. At each expansion stage, additional $s$ and $p$ functions were added by taking from the contraction the functions with the lowest exponents.

The exponents of the polarization functions follow the sequence defined by Eq.~(\ref{temper}) and the parameters $\alpha_l$ and $\beta_l$ were optimized to minimize the frozen-core ($8$ active electrons) MP2 correlation energy. While in the literature it is common to use the configuration interaction with single and double excitations (CISD) method for basis set optimization, it is not feasible for the basis sets required in this work. This is due to the high cost of CISD calculations in comparison with MP2. The parameters $\alpha_l$ and $\beta_l$ were optimized in turns using Powell's method until the convergence to within $10^{-11}\,$H in the MP2 energy was obtained. The largest basis set considered is of nonuple-zeta quality and includes basis set functions up to $l=9$. The optimizations were performed using the Dalton program package~\cite{daltonpaper} combined with an external program written especially for this purpose. Note that to carry out calculations with such high angular momentum, it is necessary to modify the source code of the Dalton package before compilation. Details of how to perform necessary changes can be obtained from authors upon request. The composition and exponents of the optimized Gaussian basis sets is given in the Supporting Information~\cite{supporting}. For brevity, we refer to the new basis sets simply as $X$Z, $X=2,\ldots,9$, further in the text. Note that the parameter $X$ coincides with the highest angular momentum present in the basis set.

The polarizability is sensitive, to a much larger degree than the energy, to the accuracy of the long-range tail of the atomic density. Therefore, in our calculations it is necessary to augment the Gaussian basis with additional functions with low exponents. In this work the exponents are not optimized, but are generated from the formula~(\ref{temper}) by setting $n=-1,-2,\ldots$ In this way, we generated singly-augmented (adding $n=-1$ up to $l\leq X$), doubly-augmented ($n=-1,-2$ up to $l\leq X$) and triply-augmented ($n=-1,-2,-3$ up to $l\leq X$) basis sets which are denoted a$X$Z, da$X$Z, and ta$X$Z further in the text. Preliminary calculations shown that further augmentation of the basis leads to only a tiny improvement of the results which does not justify the corresponding increase of the computational costs.

Finally, in the optimization of the $X$Z basis sets we kept ten inner core orbitals of the argon atom inactive. While their influence on the results is much smaller than of the valence shells, it is still non-negligible from the point of view of the adopted accuracy standards. To take the contribution of the core orbitals into account, the basis sets must be extended with a set of functions with high exponents (tight functions). Fortunately, we found that the cc-pCVXZ basis sets available in the literature~\cite{peterson02,woon93} which were optimized to take the core-valence effects into account, do not suffer from the irregularities we encountered in reproduction of the valence contributions. Therefore, we simply use the optimized tight functions from the standard cc-pCVXZ basis sets in combination with the remaining functions from the $X$Z family. We denote this extended core-valence basis sets as c$X$Z and their augmented counterparts by ac$X$Z, dac$X$Z, etc.

\section{Overview of the theoretical approach and computational details}
\label{sec:overview}

The main goal of the present work is to theoretically determine the polarizability of the argon atom, denoted $\alpha(\omega)$, including all known physical effects that bring a significant contribution. The atomic polarizability depends on the frequency $\omega$ of the external electromagnetic field that the atom is subjected to. We are interested in a range of wavelengths above (roughly) $450\,$nm which covers operating wavelengths most of the practically used gas lasers based on noble gases. After conversion to the atomic units, this gives the interval $0\leq\omega\lesssim0.1$. As the supremum of this interval is significantly lower than the first resonance frequency of the argon atom (equal to about $\omega_{\mathrm{res}}\approx 0.42$~\cite{minnhagen73}) we can use the power expansion:
\begin{align}
\label{dispexp}
 \alpha(\omega)=\alpha_0 +\alpha_2\,\omega^2+\alpha_4\,\omega^4+\ldots
\end{align}
The first term of the above formula, $\alpha_0$, is usually called the static polarizability, while the quantities $\alpha_2$, $\alpha_4$, etc. are the dispersion coefficients (or Cauchy coefficients).

The static polarizability of the argon atom has recently been determined experimentally by Gaiser and Fellmuth~\cite{gaiser18} using the dielectric-constant gas thermometry. They obtained the relative accuracy of about 2 parts per million (ppm). Within the current state of the theory, it is unlikely that any calculations can deliver a similar accuracy level. In fact, among noble gases, only for helium the quality of theoretical predictions matches (or even surpasses) that of the experiments, but this is feasible only because of a relatively simple electronic structure of two-electron systems. Already for neon, in the most accurate theoretical calculations performed thus far~\cite{lesiuk20,hellmann22}, the uncertainty estimates are several times larger than of the experimental results. Instead, we will test and validate our theoretical model by comparing with the experimental data for $\alpha_0$. 

Our focus in this work is placed on the dispersion coefficients. These quantities cannot be determined experimentally at present as accurately as the static polarizability. Therefore, in applications where the frequency dependence of the polarizability is necessary, theoretical results for the dispersion coefficients can supplement the experimental $\alpha_0$. Therefore, let us estimate the accuracy level required in $\alpha_n$ to achieve the accuracy of about $10\,$ppm for wavelengths above (roughly) $450\,$nm. Such accuracy level is acceptable from the point of view of metrology. Considering the worst-case scenario of $\omega\approx0.1$, the value of $\alpha_2$ must be determined with relative accuracy of about $\cdot10^{-3}$, while $\alpha_4$ -- about $10\%$. The higher-order dispersion coefficients, $\alpha_n$, $n\geq6$, can be neglected. In this analysis, we assumed that all $\alpha_n$, $n=0,2,4$, are of a similar magnitude. In practice, $\alpha_2$ is larger than $\alpha_0$ by roughly a factor of two and hence the relative accuracy of around $5\cdot10^{-4}$ is needed. The value of $\alpha_4$ is about $8$ times larger than $\alpha_0$, thus it has to be determined with the accuracy of about $1\%$. One can also expect that the value of $\alpha_6$ is significantly larger than $\alpha_0$ and it may contribute for short wavelengths. Therefore, in this work we additionally determine the value of $\alpha_6$ with the accuracy goal~of~$10\%$.

Regarding the magnetic susceptibility, its value is by a factor of around $2\cdot 10^{-5}$ smaller than the polarizability. Taking into account that a simple sum of the two quantities is relevant from the point of view of Eq.~(\ref{eqll}), it is sufficient to determine $\chi$ with accuracy of about $10\%$. Provided that this level of accuracy can be reached, the sum $\alpha_0+\chi_0$ would have the uncertainty comparable with the experiment of Gaiser and Fellmuth~\cite{gaiser18} for $\alpha_0$. This allows us to adopt several approximations in determination of the magnetic susceptibility. First, we neglect the frequency dependence - its influence is expect to be around $1\%$ within the relevant frequency range. Second, we omit all corrections with contribute to less than $1\%$ in the case of the static polarizability. This eliminates the relativistic, QED and several other minor corrections. We hence focus on accurate determination of the non-relativistic value which is much more straightforward and sufficient for all practical purposes.

The calculations reported in this work are based primarily on the coupled-cluster hierarchy of methods. For calculations using the Hartree-Fock, CCSD~\cite{purvis82,scuseria87}, CCSD(T)~\cite{ragha89}, and CC3~\cite{koch97} methods (including the relativistic effects), we employed the \textsc{Dalton} program package~\cite{daltonpaper} with the aforementioned modifications of the source code to enable calculations with high angular momentum functions. For CCSDT~\cite{noga87,scuseria88} calculations and computations of the finite nuclear mass corrections we used the \textsc{CFour} program~\cite{cfour}, interfaced with the \textsc{MRCC} package~\cite{kallay20}. The latter code is used for all higher-order CC methods (CCSDTQ~\cite{kucharski91,oliphant91,kucharski92,kucharski10}, CCSDTQP~\cite{musial00,musial02} and higher~\cite{kallay01,olsen00,hirata03}). In all calculations we use tight thresholds for the convergence of the CC iterations and of the response function solver ($10^{-9}$ in the norm of the residual vector). Purely spherical Gaussian basis sets are used throughout this work. The orbital-unrelaxed variant of the CC response theory is employed in all calculations.

\section{Non-relativistic polarizability}
\label{sec:nrel}

\begin{table*}[t]
\caption{\label{tab:afcsd}
Linear-response frozen-core CCSD contribution to the static polarizability and dispersion coefficients, $\Delta\alpha_n^{\mathrm{SD}}$, of the argon atom calculated using the augmented $X$Z basis sets. In the last row we provide results extrapolated to the complete basis set limit according to Eq.~(\ref{riemann}) and the corresponding error estimate (see the main text for the discussion).
}
\begin{ruledtabular}
\begin{tabular}{cccccccccc}
 & 
 \multicolumn{3}{c}{singly augmented} &
 \multicolumn{3}{c}{doubly augmented} & 
 \multicolumn{3}{c}{triply augmented} \\
 $X$
 & $\Delta\alpha_0^{\mathrm{SD}}$ 
 & $\Delta\alpha_2^{\mathrm{SD}}$ 
 & $\Delta\alpha_4^{\mathrm{SD}}$
 & $\Delta\alpha_0^{\mathrm{SD}}$ 
 & $\Delta\alpha_2^{\mathrm{SD}}$ 
 & $\Delta\alpha_4^{\mathrm{SD}}$
 & $\Delta\alpha_0^{\mathrm{SD}}$ 
 & $\Delta\alpha_2^{\mathrm{SD}}$ 
 & $\Delta\alpha_4^{\mathrm{SD}}$ \\
\hline\\[-1em]
2 & $-$0.4025 & 1.4828 & 13.4945 & $-$0.3166 & 2.3074 & 17.3982 & $-$0.2635 & 2.8722 & 20.7742 \\
3 & $-$0.2775 & 2.3570 & 15.8547 & $-$0.2184 & 2.8555 & 18.8666 & $-$0.2162 & 2.8822 & 19.0202 \\
4 & $-$0.3270 & 2.0304 & 13.1597 & $-$0.3115 & 2.2346 & 14.7719 & $-$0.3117 & 2.2346 & 14.7754 \\
5 & $-$0.3412 & 1.9645 & 12.6164 & $-$0.3320 & 2.0923 & 13.7132 & $-$0.3321 & 2.0923 & 13.7155 \\
6 & $-$0.3507 & 1.9142 & 12.2765 & $-$0.3447 & 2.0147 & 13.2029 & $-$0.3447 & 2.0146 & 13.2055 \\
7 & $-$0.3551 & 1.8991 & 12.2270 & $-$0.3512 & 1.9711 & 12.9275 & $-$0.3512 & 1.9712 & 12.9318 \\
8 & $-$0.3582 & 1.8847 & 12.1634 & $-$0.3552 & 1.9429 & 12.7499 & $-$0.3552 & 1.9432 & 12.7569 \\
9 & $-$0.3605 & 1.8689 & 12.0687 & $-$0.3577 & 1.9246 & 12.6333 & $-$0.3577 & 1.9252 & 12.6462 \\
\hline
$\infty$ & $-$0.3665(32) & 1.8290(476) & 11.8286(944)
         & $-$0.3642(4)  & 1.8781(48)  & 12.3376(401) 
         & $-$0.3642(4)  & 1.8797(32)  & 12.3654(99) \\
\end{tabular}
\end{ruledtabular}
\end{table*}

The dominant contribution to the polarizability of argon comes from the non-relativistic clamped-nucleus approximation. Therefore, this contribution must be calculated with high precision and we adopt a composite scheme based on the CC theory for this purpose. 

\subsection{Mean-field contribution}

The first contribution to the polarizability and dispersion coefficients, denoted by the symbol $\alpha_n^{\mathrm{HF}}$, comes from the (restricted closed-shell) Hartree-Fock method and was calculated using the standard coupled-perturbed response theory. We used the large $37s37p$ basis set described in the preceding section, augmented with three sets of diffuse functions, giving $40s40p$ set in total. Note that at the Hartree-Fock level only $s$ and $p$ functions are needed to expand the ground-state orbitals, but $d$ functions are additionally needed for the calculation of the polarizability. Therefore, a set of $40d$ functions was added to the basis with the same exponents as for the $p$ functions. Within the complete $40s40p40d$ basis we obtain the following values of the polarizability and dispersion coefficients at the Hartree-Fock level of theory:
\begin{align}
\begin{split}
 \alpha_0^{\mathrm{HF}} &= \phantom{0}11.4726(1), \\
 \alpha_2^{\mathrm{HF}} &= \phantom{0}25.6162(1), \\
 \alpha_4^{\mathrm{HF}} &= \phantom{0}78.9658(2), \\ 
 \alpha_6^{\mathrm{HF}} &=           297.775(6).
\end{split}
\end{align}
We accessed the accuracy of the results by randomly removing one or two functions of each angular momentum from the basis and recomputing the polarizability with the reduced set (bootstrapping). In all cases, we observed deviations at the level of $1-2\,$ppm which is negligible in the present context.

\begin{figure*}[ht]
\includegraphics[scale=0.45]{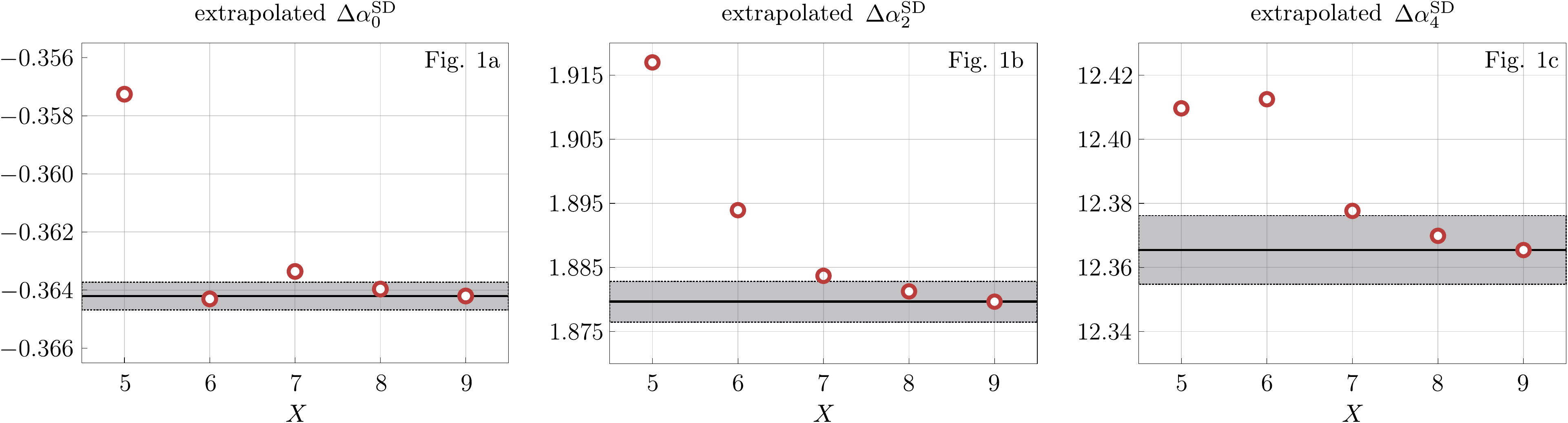}
\caption{\label{fig:ccsd-cbs}
Extrapolation of the $\Delta\alpha_n^{\mathrm{SD}}$ contributions to the complete basis set limit using the formula~(\ref{riemann}). The solid horizontal line denotes the best estimate of the corresponding quantity (see text for details) and the shaded area represents the respective error bars.
}
\end{figure*}

\subsection{Valence correlation contribution}

The second contribution to the polarizability, denoted $\Delta\alpha_n^{\mathrm{SD}}$, was obtained at the frozen-core CCSD level of theory ($8$ active valence electrons). Here we use the optimized $X$Z basis sets, $X=2,\ldots,9$ (fully uncontracted variants) and their counterparts augmented with diffuse functions. In Table~\ref{tab:afcsd} we present results of the calculations. As it is well-known, the results converge rather slowly (asymptotically as $X^{-3}$) with respect to the basis set size which is a consequence of the non-analytic behavior of the exact wavefunction at the coalescence points of the electrons (the cusp condition). To eliminate the residual basis set incompleteness error, we perform extrapolation to the complete basis set (CBS) limit using the recently proposed formalism based on the Riemann zeta function~\cite{lesiuk19}. Application of this scheme is straightforward provided that results obtained with two consecutive basis sets ($X$ and $X-1$) are available. Let us denote the quantity of interest $\mathcal{O}$ obtained with the basis set $X$ by $\mathcal{O}_X$. The CBS limit $\mathcal{O}_\infty$ is then estimated from the formula~\cite{lesiuk19}
\begin{align}
\label{riemann}
 \mathcal{O}_\infty = \mathcal{O}_X + 
 X^4\Big[\zeta(4)-\sum_{l=1}^X l^{-4}\Big]
 \big( \mathcal{O}_X - \mathcal{O}_{X-1} \big),
\end{align}
where $\zeta(s)=\sum_{n=1}^\infty n^{-s}$ is the Riemann zeta function and hence $\zeta(4)=\frac{\pi^4}{90}$. Throughout this work, this extrapolation formula is used for all components of the static polarizability and the dispersion coefficients. Note that the CBS limit estimated from Eq.~(\ref{riemann}) is still formally dependent on the variable $X$ and by observing the progression of the extrapolated values from a series of basis sets one can estimate the uncertainty of the predictions. To illustrate this, in Fig.~\ref{fig:ccsd-cbs} we provide extrapolated values of the $\Delta\alpha_n^{\mathrm{SD}}$ contributions, $n=0,2,4$, as a function of the $X$ parameter that defines the size of the basis set. The extrapolated values converge quickly to the vicinity of the limiting value; for basis sets $X\geq 7$ the differences are minor. It is reasonable to estimate that the extrapolation error is equal to the difference between the CBS limits obtained with $X=9,8$ and $X=8,7$ basis set pairs. However, to make this error estimate more conservative, we additionally multiply it by a factor of two. The shaded areas plotted in Fig.~\ref{fig:ccsd-cbs} represent the error bars obtained in this way.

While the bulk of the basis set incompleteness error, addressed in the previous paragraph, stems from truncation with respect to the angular momentum, the secondary source of error is related to the augmentation with diffuse functions. Fortunately, the results converge rapidly with increasing augmentation level, as evident from Table~\ref{tab:afcsd}. The single augmentation level is not satisfactory with the present accuracy standards, but the differences between the results obtained doubly- and triply-augmented basis sets are minor. This is especially true for the static polarizability, where the estimated CBS limits are essentially indistinguishable. However, for the dispersion coefficients we observe a small discrepancy between the CBS limits obtained with da$X$Z and ta$X$Z basis sets, signaling that the adopted extrapolation scheme~(\ref{riemann}) does not fully resolve this problem. To eliminate this issue, we assume that the results converge exponentially with respect to the augmentation levels. The CBS limits obtained with a$X$Z, da$X$Z and ta$X$Z are fitted with the functional form
\begin{align}
 \Delta\alpha_n^{\mathrm{SD}} = A_n + B_n e^{-C_n m},
\end{align}
where $m$ is the augmentation level. By extrapolating to the limit $m\rightarrow\infty$ we obtain the final estimates of the frozen-core CCSD contribution to the polarizability and dispersion coefficients:
\begin{align}
\begin{split}
 \Delta\alpha_0^{\mathrm{SD}} &=  -0.3642(4), \\
 \Delta\alpha_2^{\mathrm{SD}} &=  \phantom{-}1.8797(32), \\
 \Delta\alpha_4^{\mathrm{SD}} &=  \phantom{-}12.3670(99), \\
 \Delta\alpha_6^{\mathrm{SD}} &=  \phantom{-}67.906(26),
\end{split}
\end{align}
For the sake of brevity, in the above discussion we have not considered the quantity $\Delta\alpha_6^{\mathrm{SD}}$ explicitly. However, the value given above has been obtained using exactly the same protocol as for the lower-order coefficients.

The next contribution to the polarizability and dispersion coefficients is due to coupled-cluster triple excitations within the frozen-core approximation. We split this contribution into two parts. The first (dominant) part is calculated at the CC3 level of theory, denoted $\Delta\alpha_n^{\mathrm{CC3}}$ further in the text, while the second (presumably minor) is the difference between the CC3 and full CCSDT results, that is 
\begin{align}
 \Delta\alpha_n^{\mathrm{T}}=\Delta\alpha_n^{\mathrm{CCSDT}}-\Delta\alpha_n^{\mathrm{CC3}}.
\end{align}
The reason for adopting this two-step approach is the fact that the CC3 calculations are significantly less computationally expensive than the full CCSDT. Moreover, the CC3 method is known to capture a majority of the triple-excitation effects. We managed to calculate $\Delta\alpha_n^{\mathrm{CC3}}$ with doubly-augmented basis sets da$X$Z, $X=2,\ldots,7$, but calculations of $\Delta\alpha_n^{\mathrm{CCSDT}}$ are feasible only for $X=2,\dots,5$ at this augmentation level. Based on a set of preliminary calculations, we found that the triply-augmented basis sets ta$X$Z give almost the same results as da$X$Z and, in order to reduce the computational costs, the latter basis sets are adopted. In calculation of both triple-excitation contributions we use the contracted variants of the da$X$Z basis sets.

In the determination of the $\Delta\alpha_n^{\mathrm{T}}$ contribution we face an additional technical difficulty. The CCSD and CC3 calculations reported here were carried out with the help of the \textsc{Dalton} package which is able to determine the dispersion coefficients directly. However, the CCSDT (and higher-order) methods are not implemented in this program and we employ the \textsc{CFour} and \textsc{MRCC} packages for this calculations. Unfortunately, in the latter two codes the dispersion coefficients are not computed explicitly. Instead, one has to perform calculations of the frequency-dependent polarizability at a grid of frequencies and determine the coefficients in the expansion~(\ref{dispexp}) by fitting. This procedure is an additional source of potential error that has to be controlled. Fortunately, we are able to perform benchmark calculations at the CC3 level of theory, where the dispersion coefficients can be determined both directly and by fitting, to judge the accuracy of the procedure.

\begin{table}[b]
\caption{\label{tab:omegafit}
Comparison of the $\Delta\alpha_n^{\mathrm{CC3}}$ coefficients obtained by fitting and from the direct calculation (da$4$Z basis set). The fitting procedure includes even powers of $\omega$ up to the tenth order.
}
\begin{ruledtabular}
\begin{tabular}{ccc}
 quantity & direct & fitting \\
 \hline\\[-1em]
 $\Delta\alpha_0^{\mathrm{CC3}}$ & $-$0.0250 & $-$0.0250 \\[0.25em]
 $\Delta\alpha_2^{\mathrm{CC3}}$ & \phantom{$-$}0.2285 & \phantom{$-$}0.2285 \\[0.25em]
 $\Delta\alpha_4^{\mathrm{CC3}}$ & \phantom{$-$}1.6473 & \phantom{$-$}1.6509 \\[0.25em]
 $\Delta\alpha_6^{\mathrm{CC3}}$ & \phantom{$-$}9.8084 & \phantom{$-$}9.1503 \\
\end{tabular}
\end{ruledtabular}
\end{table}

\begin{table*}[t]
\caption{\label{tab:afct}
Linear-response frozen-core triple-excitation contribution to the static polarizability and dispersion coefficients, $\Delta\alpha_n^{\mathrm{CC3}}$ and $\Delta\alpha_n^{\mathrm{T}}$, of the argon atom calculated using the doubly-augmented da$X$Z basis sets. See the main text for the details of the adopted extrapolation and error estimation procedures.
}
\begin{ruledtabular}
\begin{tabular}{ccccccc}
 $X$ & 
 $\Delta\alpha_0^{\mathrm{CC3}}$ &
 $\Delta\alpha_2^{\mathrm{CC3}}$ &
 $\Delta\alpha_4^{\mathrm{CC3}}$ &
 $\Delta\alpha_0^{\mathrm{T}}$   &
 $\Delta\alpha_2^{\mathrm{T}}$   &
 $\Delta\alpha_4^{\mathrm{T}}$   \\
 \hline\\[-1em]
 2 & \phantom{$-$}0.0452 & 0.3335 & 1.9729
   & 0.0006 & 0.0333 & 0.2909 \\
 3 & $-$0.0175 & 0.2244 & 1.6777 
   & 0.0051 & 0.0609 & 0.4627 \\
 4 & $-$0.0250 & 0.2285 & 1.6473 
   & 0.0029 & 0.0551 & 0.3846 \\
 5 & $-$0.0133 & 0.3445 & 2.3298 
   & 0.0010 & 0.0413 & 0.3171 \\
 6 & $-$0.0069 & 0.4022 & 2.6685 
   & 0.0009 & 0.0373 & 0.3012 \\
 7 & $-$0.0054 & 0.4186 & 2.7659 
   & --- & --- & --- \\
 8 & $-$0.0047 & 0.4261 & 2.8083 
   & --- & --- & --- \\
 \hline
 $\infty$ & $-$0.0031(4) & 0.4427(68) & 2.9018(473) &
 0.0007(2) & 0.0312(62) & 0.2764(553) \\
\end{tabular}
\end{ruledtabular}
\end{table*}

As an example, we provide details of the aforementioned benchmark calculations within the da$4$Z basis set. The polarizability was calculated at the CC3 level of theory for $31$ frequencies uniformly spaced in the interval $\omega\in[0.000, 0.150]$ including the endpoints. The largest frequency corresponds to the wavelength $303.76\,$nm and hence the whole experimentally relevant range of frequencies is covered. Independently, the dispersion coefficients were calculated directly at the same level of theory. In this work we are not interested in dispersion coefficients of higher order than sixth. Nonetheless, we found that inclusion of additional coefficients proportional to $\omega^8$ and $\omega^{10}$ stabilizes the fitting procedure and improves the accuracy. Therefore, the expansion~(\ref{dispexp}) used in the fitting procedure includes all even powers of $\omega$ up to $\omega^{10}$. Incorporation of higher powers of $\omega$ does not change the results in a meaningful way and hence they were eliminated to reduce the risk of over-fitting.
In Table~\ref{tab:omegafit} we compare the $\Delta\alpha_n^{\mathrm{CC3}}$ coefficients obtained by fitting and from the direct calculation within the da$4$Z basis set. Overall, the fitting procedure yields reliable values of the required coefficients. For $n=0$ and $n=2$ the fitted results are essentially identical to those calculated directly. Only for $n=6$ we observe a substantial deviation, but this is acceptable within the present context. Based on this benchmark calculation we shall assume in the remainder of the paper that the fitting procedure is able to deliver the accuracy of at least four significant digits for the second-order coefficient, at least three significant digits for the fourth-order coefficient, and at least one significant digit for the sixth-order coefficient. 

In Table~\ref{tab:afct} we report results of the calculations of the $\Delta\alpha_n^{\mathrm{CC3}}$ and $\Delta\alpha_n^{\mathrm{T}}$ contributions. The former contribution was calculated directly, while the fitting procedure was used for the latter. The results obtained with the da2Z basis set are somewhat erratic, but starting with $X=3$ the convergence pattern towards the CBS limit becomes regular. For the $\Delta\alpha_n^{\mathrm{CC3}}$ and $\Delta\alpha_n^{\mathrm{T}}$ contributions we adopted exactly the same extrapolation scheme and error estimation method as in the preceding CCSD calculations. The final estimates of the $\Delta\alpha_n^{\mathrm{CC3}}$ and $\Delta\alpha_n^{\mathrm{T}}$ contributions and the corresponding error bars are given in Table~\ref{tab:afct}. While the results for the $\Delta\alpha_6^{\mathrm{CC3}}$ contribution are not given explicitly, the same procedure as for the lower-order coefficients gives:
\begin{align}
 \Delta\alpha_6^{\mathrm{CC3}} = 16.85(30).
\end{align}
The contribution $\Delta\alpha_6^{\mathrm{T}}$ is neglected, see the discussion in Sec.~\ref{sec:overview}. The same is true for contributions of higher excitations to the sixth-order dispersion coefficient.

Next, we consider the contributions to the polarizability originating from quadruple excitations with respect to the reference determinant. The full CCSDTQ computations scale as $N^{10}$ with the system size and are very costly. To reduce this cost, it is customary to employ non-iterative models that account for quadruple excitations, such as CCSDT(Q)~\cite{bomble05,kallay05} or CCSDT[Q]~\cite{kucharski89}. Unfortunately, as the electronic wavefunction is not well-defined in these methods, they cannot be used for determination of dynamic response properties or excitation spectra. Another option is to employ the CC4 model~\cite{kallay05} which is free from this drawback and has recently been shown to deliver very accurate excitation energies. However, to the best of our knowledge, calculation of the dynamic polarizabilities at the CC4 level of theory is not implemented in any electronic structure package at present. Therefore, in determination of the quadruple excitations contribution to the polarizability and dispersion coefficients, denoted $\Delta\alpha_n^{\mathrm{Q}}$, the full CCSDTQ method is used in this work.

In Table~\ref{tab:afcq} we report the calculations of the $\Delta\alpha_n^{\mathrm{Q}}$ contribution. Due to the aforementioned high cost of these computations, da$X$Z basis sets only up to $X=4$ were feasible. Similarly as for other contributions, $\Delta\alpha_n^{\mathrm{Q}}$ was calculation for a finite set of frequencies followed by an analytical fitting procedure. In general, the behavior of the results is similar as in the case of the $\Delta\alpha_n^{\mathrm{T}}$ correction, cf.~Table~\ref{tab:afct}, but the convergence with respect to the basis set size is noticeably faster. We employ the extrapolation formula (\ref{riemann}) from the $X=3,4$ pair to eliminate the basis set incompleteness error, as shown in Table~\ref{tab:afcq}. In order to estimate the error of these results, we repeat the same procedure for extrapolation of the $\Delta\alpha_n^{\mathrm{T}}$ correction and compare with more reliable results from Table~\ref{tab:afct}. The error bars given in Table~\ref{tab:afcq} were obtained under the assumption that the relative error in the $\Delta\alpha_n^{\mathrm{Q}}$ contribution extrapolated from the $X=3,4$ basis set pair is no larger than for the $\Delta\alpha_n^{\mathrm{T}}$ contribution calculated in the same way.

\begin{table}[b]
\caption{\label{tab:afcq}
Coupled-cluster quadruple excitation contributions to the static polarizability and dispersion coefficients, $\Delta\alpha_n^{\mathrm{Q}}$, of the argon atom calculated using the doubly-augmented da$X$Z basis sets.
}
\begin{ruledtabular}
\begin{tabular}{ccccccc}
 $X$ & 
 $\Delta\alpha_0^{\mathrm{Q}}$ &
 $\Delta\alpha_2^{\mathrm{Q}}$ &
 $\Delta\alpha_4^{\mathrm{Q}}$ \\
 \hline\\[-1em]
2 & $-$0.0114 & $-$0.0593 & $-$0.3400 \\
3 & $-$0.0128 & $-$0.0866 & $-$0.5383 \\
4 & $-$0.0085 & $-$0.0477 & $-$0.2858 \\
 \hline
 $\infty$ & $-$0.0045(10) & $-$0.0122(45) & $-$0.0549(64) \\
\end{tabular}
\end{ruledtabular}
\end{table}

It is worth pointing out an unusual feature of the triple and quadruple excitation contributions to the static polarizability. The total triple excitation contribution (that is, the sum of the CC3 and CCSDT contributions), equal to roughly $-0.0024$, is smaller in absolute terms than the quadruple excitation contribution, $-0.0045$, see Tables~\ref{tab:afct}~and~\ref{tab:afcq}. This unexpected phenomenon is a consequence of a peculiar behavior of the $\Delta\alpha_0^{\mathrm{CC3}}$ which accidentally crosses zero in the vicinity of $\omega=0$. A similar phenomenon does not occur for the quadruple excitations, explaining the unusual ratio of the two corrections. Moreover, this feature is not observed for the dispersion coefficients. In fact, both for the second- and fourth-order coefficients, the quadruple excitation contribution is about $50$ times smaller than the triple excitation effects, in line with the expectations based on the conventional wisdom.

Finally, we study the contribution of higher-order excitations to the polarizability and dispersion coefficients. The contributions of pentuple, $\Delta\alpha_n^{\mathrm{P}}$, and hextuple, $\Delta\alpha_n^{\mathrm{H}}$, excitations were calculated within the da2Z basis set. Unfortunately, these calculations are unfeasible with a larger basis set and hence it is not possible to perform an extrapolation. Therefore, we assign a conservative uncertainty estimate of $50\%$ to the values calculated within the da2Z basis. The following results were obtained using the fitting procedure described in the preceding paragraphs:
\begin{align}
 \begin{split}
  &\Delta\alpha_0^{\mathrm{P}} = -0.0007(3),\;\;\;\;\;\;
  \phantom{00}\Delta\alpha_0^{\mathrm{H}} = -0.00002(1), \\
  &\Delta\alpha_2^{\mathrm{P}} = -0.0050(24),\;\;\;\;\;\;
  \phantom{0}\Delta\alpha_2^{\mathrm{H}} = -0.00009(5), \\
  &\Delta\alpha_4^{\mathrm{P}} = -0.0386(193),\;\;\;\;\;\;
  \Delta\alpha_4^{\mathrm{H}} = \phantom{-}0.0002(1)
 \end{split}
\end{align}
It is also worth pointing out the rapid convergence of the results with respect to the maximum excitation level included in the coupled-cluster wavefunction. Taking the $\Delta\alpha_2$ coefficient as an example, the best estimate of the single and double excitations contribution is $-1.8797(32)$, triple excitations: $0.4739(92)$, quadruple excitations: $-0.0122(45)$, pentuple excitations: $-0.0050(24)$, and finally hextuple excitations: $-0.00009(5)$. In this light, the contributions of higher-order excitations can be neglected, as they are most likely smaller than the combined uncertainty of other terms. Therefore, the contribution of septuple and higher excitations is not considered in this work.

\subsection{Core correlation contribution}

In the calculations discussed in the previous subsection, we adopted the frozen-core approximation, neglecting the correlation contribution from the $1s^2\,2s^2\,2p^6$ core orbitals. The influence of the core correlation is expected to be small, but nonetheless non-negligible within the present accuracy goals.

\begin{table}[t]
\caption{\label{tab:cc3core}
Contribution the core-core and core-valence correlations to the static polarizability and dispersion coefficients or argon calculated at the CC3 level of theory, $\Delta\alpha_n^{\mathrm{AE\mbox{-}CC3}}$, using the dac$X$Z basis sets.
}
\begin{ruledtabular}
\begin{tabular}{ccccccc}
 $X$ & 
 $\Delta\alpha_0^{\mathrm{AE\mbox{-}CC3}}$ &
 $\Delta\alpha_2^{\mathrm{AE\mbox{-}CC3}}$ &
 $\Delta\alpha_4^{\mathrm{AE\mbox{-}CC3}}$ \\
 \hline\\[-1em]
2 & $-$0.0052 & $-$0.0294 & $-$0.1526 \\
3 & $-$0.0321 & $-$0.1274 & $-$0.5662 \\
4 & $-$0.0353 & $-$0.1383 & $-$0.6027 \\
5 & $-$0.0365 & $-$0.1348 & $-$0.5572 \\
6 & $-$0.0370 & $-$0.1315 & $-$0.5372 \\
 \hline
 $\infty$ & $-$0.0379(9) & $-$0.1264(50) & $-$0.5060(312) \\
\end{tabular}
\end{ruledtabular}
\end{table}

To eliminate this source of uncertainty, we carried out additional calculations at the CC3, CCSDT and CCSDTQ levels of theory with all electrons correlated. The corrections accounting for the core-core and core-valence correlations are defined as the difference between the results obtained with all electrons correlated and with frozen $1s^2\,2s^2\,2p^6$ orbitals. These corrections are denoted as, for example, $\Delta\alpha_n^{\mathrm{AE\mbox{-}CC3}}$ in the case of the core orbitals correction calculated using the CC3 method.

In Table~\ref{tab:cc3core} we report results of the calculations of the $\Delta\alpha_n^{\mathrm{AE\mbox{-}CC3}}$ correction using the modified dac$X$Z basis sets that include the tight functions with large Gaussian exponents for better description of the core region of the wavefunction. The results are extrapolated to the complete basis set limit using the formula~(\ref{riemann}). The corresponding uncertainty is estimated as the difference between the extrapolated value and the result obtained with the largest basis set available.

It is also necessary to estimate the contribution of higher-order excitations to the core-core and core-valence correlation correction. Unfortunately, all-electron calculations using the CCSDT and CCSDTQ method are extremely costly. This is a result of a larger number of active electrons in comparison with valence-only computations ($8$ vs. $18$ active particles). Additionally, the dac$X$Z basis sets include the aforementioned tight functions and hence their total size is significantly increased in comparison with their valence counterparts. Because of these obstacles, we managed to perform all-electron CCSDT calculations only within $X=2,3$ basis sets, while for the CCSDTQ method we are limited solely to $X=2$.

Fortunately, the $\Delta\alpha_n^{\mathrm{AE\mbox{-}T}}$ and $\Delta\alpha_n^{\mathrm{AE\mbox{-}Q}}$ corrections are small and do not have to be calculated very accurately. In the former case two basis sets are available and hence the extrapolation towards the CBS is possible. However, the $X=2$ is not reliable enough to make such extrapolation beneficial. In fact, considering the CC3 data included in Table~\ref{tab:cc3core}, extrapolation from the $X=2,3$ basis set pair overestimates the limit by roughly $50\%$. On the other hand, the $X=3$ result has an error smaller than $15\%$. Therefore, a more accurate results is most likely obtained by simply taking the value of $\Delta\alpha_n^{\mathrm{AE\mbox{-}T}}$ obtained within the $X=3$ basis and assign a large uncertainty of $15\%$. This gives:
\begin{align}
\begin{split}
 \Delta\alpha_0^{\mathrm{AE\mbox{-}T}} &= -0.0051(8),\\
 \Delta\alpha_2^{\mathrm{AE\mbox{-}T}} &= -0.0516(78),\\
 \Delta\alpha_4^{\mathrm{AE\mbox{-}T}} &= -0.3167(476),
\end{split}
 \end{align}
Finally, we consider the $\Delta\alpha_n^{\mathrm{AE\mbox{-}Q}}$ correction, where only one basis set is available. To estimate the CBS limit of this correction we assume that it converges at the same rate as the CC3 contribution. The limit is then obtained by scaling the $\Delta\alpha_n^{\mathrm{AE\mbox{-}Q}}$ correction obtained within the $X=2$ basis by the ratio of the $\Delta\alpha_n^{\mathrm{AE\mbox{-}CC3}}$ contributions as follows
\begin{align}
\Delta\alpha_n^{\mathrm{AE\mbox{-}Q}}(\mathrm{CBS}) =
\frac{\Delta\alpha_n^{\mathrm{AE\mbox{-}CC3}}(\mathrm{CBS})}{\Delta\alpha_n^{\mathrm{AE\mbox{-}CC3}}(X\mathrm{=}2)}\,\Delta\alpha_n^{\mathrm{AE\mbox{-}Q}}(X\mathrm{\mathrm{=}2}).
\end{align}
This leads to the following estimates:
\begin{align}
\begin{split}
 \Delta\alpha_0^{\mathrm{AE\mbox{-}Q}} &= -0.0006(3),\\
 \Delta\alpha_2^{\mathrm{AE\mbox{-}Q}} &= \phantom{-}0.0014(7),\\
 \Delta\alpha_4^{\mathrm{AE\mbox{-}Q}} &= \phantom{-}0.0554(277),
\end{split}
\end{align}
where we assigned an uncertainty of $50\%$ to the values obtained by scaling. The error of the $\Delta\alpha_n^{\mathrm{AE\mbox{-}Q}}$ contribution obtained is this way is large, but in absolute terms this has little influence on the overall uncertainty of our predictions.

\section{Relativistic corrections to the polarizability}
\label{sec:relativ}

\begin{table*}[t]
\caption{\label{tab:a0rel}
Relativistic corrections to the static polarizability of the argon atom calculated using the doubly-augmented dac$X$Z basis sets at the all-electron CC3 level of theory. In the last row we provide results extrapolated to the complete basis set limit according to Eq.~(\ref{riemann}) and the corresponding error estimate (see the main text for the discussion).
}
\begin{ruledtabular}
\begin{tabular}{cccccccccc}
 & 
 \multicolumn{4}{c}{one-electron corrections} &
 \multicolumn{2}{c}{two-electron corrections} \\
 $X$
 & $\Delta\alpha_0^{\mathrm{MV}}$
 & $\Delta\alpha_0^{\mathrm{D1}}$
 & total Cowan-Griffin
 & $\Delta\alpha_0^{\mathrm{DKH2}}$
 & $\Delta\alpha_0^{\mathrm{D2}}$ 
 & $\Delta\alpha_0^{\mathrm{B}}$ \\
\hline\\[-1em]
2 & $-$0.1380 & 0.1572 & 0.0192 & 0.0198 & 0.0002 & 0.0061 \\
3 & $-$0.1474 & 0.1667 & 0.0192 & 0.0198 & 0.0003 & 0.0066 \\
4 & $-$0.1452 & 0.1638 & 0.0186 & 0.0191 & 0.0002 & 0.0065 \\
5 & $-$0.1446 & 0.1629 & 0.0183 & 0.0188 & 0.0002 & 0.0065 \\
\hline
$\infty$ & $-$0.1439(7) & 0.1618(11) & 0.0179(3) & 0.0184(4)
  & 0.0001(1) & 0.0065(1) \\
\end{tabular}
\end{ruledtabular}
\end{table*}

\begin{table}[b]
\caption{\label{tab:andkh}
Relativistic corrections $\Delta\alpha_n^{\mathrm{DKH2}}$ obtained at the CC3 level of theory within the dac$X$Z basis sets (DKH2 effective Hamiltonian).
}
\begin{ruledtabular}
\begin{tabular}{ccccccc}
 $X$ & 
 $\Delta\alpha_0^{\mathrm{DKH2}}$ &
 $\Delta\alpha_2^{\mathrm{DKH2}}$ &
 $\Delta\alpha_4^                                                                                                                                                                                                                 {\mathrm{DKH2}}$ \\
 \hline\\[-1em]
2 & 0.0198 & 0.1859 & 1.3507 \\
3 & 0.0198 & 0.1832 & 1.3331 \\
4 & 0.0191 & 0.1736 & 1.2453 \\
5 & 0.0188 & 0.1701 & 1.2173 \\
 \hline
 $\infty$ & 0.0184(4) & 0.1658(43) & 1.1827(346)
\end{tabular}
\end{ruledtabular}
\end{table}

To reach the required accuracy level in theoretical determination of the polarizability of argon, relativistic corrections have to be considered. Indeed, even for the neon atom which is much lighter, the relativistic effects constitute about two parts per thousand of the total value. Our framework for calculation of the relativistic contributions to the static polarizability and dispersion coefficients is based on two alternative methods. The first one is the standard first-order perturbation theory based on the Breit-Pauli Hamiltonian~\cite{bethe75}
\begin{align}
\label{bph}
 \hat{H}_{\mathrm{BP}} = \hat{P}_4 + \hat{D}_1 + \hat{D}_2 + \hat{B},
\end{align}
where the operators appearing above are defined as
\begin{align}
 \label{p4}
 \hat{P}_4 = -\frac{1}{8c^2}\,\sum_i \nabla_i^4,
\end{align}
\begin{align}
 \label{d1}
 \hat{D}_1 = \frac{\pi}{2c^2} Z\, \sum_i \delta(\textbf{r}_{ia}),
\end{align}
\begin{align}
 \label{d2}
 \hat{D}_2 = \frac{\pi}{c^2} \sum_{i>j}\delta(\textbf{r}_{ij}),
\end{align}
\begin{align}
 \label{bb}
 \hat{B} = \frac{1}{2c^2}\sum_{i>j} \left[\frac{\nabla_i\cdot\nabla_j}{r_{ij}}
+\frac{\textbf{r}_{ij}\cdot(\textbf{r}_{ij}\cdot\nabla_j)\nabla_i}{r_{ij}^3} \right],
\end{align}
where $Z$ is the nuclear charge and $c$ denotes the speed of light in vacuum. We avoid the use of the fine-structure constant  $\alpha$ in this work as it may easily be confused with the polarizability. The corrections to the polarizability and dispersion coefficients resulting from the operators~(\ref{bph}) shall be denoted by the symbol $\Delta\alpha_n^{\mathrm{X}}$, where X in the superscript denotes the perturbing operator. Following the usual convention, we refer to these corrections as mass-velocity (X=MV), one-electron Darwin (X=D1), two-electron Darwin (X=D2) and Breit (X=B), in the same order as appearing in Eq.~(\ref{bph}). The sum of the first two corrections (MV and D1) is referred to as the Cowan-Griffin (CG) correction~\cite{cowan76}. Another frequently used name for the $\Delta\alpha_n^{\mathrm{B}}$ term is the orbit-orbit correction, but we refer to this quantity as the Breit correction for consistency with previous works. The expectation values of the operators in Eqs.~(\ref{p4})-(\ref{bb}) are calculated analytically at the CCSD(T) level of theory as described in Ref.~\cite{coriani04}.

The second approach to determination of the relativistic corrections is based on the Douglas-Kroll-Hess~\cite{douglas74,hess85,reiher06} theory of the second order (DKH2). In this method, the one-electron part of the Hamiltonian is replaced by an effective operator arising from a specific decoupling transformation applied to the Dirac equation for one-electron systems. In the DKH2 variant the decoupling is carried out to the second order in the external potential. The practical advantage of the DKH2 Hamiltonian is the fact that is can easily be used together with any method that is able to calculate polarizabilities. The same is not true for the Breit-Pauli Hamiltonian; to the best of our knowledge, analytic calculation of the full BP correction to the polarizability (and dispersion coefficients) is not implemented in any electronic structure package. On the other hand, the DKH2 completely neglects the two-electron corrections (D2 and B), but in comparison with the Cowan-Griffin approximation it includes terms of orders higher than $1/c^2$. The relativistic correction obtained using the DKH2 method is denoted by the symbol $\Delta\alpha_n^{\mathrm{DKH2}}$.

The Breit-Pauli correction to the static polarizability was calculated using the finite-field approach. The electric field of a small finite strength was added to the Hamiltonian and the second derivative of the Breit-Pauli corrections was extracted using the simplest finite difference formula. The strength of the electric field within range $[0.00,0.01]$ were considered and for a wide interval around ca. $0.075$ the results were stable to four significant digits. This strength of the field was applied in all calculations reported here. In Table~\ref{tab:a0rel} we show relativistic corrections to the static polarizability of argon calculated using the Breit-Pauli and DKH2 approaches (all-electron CC3 method within dac$X$Z basis sets, $X=2,\ldots,5$). The results were extrapolated to the CBS limit using the formula~(\ref{riemann}). The error was estimated as a difference between the value extrapolated using the $X=4,5$ basis set pair and the raw results obtained within $X=5$ basis. Only for the D2 correction a modification of this procedure is required -- it is known~\cite{kutz08} that this correction converges to the CBS limit as $X^{-1}$, and hence proper changes to Eq.~(\ref{riemann}) were introduced similarly as in Refs.~\cite{middendorf12,bischoff10,otto97,przybytek10,przybytek17,cencek12}.

First, let us consider the differences between the Cowan-Griffin and DKH2 corrections. The error of both methods is of the order $1/c^4$, and hence we expect them to give a similar answer, provided that the perturbation theory remains valid for argon. One can see from Table~\ref{tab:a0rel} that both methods agree within their mutual error estimates. Based on that, we conclude that DKH2 is a reliable method for calculation of the one-electron relativistic corrections and we apply it also in calculations of the dispersion coefficients. The obtained results of the $\Delta\alpha_n^{\mathrm{DKH2}}$ corrections are given in Table~\ref{tab:andkh} with the same extrapolation and error estimation method as for the static polarizability.

From Table~\ref{tab:a0rel} we can also judge the importance of two-electron relativistic effects in the present context. The two-electron Darwin correction is entirely negligible. Its contribution is smaller than the uncertainty of other corrections. We tacitly assume that the same is true for the dispersion coefficient and hence omit it in further analysis. However, the situation is entirely different in the case of the Breit correction. Indeed, due to significant cancellation between the mass-velocity and one-electron Darwin corrections (which have opposite signs), the Breit correction is only about three times smaller than the total Cowan-Griffin correction. Such phenomena appears to be a common feature in calculations for many-electron systems. Moreover, the contribution of the one-electron relativistic effects increases in magnitude (on a relative basis) for the dispersion coefficients in comparison with the static polarizability (roughly $0.6\%$ and $1.5\%$ for $\alpha_2$ and $\alpha_4$, respectively, while only $0.2\%$ for $\alpha_0$). We can expect that the same is true for the Breit correction and hence the omission of this quantity in determination of the dispersion coefficient would significantly increase the overall error of our results. As mentioned above, no implementation of the Breit correction to the dispersion coefficients has been reported yet and the standard finite-field approach is not applicable to the frequency-dependent quantities.

\begin{table}[b]
\caption{\label{tab:breit0}
The relativistic Breit corrections to the static polarizability ($\Delta\alpha_0^{\mathrm{B}}$) obtained at various levels of theory (all electrons correlated where applicable) using the dac$X$Z basis sets.
}
\begin{ruledtabular}
\begin{tabular}{ccccccc}
 $X$ & 
 Hartree-Fock &
 MP2 method &
 CC3 method \\
 \hline\\[-1em]
2 & 0.0062 & 0.0061 & 0.0061 \\
3 & 0.0066 & 0.0067 & 0.0066 \\
4 & 0.0066 & 0.0066 & 0.0065 \\
5 & 0.0066 & 0.0066 & 0.0065 \\
\end{tabular}
\end{ruledtabular}
\end{table}

In order to circumvent this problem, let us first analyze the results from Table~\ref{tab:a0rel} more closely. A striking feature of the results obtained for the Breit correction is the fast convergence with respect to the basis set size. Indeed, even within the smallest dac$2$Z basis, the accuracy of the calculated Breit correction would be acceptable (with wider error bars). The apparent insensitivity of the Breit correction to the quality of the basis set suggests that the dynamic correlation effects, which typically require high angular momenta to achieve convergence, may not be important for this quantity. To verify this hypothesis we recomputed the Breit correction to the static polarizability using the Hartee-Fock theory which includes no dynamic correlation, as well as the MP2 theory which is the simplest correlated method. In Table~\ref{tab:breit0} the obtained results are compared with the CC3 data reproduced from Table~\ref{tab:a0rel} for ease of comparison. Results given in Table~\ref{tab:breit0} confirm our hypothesis that the correlation contribution to the Breit correction is tiny and the Hartee-Fock method provides entirely satisfactory accuracy.

We assume that the unimportance of the correlation contributions to the Breit correction holds true also for the dispersion coefficients. In Appendix we develop analytic equations that allow to calculate this correction to the frequency-dependent polarizability at the coupled Hartree-Fock level of theory. From these calculations we obtain
\begin{align}
 \begin{split}
  &\Delta\alpha_2^{\mathrm{B}} = 0.0530(26) \\
  &\Delta\alpha_4^{\mathrm{B}} = 0.2750(138).
 \end{split}
\end{align}
The uncertainty estimates are based on small contribution of the electron correlation to the Breit correction, which amount to only about $1\%$ for the static polarizability. In the case of $\Delta\alpha_2^{\mathrm{B}}$ and $\Delta\alpha_4^{\mathrm{B}}$ we conservatively assumed that they contribute by no more than $5\%$. As expected, the Breit correction to the dispersion coefficients is sizeable, constituting about a quarter of the total relativistic contribution. Therefore, omission of this term would significantly increase our final error.

Having included all effects of the order of $1/c^2$ we should consider the possible significance of the 
relativistic effects of the order of $1/c^4$. These effects originate from    higher-order terms in  
the Foldy-Wouthuysen transformation of the  Dirac equation and from the second-order contribution  
from the Breit-Pauli Hamiltionan~\cite{pachucki06}.
The required calculations are very complicated even for the ground state of helium atom and so far have  not been performed  for the polarizability of helium.  
To gauge the magnitude of these $1/c^4$ effects 
in argon we considered the effect of the second-order 
spin-orbit interaction on the atomic polarizability. This interaction vanishes in the first-order of perturbation theory and hence was not included in the Breit-Pauli Hamiltonian, Eq.~(\ref{bph}). Nonetheless, the spin-dependent terms enter in higher orders by coupling triplet electronic excitations to the singlet ground state. While such terms are expected to be small, there is no reason to neglect them \emph{a priori}. Unfortunately, rigorous evaluation of the contribution of the second-order spin-orbit  interaction in argon 
is  computationally unfeasible at present. Therefore, to estimate the magnitude of the spin-orbit contributions we performed fully relativistic Hartree-Fock calculations based on the four-component Dirac-Coulomb Hamiltonian as implemented in the \textsc{Dirac} program~\cite{saue20,dirac23}. In order to extract the spin-dependent contributions to the static polarizability, two sets of calculations were performed. The first set was based on the conventional Dirac-Coulomb Hamiltonian, while in the second the spin-dependent terms were eliminated using the method of Dyall~\cite{dyall94}. The spin-dependent contribution to the static polarizability, denoted $\Delta\alpha_0^{\mathrm{SO}}$ further in the text, was obtained as a difference of the corresponding results from two sets. In the calculations we used the uncontracted basis sets from Sec.~\ref{sec:nrel} to expand the large component of the spinor. While these basis sets were optimized in the non-relativistic framework and hence are sub-optimal in the four-component calculations, this is acceptable for relatively light systems such as argon atom. The small-component basis was generated automatically using the restricted kinetic balance prescription.

The spin-dependent contributions to the static polarizability converge rapidly with respect to the size of the basis set. For example, the results obtained with da$4$z and da$5$z differ by merely one part per thousand, and the difference between da$5$z and da$6$z is by an order of magnitude smaller. Therefore, we adopt the value obtained within the da$6$z basis as the final result. The major contribution to the uncertainty of this quantity comes from the neglected correlation effects. To account for this, we adopt a conservative $20\%$ error estimate. This gives the final spin-dependent contribution to the static polarizability equal to
\begin{align}
    \Delta\alpha_0^{\mathrm{SO}} = 0.0012(2).
\end{align}
This correction is smaller than the combined uncertainty of other contributions.  In the case of the dispersion coefficients, these uncertainties are significantly larger on a relative basis, and hence the spin-dependent terms can be neglected.

\section{Quantum electrodynamics corrections to the polarizability}
\label{sec:qed}

The next contributions to the polarizability and dispersion coefficients originates from the quantum electrodynamics (QED) effects, $\Delta\alpha_n^{\mathrm{QED}}$. In this work we apply the following correction~\cite{caswell86,pachucki93,pachucki98}
\begin{align}
\label{anqed}
 \Delta\alpha_n^{\mathrm{QED}} = \frac{8}{3\pi\,c} \Big(\frac{19}{30}+2\ln c-\ln k_0\Big)\Delta\alpha_n^{\mathrm{D1}},
\end{align}
where $\Delta\alpha_n^{\mathrm{D1}}$ is the relativistic D1 correction calculated in the previous section and $\ln k_0$ is the so-called Bethe logarithm~\cite{bethe75,schwartz61} (related to the mean-excitation energy of the system). In comparison with the rigorous non-relativistic quantum electrodynamics (NRQED) theory~\cite{caswell86,pachucki93,pachucki98}, several approximations were adopted to arrive at Eq.~(\ref{anqed}). First, the two-electron QED relativistic corrections were neglected. There are two corrections of this type; the first is essentially the D2 relativistic correction scaled by a small numerical factor. Taking into account that the D2 correction to the polarizability is already negligible, there is little point in including the corresponding QED correction. The second two-electron QED correction is the so-called Araki-Sucher term~\cite{araki57,sucher58}. While this contribution can be calculated within the Gaussian basis set~\cite{balcerzak17,lesiuk19a,jaquet20,czachor20}, it is typically even smaller than the D2 correction and hence entirely omissible. Another approximation used in Eq.~(\ref{anqed}) is neglect of the  external  electric field dependence of the Bethe logarithm. As discussed at length in Ref.~\cite{lesiuk20}, $\ln k_0$ is sensitive primarily to the electronic wavefunction in the region close to the nucleus. This regime is dominated by the strong electric field generated by the nucleus and hence the influence of the (perturbatively small) external electric field is very small. The excellent agreement between theory and experiment for the polarizability of the neon atom~\cite{lesiuk20,hellmann22}, where the same approximation was adopted in the calculations, confirms that the field dependence of the Bethe logarithm is indeed tiny. The same conclusion was reached in calculations for the helium atom where  the electric-field derivative of the Bethe logarithm  was calculated rigorously~\cite{lach04,puchalski20}.

The Bethe logarithm for argon was calculated at the Hartee-Fock level of theory using the same formalism as in our previous work devoted to the neon atom~\cite{lesiuk20}. Details of these calculation will be reported in a separate publication. The value of the Bethe logarithm for argon adopted here reads
\begin{align}
 \ln k_0=8.7610.
\end{align}
Based on comparison with more accurate calculations for few-electron atoms, we estimate that the accuracy of this quantity is $1$-$2\%$ which does not contribute significantly to the overall error.

With all aforementioned approximations taken into account, calculation of the $\Delta\alpha_n^{\mathrm{QED}}$ correction amounts to scaling the appropriate $\Delta\alpha_n^{\mathrm{D1}}$ by a numerical factor of approximately $-0.0437$. In the case of the static polarizability we use the $\Delta\alpha_0^{\mathrm{D1}}$ calculated in the previous section. For the dispersion coefficients the $\Delta\alpha_n^{\mathrm{D1}}$ terms were calculated using the theory developed in the Appendix~A. This leads to the following contributions
\begin{align}
\begin{split}
 \Delta\alpha_0^{\mathrm{QED}} &= -0.0071(7), \\
 \Delta\alpha_2^{\mathrm{QED}} &= -0.0223(22), \\
 \Delta\alpha_4^{\mathrm{QED}} &= -0.1039(104),
\end{split}
\end{align}
where we adopted a conservative $10\%$ error bars to account for all approximations in Eq.~(\ref{anqed}).

It is also instructive to estimate the magnitude of the higher-order QED effects, $\Delta\alpha_n^{\mathrm{QED+}}$. It is well-known that the dominant QED contribution of the order $1/c^4$ is the so-called one-loop term~\cite{eides01}. In the present case it takes the form
\begin{align}
\label{anqedplus}
\Delta\alpha_n^{\mathrm{QED+}} = \frac{2Z}{c^2}\Big(\frac{427}{96}-2\ln2\Big)\Delta\alpha_n^{\mathrm{D1}}.
\end{align}
With the knowledge of the $\Delta\alpha_n^{\mathrm{D1}}$ calculated previously, the one-loop term can be obtain by scaling with the numerical factor of roughly $0.0059$. This gives the estimates:
\begin{align}
\begin{split}
 \Delta\alpha_0^{\mathrm{QED+}} &= 0.0010(2),\\
 \Delta\alpha_2^{\mathrm{QED+}} &= 0.0030(7) \\
 \Delta\alpha_4^{\mathrm{QED+}} &= 0.0140(35).
\end{split}
\end{align}
We adopt a wide error bars of $25\%$ to account for the missing $1/c^4$ QED terms.

\section{Finite nuclear mass and size corrections to the polarizability}
\label{sec:nuclear}

In all preceding calculations, the nucleus of the argon atom was effectively treated as a stationary point charge with infinite mass. For completeness, we here consider two corrections that go beyond this simple picture. First, we consider the finite nuclear size (FNS) correction $\Delta\alpha_n^{\rm FNS}$ which takes into account that the nucleus has a finite dimension. For many-electron atoms this correction to the static polarizability is calculated from the formula~\cite{puchalski10}
\begin{align}
 \label{a0fns}
 \Delta\alpha_0^{\rm FNS} = \frac{4}{3} \frac{\langle r_c^2\rangle}{\lambdabar^2} 
 \,\Delta\alpha_0^{\mathrm{D1}},
\end{align}
where $\langle r_c^2\rangle$ is the averaged square of the nuclear charge radius and $\lambdabar\approx 386.2\,$ fm is the reduced Compton wavelength of the electron. We employ the value $\langle r_c^2\rangle=11.512\,$fm$^2$ for the $^{40}$Ar isotope which was obtained in Ref.~\cite{devries87} using the two-parameter Fermi model of the nuclear charge distribution. The uncertainty of this quantity reported in Ref.~\cite{devries87} is negligible in the present context. Using the value $\Delta\alpha_0^{\mathrm{D1}}$ from Table~\ref{tab:a0rel} we find
\begin{align}
 \Delta\alpha_0^{\rm FNS} = 1.7\cdot 10^{-5}.
\end{align}
This correction is negligible in comparison with other sources of error. Since there is no reason to believe that the FNS correction is substantially larger for the dispersion coefficients, it has been neglected in our analysis.

Next, we consider the finite nuclear mass (FNM) correction. In the case of the static polarizability, it can be determined from the formula for the diagonal Born-Oppenheimer correction (DBOC)~\cite{born55}
\begin{align}
\label{a0fnm}
 \Delta\alpha_0^{\rm FNM} = \frac{1}{2M_{\mathrm{nuc}}} \partial_\varepsilon^2\big|_{\varepsilon=0}\langle \Psi_0| \nabla_{\mathrm{nuc}}^2|\Psi_0\rangle,
\end{align}
where $\Psi_0$ is the ground-state wavefunction, $M_{\mathrm{nuc}}$ is the nuclear mass, $\varepsilon$ denotes the strength of the external electric field, and $\nabla_{\mathrm{nuc}}$ is the gradient operator with respect to the coordinates of the nucleus. As we expect the contribution of the FNM correction to be relatively small, it is sufficient to calculate $\Delta\alpha_0^{\rm FNM}$ using the simplest correlated of theory which is the MP1 method described in Refs.~\cite{gauss06,tajti07}. The derivative with respect to the electric field in Eq.~(\ref{a0fnm}) is calculated using the finite-difference approach with the same settings as described in Sec.~\ref{sec:relativ}. The final value of the $\Delta\alpha_0^{\rm FNM}$ correction adopted here was obtained by extrapolating the results from the da$4$Z/da$5$Z basis set pair according to Eq.~(\ref{riemann}). It reads:
\begin{align}
 \Delta\alpha_0^{\rm FNM} = 1.9(3)\cdot10^{-4},
\end{align}
where the error estimate is equal to the difference between the extrapolated value and the result obtained within the da$5$Z basis set. This correction is essentially negligible in comparison with other sources of error and the same conclusion is most likely true for the dispersion coefficients, as well. Therefore, we neglect the FNM mass effects in determination of the polarizability dispersion.

\begin{table}[t]
\caption{\label{tab:chi0fc}
Valence coupled-cluster calculations of the mean square electron-nucleus distance, $\langle r^2\rangle$, for argon atom obtained within the da$X$Z basis set family.
}
\begin{ruledtabular}
\begin{tabular}{ccccccc}
 $X$ & 
 $\Delta\langle r^2\rangle_{\mathrm{SD(T)}}$ &
 $\Delta\langle r^2\rangle_{\mathrm{T}}$ &
 $\Delta\langle r^2\rangle_{\mathrm{Q}}$ &
 $\Delta\langle r^2\rangle_{\mathrm{P}}$ \\
 \hline\\[-1em]
2 & \phantom{$-$}0.2298 & \phantom{$-$}0.0035 & 0.0004 & $-$0.0006 \\
3 & \phantom{$-$}0.0909 & \phantom{$-$}0.0007 & 0.0007 & $-$0.0010 \\
4 & \phantom{$-$}0.0156 & $-$0.0008           & 0.0020 & --- \\
5 & $-$0.0104           & $-$0.0017 & --- & --- \\
6 & $-$0.0227           & $-$0.0019 & --- & --- \\
7 & $-$0.0290           & --- & --- & --- \\
8 & $-$0.0326           & --- & --- & --- \\
9 & $-$0.0349           & --- & --- & --- \\
 \hline
 $\infty$ & $-$0.0408(4) & $-$0.0022(3) & 0.0032(12) & $-$0.0012(2) \\
\end{tabular}
\end{ruledtabular}
\end{table}

\section{Magnetic susceptibility}
\label{sec:chi}

As discussed in Sec.~\ref{sec:overview}, the magnetic susceptibility does not have to be determined as accurately as the polarizability and relative accuracy of around $10\%$ is entirely sufficient. Therefore, in our treatment we neglect the frequency dependence of this quantity and concentrate solely of the static magnetic susceptibility, $\chi_0$. It is worth pointing out that for isolated atoms the frequency dependence of $\chi_0$ comes only from paramagnetic terms (which are minor in absolute terms) and hence it is highly unlikely that the frequency contribution to $\chi_0$ exceeds $1\%$ for argon, see Ref.~\cite{lesiuk20}.

Additionally, we neglect relativistic, QED and finite nuclear mass/size corrections to $\chi_0$. Note that calculation of these corrections is a significant challenge and has not been attempted thus far (without additional approximations to the theoretical formalism) even for the helium atom. Therefore, such calculations are beyond the scope of the present work and here we focus solely on the ``non-relativistic'' value of $\chi_0$. Parenthetically, we note that the use of the term ``non-relativistic'' may be viewed as a misnomer in this context, because the magnetic susceptibility in itself is of the order $1/c^2$ and hence vanishes in the non-relativistic limit, $c\rightarrow\infty$. However, the use of this name appears to be common in the literature and hence we follow this naming convention.

Neglecting terms of higher order in $1/c$ and assuming that the nucleus has an infinite mass, the atomic magnetic susceptibility is related to the mean square electron-nucleus distance through the following formula~\cite{bethe75}
\begin{align}
\label{chi0}
\chi_0 = -\frac{1}{6c^2}\,\langle \sum_i r_i^2 \rangle,
\end{align}
which has roots in the Langevin theory of diamagnetism~\cite{langevin1905}. In this section we focus on accurate determination of the value of $\langle\sum_i r_i^2 \rangle$ for argon. For brevity, we adopt a shorthand notation $\langle r^2\rangle\equiv\langle\sum_i r_i^2 \rangle$. As shall become apparent, in our calculations we include several corrections which are smaller than our stated accuracy goal and hence could possibly be neglected. Nonetheless, our motivation is to establish how accurately the non-relativistic value of $\chi_0$ can be determined at present. This provides an outlook as to how accurately the relativistic (and other) corrections must be computed in subsequent papers. In the calculations of $\langle r^2\rangle$ we adopt a similar strategy as for the non-relativistic contribution to the polarizability with only minor modifications. In particular, the same basis sets are used, including the augmented and core-valence functions, and the calculations are split into valence-only (frozen $1s^22s^22p^6$ core orbitals) and all-electron components.

First, we consider the Hartree-Fock contribution, denoted $\langle r^2\rangle_{\mathrm{HF}}$ further in the text. As the HF equations for atoms can be solved using a grid based approach with extremely high accuracy, there is little point in attempting to reproduce these results within a Gaussian basis. Therefore, we take $\langle r^2\rangle_{\mathrm{HF}}=26.0344$ from Ref.~\cite{saito09} which is essentially exact for our purposes to all digits given.

The second major contribution to $\langle r^2\rangle$ was calculated at the frozen-core CCSD(T) level of theory, $\Delta\langle r^2\rangle_{\mathrm{SD(T)}}$, using the da$X$Z basis sets. Similarly as for the polarizability, we found that further augmentation of the basis sets leads to tiny changes in the results which are not worth a significant increase of the computational time. Note that the CCSD(T) method is used here rather than the CC3 theory employed for the polarizability. This choice is justified by the observation that both CC3 and CCSD(T) have a similar accuracy, yet the latter is usually significantly less expensive due to the non-iterative treatment of the triple excitations. Such shortcut was not available in the case of the (dynamic) polarizability as this quantity is not well defined within the CCSD(T) model. In Table~\ref{tab:chi0fc} we report values of the $\Delta\langle r^2\rangle_{\mathrm{SD(T)}}$ correction calculated with basis sets $X=2,\ldots,9$. The CBS limit of this quantity is obtained by extrapolation using the formula~(\ref{riemann}) with $X=8,9$. The uncertainty is estimated as twice the difference between the extrapolated values from $X=8,9$ and $X=7,8$ basis set pairs, analogously as for the polarizability, see Sec.~\ref{sec:nrel}.

Next, we consider corrections to the magnetic susceptibility accounting for higher-order excitations with respect to the reference determinant. They are denoted by the symbols $\Delta\langle r^2\rangle_{\mathrm{T}}$ (the difference between CCSDT and CCSD(T) results), $\Delta\langle r^2\rangle_{\mathrm{Q}}$ (the difference between CCSDTQ and CCSDT), and so on. We consider corrections up to pentuple excitations, $\Delta\langle r^2\rangle_{\mathrm{P}}$, and higher-order corrections are neglected based on their small magnitude. For example, the $\Delta\langle r^2\rangle_{\mathrm{H}}$ correction calculated within the da2Z basis set amounts to only about $-1\cdot10^{-5}$. Even if one conservatively assumes that within this small basis the $\Delta\langle r^2\rangle_{\mathrm{H}}$ correction is underestimated by a factor of $20$, the resulting value is still smaller than the uncertainties of other contributions and hence can be safely neglected without increasing the overall error. Noting the rapid convergence of the results with respect to the excitation level, the same is true for contributions of even higher excitations.

The calculated higher-order contributions to the magnetic susceptibility are given in Table~\ref{tab:chi0fc}. The CBS limits are obtained by the standard extrapolation, Eq.~(\ref{riemann}), using the largest two basis sets available for a given quantity. However, because in calculation of these corrections we are unable to employ basis sets as large as for $\Delta\langle r^2\rangle_{\mathrm{SD(T)}}$, a more conservative uncertainty estimate is used. Namely, the error of the CBS limit is computed as a difference between the extrapolated value and the result obtained within the largest basis set feasible for a given quantity.

Next, we consider core-valence contribution to $\langle r^2\rangle$, defined as the difference between results obtained with all occupied orbitals correlated and with frozen $1s^22s^22p^6$ core orbitals. In determination of this correction we adopt analogous strategy as in the valence calculations, with the exception that dac$X$Z basis sets supplemented with additional tight functions are used. The core-valence corrections are denoted by the symbols $\Delta\langle r^2\rangle_{\mathrm{AE-SD(T)}}$, $\Delta\langle r^2\rangle_{\mathrm{AE-T}}$, and so on. The results of the calculations obtained in the same way as for the valence contribution are given in Table~\ref{tab:chi0ae}.

A somewhat surprising phenomena encountered when comparing results from Tables~\ref{tab:chi0fc}~and~\ref{tab:chi0ae} is the fact that the valence contribution calculated at the CCSD(T) level of theory is smaller (in absolute terms) than the corresponding core-valence contribution. This feature is observed only in the CCSD(T) calculations and absent in any other CC variant. Moreover, even looking at the Hartree-Fock reference function, the contribution of the $1s^22s^22p^6$ core orbitals is about two orders of magnitude smaller than of the valence shells. This unusual behavior of the correlation contribution at the CCSD(T) level of theory is somewhat unfortunate as the core corrections cannot be calculated with basis sets as large as in the valence calculations. As a result, the uncertainty of the $\Delta\langle r^2\rangle_{\mathrm{AE-SD(T)}}$ component actually dominates our error budget for the magnetic susceptibility. Larger core-valence basis sets need to be optimized in the future if a significant error reduction is desired. It is also worth pointing out that the $\Delta\langle r^2\rangle_{\mathrm{AE-T}}$ and $\Delta\langle r^2\rangle_{\mathrm{AE-Q}}$ corrections are essentially negligible at present.

\begin{table}[t]
\caption{\label{tab:chi0ae}
All-electron coupled-cluster calculations of the mean square electron-nucleus distance, $\langle r^2\rangle$, for argon atom obtained within the dac$X$Z basis set family.
}
\begin{ruledtabular}
\begin{tabular}{ccccccc}
 $X$ & 
 $\Delta\langle r^2\rangle_{\mathrm{AE-SD(T)}}$ &
 $\Delta\langle r^2\rangle_{\mathrm{AE-T}}$ &
 $\Delta\langle r^2\rangle_{\mathrm{AE-Q}}$ \\
 \hline\\[-1em]
2 & $-$0.0092 & $-$0.0003 & 0.0004 \\
3 & $-$0.0389 & $-$0.0005 & 0.0001 \\
4 & $-$0.0555 & $-$0.0001 & --- \\
5 & $-$0.0640 & ---       & --- \\
6 & $-$0.0681 & ---       & --- \\
 \hline
 $\infty$ & $-$0.0743(62) & 0.0002(3) & 0.0000(2) \\
\end{tabular}
\end{ruledtabular}
\end{table}

By summing all calculated contributions we obtain the final estimate of the mean square electron-nucleus distance in argon atom equal to
\begin{align}
 \langle r^2\rangle = 25.9193(64),
\end{align}
where the final error is calculated by adding squares of errors of individual contributions and taking the square root. According to Eq.~(\ref{chi0}), this translates to the following value of the magnetic susceptibility of argon
\begin{align}
\label{chi0-final-1}
 \chi_0 = -2.3004(6)\cdot10^{-4}.
\end{align}
We would like to stress that above result is based on purely ``non-relativistic'' formula~(\ref{chi0}) and the corresponding error estimate takes into account only the uncertainties in $\langle r^2\rangle$. Other corrections to $\chi_0$ such as relativistic, quantum electrodynamics, etc. are completely neglected and not included in the above error bars. Nonetheless, assuming the magnitude of these corrections is similar as for the static polarizability, one can conclude that the value given above is accurate to at least 1\%. As discussed in Sec.~\ref{sec:overview}, this level of accuracy is sufficient from the point of view of refractive coefficient measurements. In the subsequent section, the result given above is compared with the available literature data.

\section{Final results and discussion}
\label{sec:final}

\begin{table}[t]
\caption{\label{tab:budget}
The final error budget of the calculations of the static polarizability and dispersion coefficients for the argon atom.}
\begin{ruledtabular}
\begin{tabular}{lD{.}{.}{1.8}D{.}{.}{1.8}D{.}{.}{1.8}}
 & \multicolumn{1}{c}{$n=0$} & 
   \multicolumn{1}{c}{$n=2$} & 
   \multicolumn{1}{c}{$n=4$} \\
\hline\\[-2.8ex]
\multicolumn{4}{c}{non-relativistic valence $\big(3s^23p^6\big)$ contributions} \\
\hline\\[-2.8ex]
$\alpha_n^{\mathrm{HF}}$        & 11.4726(1)  & 25.6162(1)    & 78.9658(2) \\
$\Delta\alpha_n^{\mathrm{SD}}$  & -0.3642(4)  & 1.8797(32)    & 12.3670(99) \\
$\Delta\alpha_n^{\mathrm{CC3}}$ & -0.0031(4)  & 0.4427(68)    & 2.9018(473) \\
$\Delta\alpha_n^{\mathrm{T}}$   & 0.0007(2)   & 0.0312(62)    & 0.2764(553) \\
$\Delta\alpha_n^{\mathrm{Q}}$   & -0.0045(10) & -0.0122(45)   & -0.0549(64) \\
$\Delta\alpha_n^{\mathrm{P}}$   & -0.0007(3)  & -0.0050(24)   & -0.0386(193) \\
$\Delta\alpha_n^{\mathrm{H}}$   & 0.0000(1)   & 0.0000(1)     & 0.0002(1) \\
\hline\\[-2.8ex]
\multicolumn{4}{c}{non-relativistic core $\big(1s^22s^22p^6\big)$ correlation contributions} \\
\hline\\[-2.8ex]
$\Delta\alpha_n^{\mathrm{CC3}}$ & -0.0379(9)  & -0.1264(50)   & -0.5060(312) \\
$\Delta\alpha_n^{\mathrm{T}}$   & -0.0051(8)  & -0.0516(78)   & -0.3167(476) \\
$\Delta\alpha_n^{\mathrm{Q}}$   & -0.0006(3)  & 0.0014(7)     & 0.0554(277) \\
\hline
\multicolumn{4}{c}{relativistic and QED corrections} \\
\hline\\[-2.8ex]
$\Delta\alpha_n^{\mathrm{DKH2}}$ & 0.0184(4)  & 0.1658(43)    & 1.1827(346) \\
$\Delta\alpha_n^{\mathrm{D2}}$   & 0.0001(1)  & 
\multicolumn{1}{c}{---} & \multicolumn{1}{c}{---} \\
$\Delta\alpha_n^{\mathrm{B}}$    & 0.0065(1)  & 0.0530(26)    & 0.2750(138) \\
$\Delta\alpha_0^{\mathrm{SO}}$   & 0.0012(2) &
\multicolumn{1}{c}{---} & \multicolumn{1}{c}{---} \\
$\Delta\alpha_n^{\mathrm{QED}}$  & -0.0071(7) & -0.0223(22)   & -0.1039(104) \\
$\Delta\alpha_n^{\mathrm{QED+}}$ & 0.0010(2)  & 0.0030(7)     & 0.0140(35) \\
\hline
\multicolumn{4}{c}{other minor corrections} \\
\hline\\[-2.8ex]
$\Delta\alpha_n^{\mathrm{FNS}}$ & 0.0000(1)   & 0.0000(1)     & 0.0000(1) \\
$\Delta\alpha_n^{\mathrm{FNM}}$ & 0.0002(1)   & 
\multicolumn{1}{c}{---} & \multicolumn{1}{c}{---} \\
\hline\\[-3.0ex]
total                           & 11.0775(19) & 27.976(15)    & 95.02(11) \\
\hline\\[-2.5ex]
rel. accuracy                   & 1.7\cdot 10^{-4} & 5.5\cdot 10^{-4} & 1.1\cdot 10^{-3} \\
\vspace{-0.5cm}
\end{tabular}
\end{ruledtabular}
\end{table}

In Table~\ref{tab:budget} we present a summary of the theoretical results obtained in this work for the static polarizability and dispersion coefficients for argon. The final estimates (denoted ``total'' in Table~\ref{tab:budget}) are obtained by summing all relevant contributions. The total error is obtained by calculating sum of squares of individual uncertainties and taking the square root. This approach is justified by the standard error propagation formulas under the assumption that all contributions to the final results are independent variables in the statistical sense. 

In the case of the sixth-order dispersion coefficient, we used a simplified computational scheme where only the Hartree-Fock, valence CCSD and valence CC3 contributions are included. By summing these quantities we obtain the final estimate
\begin{align}
 \alpha_6 = 382.5.
\end{align}
Because the accuracy of $\alpha_6$ is not critical, we do not attempt a rigorous error estimation for this quantity. However, by analysing the impact of analogous approximations on the lower-order dispersion coefficients, it is safe to assume that the value of $\alpha_6$ given above has the relative error no larger than $10\%$. According to the discussion from Sec.~\ref{sec:overview}, this is entirely sufficient from the point of view of metrology.

\begin{table}[t]
\caption{\label{tab:literature}
Comparison with other theoretical and experimental literature values of $\alpha_n$. 
The error estimation is not present in cases where it has not been provided by the original authors.
All values are given in the atomic units.}
\begin{ruledtabular}
\begin{tabular}{lD{.}{.}{1.8}D{.}{.}{1.5}D{.}{.}{1.6}}
 & \multicolumn{1}{c}{$\alpha_0$} & \multicolumn{1}{c}{$\alpha_2$} & \multicolumn{1}{c}{$\alpha_4$} 
\\
\hline
\multicolumn{4}{c}{experimental or semi-empirical} \\
\hline
 Kumar and Thakkar~\cite{kumar10} 
 & 11.08(11)  & 27.89(28)  & 95.62(96)  \\
 Orcutt and Cole~\cite{orcutt67} 
 & 11.0753(54)   & \multicolumn{1}{c}{---} & \multicolumn{1}{c}{---} \\
 Buckley \emph{at al.}~\cite{buckley00} 
 & 11.0774(10)   & \multicolumn{1}{c}{---} & \multicolumn{1}{c}{---} \\
 Gaiser and Fellmuth~\cite{gaiser18} 
 & 11.077183(27) & \multicolumn{1}{c}{---} & \multicolumn{1}{c}{---} \\
\hline
\multicolumn{4}{c}{theoretical} \\
\hline
 Paw\l owski \emph{et al.}$^a$~\cite{pawlowski05} 
 & 11.102 & 27.996 & 94.846 \\
 Lupinetti \emph{et al.}$^b$~\cite{lupinetti05} 
 & 11.07 & \multicolumn{1}{c}{---} & \multicolumn{1}{c}{---} \\
\hline
this work & 11.0775(19) & 27.976(15) & 95.02(11) \\
\vspace{-0.5cm}
\end{tabular}
\end{ruledtabular}
\vspace{-0.3cm}
\begin{flushleft}
 $^a\,$CC3 level of theory, sextuple-zeta GTO basis; \\
 $^b\,$finite-field CCSD(T) calculations; 
\end{flushleft}
\end{table}

In order to verify the accuracy of the theoretical predictions, we first compare the final results obtained for the static polarizability with the experimental data. As expected, our value for $\alpha_0$ is significantly less accurate than the latest experiment of Gaiser and Fellmuth~\cite{gaiser18}. Nonetheless, the experimental value is within the error bars estimated by us. In fact, the relative error with respect to the data of Gaiser and Fellmuth~\cite{gaiser18} is about five times smaller than the uncertainty estimated from theory. This suggests than our error estimation protocol is conservative and leads to overestimation of the uncertainty, but may also be in part due to fortuitous error cancellation. Therefore, we are reluctant to arbitrary decrease our uncertainty estimates basing solely on this comparison.

Concerning the dispersion coefficients, the results provided by us appear to be the most accurate reported thus far. We improve the accuracy by more than an order of magnitude in comparison with the available data. Unfortunately, more accurate theoretical and/or experimental values for this quantities are not available. However, in the recent work by Egan~\emph{et al.}~\cite{egan19} the molar polarizability of argon was determined for a single laser frequency corresponding to the wavelength $\lambda_E=632.9908(2)\,$nm ($\omega_E=0.071\,981$ in the atomic units, red He-Ne
laser). After converting to the unit system used in the present work their result reads
\begin{align}
 \label{egan}
 \alpha_{\mathrm{exp.}}(\omega_E) = 11.224\,31(17).
\end{align}
To compare this value with the results obtained in the present work, we use the expansion~(\ref{dispexp}). For the static polarizability we adopt the value of Gaiser and Fellmuth~\cite{gaiser18}, while the dispersion coefficients are taken from Table~\ref{tab:budget}, and the value from Eq.(\ref{chi0-final-1}) is used for the magnetic susceptibility. Note that the results of Egan~\emph{et al.}~\cite{egan19} are based on laser refractometry experiments and hence the sum of static polarizability and magnetic susceptibility must be used for the frequency-independent component to allow for a meaningful comparison. The contributions of the sixth- and higher-order dispersion coefficients of the polarizability, as well as of the frequency dependence of the magnetic susceptibility, are negligible for the laser frequency under consideration. This leads to the following theoretical estimate:
\begin{align}
 \label{alpha-633}
 \alpha_{\mathrm{theory}}(\omega_E) = 11.224\,45(11).
\end{align}
As one can see, the theoretical and experimental results are in agreement. While the experimental value lies slightly outside the error bars of $\alpha_{\mathrm{theory}}(\omega_E)$, they are mutually within their combined uncertainty. It is also worth pointing out that the relative uncertainty of the theoretical data, roughly $10\,$ppm, is of comparable magnitude as of the experiment. This comparison proves that by combining the static polarizability determined by Gaiser and Fellmuth~\cite{gaiser18} with the dispersion and magnetic susceptibility derived from theory, one obtains the most reliable data for the polarizability at a finite frequency available in the literature. According to our analysis from Sec.~\ref{sec:overview} the data reported in this work is accurate enough to apply the same procedure to other experimentally-relevant laser wavelengths above roughly $450\,$nm. Therefore, we believe that the main results of this work, besides establishing a rigorous benchmark for other theoretical methods, will find their use in metrology and related fields.

\begin{table}[t]
\caption{\label{tab:literature-mag}
Comparison with other theoretical and experimental literature values of static magnetic susceptibility of argon. The error estimation is not present in cases where it has not been provided by the original authors. All values are given in the atomic units.}
\begin{ruledtabular}
\begin{tabular}{ll}
 & \multicolumn{1}{c}{$\chi_0$} \\
\hline
\multicolumn{2}{c}{experimental} \\
\hline\\[-3.0ex]
 Havens~\cite{havens33}
 & $-2.15(2)\cdot 10^{-4}$ \\
 Mann~\cite{mann36} 
 & $-2.19(2)\cdot 10^{-4}$ \\
 Abonnenc~\cite{abonnenc39}
 & $-2.15\cdot 10^{-4}$ \\
 \multirow{2}{*}{Barter \emph{el al.}~\cite{barter60}} 
 & $-2.16(2)^a\cdot 10^{-4}$ \\
 & $-2.16(15)^b\cdot 10^{-4}$ \\
\hline
\multicolumn{2}{c}{theoretical} \\
\hline\\[-3.0ex]
 Yoshizawa and Hada$^c$~\cite{yoshizawa09}
 & $-2.22\cdot 10^{-4}$ \\
 Ruud \emph{et al.}$^d$~\cite{ruud94} / Jaszu\'{n}ski \emph{et al.}$^d$~\cite{jaszunski95} 
 & $-2.31\cdot 10^{-4}$ \\
 Reinsch and Meyer$^e$~\cite{reinsch76} & $-2.32\cdot 10^{-4}$ \\
 Levy and Perdew$^f$~\cite{levy85} / Desclaux$^f$~\cite{desclaux73}   & $-2.30\cdot 10^{-4}$ \\
\hline
this work & $-2.30(2)\cdot10^{-4}$ \\
\vspace{-0.5cm}
\end{tabular}
\end{ruledtabular}
\vspace{-0.3cm}
\begin{flushleft}
 $^a\,$original error estimate from Ref.~\cite{barter60}; \\
 $^b\,$revised error estimate proposed in Ref.~\cite{rourke21}; \\
 $^c\,$MP2-DKH2$(V+\mathbf{A})$ method, $23s16p16d16f10g$ GTO basis; \\
 $^d\,$MCSCF calculations with $3s3p3d4s4p$ active orbitals; \\
 $^e\,$PNO-CEPA calculations, $14s11p4d$ GTO basis; \\
 $^f\,$numerical relativistic Dirac-Fock;
\end{flushleft}
\end{table}

Regarding the magnetic susceptibility, our final result reads
\begin{align}
\label{chi0-final-2}
 \chi_0 = -2.30(2)\cdot10^{-4},
\end{align}
where we have adopted a global 1\% uncertainty estimate to account for the missing relativistic, quantum electrodynamics, etc. corrections. In Table~\ref{tab:literature-mag} we compare this value with the experimental and theoretical data available in the literature. The most frequently cited experimental result is given in the work of Barter \emph{et al.}~\cite{barter60}, $-2.16(2)\cdot 10^{-4}$. However, it has to be pointed out that this result is not an independent measurement, but rather an arithmetic average of 3 previous experimental values~\cite{havens33,mann36,abonnenc39} used to calibrate the apparatus. It has recently been suggested~\cite{rourke21} that an issue with purity of argon gas in these three experiments could have been an additional source of error not accounted for in the uncertainty estimates. This led to the revised error estimate, $-2.16(15)\cdot 10^{-4}$, which we adopt in this work.

From Table~\ref{tab:literature-mag}, we see that all theoretical calculations reported in the literature, with the exception of the paper of Yoshizawa and Hada~\cite{yoshizawa09}, lead to a value $\chi_0=-2.30 \cdot 10^{-4}$ or lower. By comparison, the experimental results cluster around $\chi_0=-2.15\cdot 10^{-4}$, a difference of roughy $6-7\%$ in relative terms. In analogy with the current state of data for helium and neon, we strongly recommend that the current theoretical result~(\ref{chi0-final-2}) is used as an interim reference value. In future works, we plan to calculate the magnetic susceptibility of all noble gases with significantly higher accuracy, including all relevant physical effects beyond Eq.~(\ref{chi0}). We believe that this will establish a solid reference value for most applications. However, in order to validate and double-check the results, new independent measurements of the magnetic susceptibility of noble gases with modern setup and rigorous error control would be extremely valuable. The same is true for verification of theoretical results by a set of independent calculations, preferably within a different framework.

\section{Conclusions}
\label{sec:conclusion}

In this work we have reported first-principles theoretical calculations of the dipole polarizability  and magnetic susceptibility of the argon atom. Frequency-dependence of the latter is neglected, while for the former it is taken into account by means of power series expansion in terms of the so-called dispersion coefficients (Cauchy coefficients). This approach is sufficient in terms of accuracy for experimentally relevant wavelengths below the first resonant frequency.

In the reported calculations, we include all non-negligible physical effects including the relativistic, quantum electrodynamics, finite nuclear mass, and finite nuclear size corrections. The dominant non-relativistic clamped-nuclei contribution is computed using a hierarchy of coupled-cluster methods combined with Gaussian basis sets up to nonuple-zeta quality optimized specifically for this task. Relativistic effects are determined using either Breit-Pauli Hamiltonian or DKH effective approach, with excellent agreement between these two methods. Other minor corrections are calculated with help of the first-order perturbation theory.

The final results, with inclusion of all relevant physical effects, are $\alpha_0=11.0775(19)$ for the static polarizability and $\alpha_2=27.976(15)$ and $\alpha_4=95.02(11)$ for the second and fourth dispersion coefficients, respectively. We additionally determined the sixth-order dispersion coefficient, $\alpha_6 = 382.5$, but with a significantly larger uncertainty of about $10\%$.
Our result obtained for the static polarizability agrees (within the estimated uncertainty) with the most recent experimental data~\cite{gaiser18}, but is less accurate. The dispersion coefficients determined in this work appear to be most accurate in the literature, improving by more than an order of magnitude upon previous estimates. By combining the experimentally determined value of the static polarizability with the dispersion coefficients from our calculations, the polarizability of argon can be calculated with accuracy of around $10\,$ppm for wavelengths above roughly $450\,$nm. 

Additionally, in this work we calculate the static magnetic susceptibility of argon which relates the refractive index of dilute argon gas with its pressure. While our result for this quantity are less accurate than in the case of the polarizability, it provides a starting point for more rigorous calculations in the future. In subsequent papers, we shall report relativistic calculations of the magnetic susceptibility of noble atoms.

The results reported in this work increase the current knowledge of several fundamental properties of atomic argon. This is important from the point of view of quantum metrology, especially for a new pressure standard based on thermophysical properties of gaseous argon.

\begin{acknowledgments}
We thank Christian G\"{u}nz (PTB) and Allan Harvey (NIST) for insightful comments on the manuscript and to M. Przybytek (UW) for providing several types of integrals required in the relativistic calculations. This project (QuantumPascal project 18SIB04) has received funding from the EMPIR programme cofinanced by the Participating States and from the European Union’s Horizon 2020 research
and innovation program. The authors also acknowledge support from the National Science Center, Poland, within the Project No. 2017/27/B/ST4/02739. This research was supported in part by PLGrid Infrastructure through the computational grant \textsc{plgtdmcc}.
\end{acknowledgments}

\appendix
\section{Two-electron relativistic corrections to the dynamic polarizability}

In the following, the exact wavefunction is denoted by the symbol $|\Psi_0\rangle$ and the electronic Hamiltonian of the system by $H$. The exact ground state energy is denoted by $E_0$. The dynamic dipole polarizability at a real frequency $\omega$ (away from the resonant frequencies of the system) of the ground state is defined as
\begin{align}
 \alpha(\omega) = -\frac{1}{3}
 \langle \Psi_0|\mathbf{r}\,\frac{Q}{H-E_0+\omega}\,\mathbf{r}|\Psi_0\rangle + \mbox{g.h.c.},
\end{align}
where $Q=1-|\Psi_0\rangle\langle\Psi_0|$ is the projection operator onto the subspace orthogonal to $\Psi_0$ and $\mathbf{r}=\sum_i \mathbf{r}_i$ is the electronic dipole operator. Note that $Q$ commutes with the Hamiltonian of the system and any analytic function of $H$. The symbol ``g.h.c.'' denotes the generalized hermitian conjugation which amounts to exchanging wavefunctions in bra and ket, and reversing the sign of the frequency, i.e. $\omega\rightarrow-\omega$.

Let us define the first-order response function $\Psi_1$ by the formula
\begin{align}
 |\Psi_1\rangle = -\frac{Q}{H-E_0+\omega}\,\mathbf{r}|\Psi_0\rangle.
\end{align}
It can be obtained by solving the following equation
\begin{align}
\label{psi1}
 \big(H-E_0+\omega\big)|\Psi_1\rangle+\mathbf{r}\,|\Psi_0\rangle=0,
\end{align}
where for any operator $X$, the symbol $\langle X\rangle$ stands for the expectation value $\langle\Psi_0|X|\Psi_0\rangle$. With help of the response function the polarizability can be rewritten as
\begin{align}
 \alpha(\omega) = \frac{1}{3}\langle \Psi_0|\,\mathbf{r}\,|\Psi_1\rangle + \mbox{g.h.c.}
\end{align}
Assume that the Hamiltonian is modified by adding a small perturbation, i.e. $H\rightarrow H+\lambda V$, where $V$ is an operator and $\lambda$ controls the strength of the perturbation. When the perturbation is switched on, all quantities defined above become dependent on $\lambda$, but we do not write this explicitly. We are interested in the derivative of the polarizability with respect to $\lambda$ for $\lambda=0$, i.e. $\partial_\lambda\big|_{\lambda=0}\,\alpha(\omega)$.

The response of the exact wavefunction to the perturbation, $ |\Psi_V\rangle\equiv \partial_\lambda\big|_{\lambda=0}\,|\Psi_0\rangle$, is found by solving
\begin{align}
 \big( H-E_0 \big)\,|\Psi_V\rangle + \big(V-\langle V\rangle\big) |\Psi_0\rangle=0,
\end{align}
subject to the orthogonality condition $\langle \Psi_0|\Psi_V\rangle=0$. The derivative of the polarizability can be formally expressed as
\begin{align}
\begin{split}
 \partial_\lambda\big|_{\lambda=0}\,\alpha(\omega) &= 
 \frac{1}{3}\langle \Psi_0|\mathbf{r}|\partial_\lambda\big|_{\lambda=0}\Psi_1\rangle \\
 &+ \frac{1}{3}\langle \Psi_V|\mathbf{r}|\Psi_1\rangle + \mbox{g.h.c.}
\end{split}
\end{align}
The above expression is somewhat over-complicated as it involves the derivative of the response function with respect to $\lambda$. To eliminate this quantity we first note that according to Eq.~(\ref{psi1})
\begin{align}
\begin{split}
 &\langle \Psi_0|\mathbf{r}|\partial_\lambda\big|_{\lambda=0}\Psi_1\rangle = \\
&-\langle \Psi_1|\big(H-E_0+\omega\big)|\partial_\lambda\big|_{\lambda=0}\Psi_1\rangle.
\end{split}
\end{align}
Next, by differentiation of Eq.~(\ref{psi1}) with respect to $\lambda$ one can show that
\begin{align}
\begin{split}
 &-\langle \Psi_1|\big(H-E_0+\omega\big)|\partial_\lambda\big|_{\lambda=0}\Psi_1\rangle = \\
&\langle \Psi_1| \big(V-\langle V\rangle\big)|\Psi_1\rangle
+ \langle \Psi_1|\mathbf{r}|\Psi_V\rangle.
\end{split}
\end{align}
We are left with the final formula
\begin{align}
\begin{split}
 &\partial_\lambda\big|_{\lambda=0}\,\alpha(\omega) = 
 \frac{2}{3}\langle \Psi_1|\mathbf{r}|\Psi_V\rangle \\
& +\frac{1}{3}\langle \Psi_1| \big(V-\langle V\rangle\big)|\Psi_1\rangle + \mbox{g.h.c.}
\end{split}
\end{align}
In this formulation we have adopted no approximations thus far. However, in actual calculations we use the Hartree-Fock determinant as $|\Psi_0\rangle$. By applying the Slater-Condon rules and noting that $\mathbf{r}$ is a one-electron operator, one can show that the response function $|\Psi_1\rangle$ can be represented as a linear combination of singly-excited determinants. 
In our calculations the operator $V$ is a two-electron quantity, see Eq.~(\ref{bb}), and hence the perturbed wavefunction $|\Psi_V\rangle$ is expanded in terms of all singly- and doubly-excited determinants.

\bibliography{ar_polar}

\begin{thebibliography}{106}%
\makeatletter
\providecommand \@ifxundefined [1]{%
 \@ifx{#1\undefined}
}%
\providecommand \@ifnum [1]{%
 \ifnum #1\expandafter \@firstoftwo
 \else \expandafter \@secondoftwo
 \fi
}%
\providecommand \@ifx [1]{%
 \ifx #1\expandafter \@firstoftwo
 \else \expandafter \@secondoftwo
 \fi
}%
\providecommand \natexlab [1]{#1}%
\providecommand \enquote  [1]{``#1''}%
\providecommand \bibnamefont  [1]{#1}%
\providecommand \bibfnamefont [1]{#1}%
\providecommand \citenamefont [1]{#1}%
\providecommand \href@noop [0]{\@secondoftwo}%
\providecommand \href [0]{\begingroup \@sanitize@url \@href}%
\providecommand \@href[1]{\@@startlink{#1}\@@href}%
\providecommand \@@href[1]{\endgroup#1\@@endlink}%
\providecommand \@sanitize@url [0]{\catcode `\\12\catcode `\$12\catcode
  `\&12\catcode `\#12\catcode `\^12\catcode `\_12\catcode `\%12\relax}%
\providecommand \@@startlink[1]{}%
\providecommand \@@endlink[0]{}%
\providecommand \url  [0]{\begingroup\@sanitize@url \@url }%
\providecommand \@url [1]{\endgroup\@href {#1}{\urlprefix }}%
\providecommand \urlprefix  [0]{URL }%
\providecommand \Eprint [0]{\href }%
\providecommand \doibase [0]{http://dx.doi.org/}%
\providecommand \selectlanguage [0]{\@gobble}%
\providecommand \bibinfo  [0]{\@secondoftwo}%
\providecommand \bibfield  [0]{\@secondoftwo}%
\providecommand \translation [1]{[#1]}%
\providecommand \BibitemOpen [0]{}%
\providecommand \bibitemStop [0]{}%
\providecommand \bibitemNoStop [0]{.\EOS\space}%
\providecommand \EOS [0]{\spacefactor3000\relax}%
\providecommand \BibitemShut  [1]{\csname bibitem#1\endcsname}%
\let\auto@bib@innerbib\@empty
\bibitem [{\citenamefont {Gaiser}\ \emph {et~al.}(2020)\citenamefont {Gaiser},
  \citenamefont {Fellmuth},\ and\ \citenamefont {Sabuga}}]{gaiser20}%
  \BibitemOpen
  \bibfield  {author} {\bibinfo {author} {\bibfnamefont {C.}~\bibnamefont
  {Gaiser}}, \bibinfo {author} {\bibfnamefont {B.}~\bibnamefont {Fellmuth}}, \
  and\ \bibinfo {author} {\bibfnamefont {W.}~\bibnamefont {Sabuga}},\
  }\href@noop {} {\bibfield  {journal} {\bibinfo  {journal} {Nat. Phys.}\
  }\textbf {\bibinfo {volume} {16}},\ \bibinfo {pages} {177} (\bibinfo {year}
  {2020})}\BibitemShut {NoStop}%
\bibitem [{\citenamefont {Gaiser}\ \emph {et~al.}(2022)\citenamefont {Gaiser},
  \citenamefont {Fellmuth},\ and\ \citenamefont {Sabuga}}]{gaiser22}%
  \BibitemOpen
  \bibfield  {author} {\bibinfo {author} {\bibfnamefont {C.}~\bibnamefont
  {Gaiser}}, \bibinfo {author} {\bibfnamefont {B.}~\bibnamefont {Fellmuth}}, \
  and\ \bibinfo {author} {\bibfnamefont {W.}~\bibnamefont {Sabuga}},\
  }\href@noop {} {\bibfield  {journal} {\bibinfo  {journal} {Ann. Phys.}\
  }\textbf {\bibinfo {volume} {534}},\ \bibinfo {pages} {2200336} (\bibinfo
  {year} {2022})}\BibitemShut {NoStop}%
\bibitem [{\citenamefont {Mohr}\ \emph {et~al.}(2018)\citenamefont {Mohr},
  \citenamefont {Newell}, \citenamefont {Taylor},\ and\ \citenamefont
  {Tiesinga}}]{mohr18}%
  \BibitemOpen
  \bibfield  {author} {\bibinfo {author} {\bibfnamefont {P.~J.}\ \bibnamefont
  {Mohr}}, \bibinfo {author} {\bibfnamefont {D.~B.}\ \bibnamefont {Newell}},
  \bibinfo {author} {\bibfnamefont {B.~N.}\ \bibnamefont {Taylor}}, \ and\
  \bibinfo {author} {\bibfnamefont {E.}~\bibnamefont {Tiesinga}},\ }\href@noop
  {} {\bibfield  {journal} {\bibinfo  {journal} {Metrologia}\ }\textbf
  {\bibinfo {volume} {55}},\ \bibinfo {pages} {125} (\bibinfo {year}
  {2018})}\BibitemShut {NoStop}%
\bibitem [{\citenamefont {Fischer}(2019)}]{fischer19}%
  \BibitemOpen
  \bibfield  {author} {\bibinfo {author} {\bibfnamefont {J.}~\bibnamefont
  {Fischer}},\ }\href@noop {} {\bibfield  {journal} {\bibinfo  {journal} {Ann.
  Phys.}\ }\textbf {\bibinfo {volume} {531}},\ \bibinfo {pages} {1800304}
  (\bibinfo {year} {2019})}\BibitemShut {NoStop}%
\bibitem [{\citenamefont {{Machin}}(2019)}]{machin19}%
  \BibitemOpen
  \bibfield  {author} {\bibinfo {author} {\bibfnamefont {G.}~\bibnamefont
  {{Machin}}},\ }\href@noop {} {\bibfield  {journal} {\bibinfo  {journal} {IEEE
  Instru. Meas. Mag.}\ }\textbf {\bibinfo {volume} {22}},\ \bibinfo {pages}
  {17} (\bibinfo {year} {2019})}\BibitemShut {NoStop}%
\bibitem [{\citenamefont {Gaiser}\ \emph {et~al.}(2014)\citenamefont {Gaiser},
  \citenamefont {Fellmuth},\ and\ \citenamefont {Zandt}}]{fellmuth14}%
  \BibitemOpen
  \bibfield  {author} {\bibinfo {author} {\bibfnamefont {C.}~\bibnamefont
  {Gaiser}}, \bibinfo {author} {\bibfnamefont {B.}~\bibnamefont {Fellmuth}}, \
  and\ \bibinfo {author} {\bibfnamefont {T.}~\bibnamefont {Zandt}},\
  }\href@noop {} {\bibfield  {journal} {\bibinfo  {journal} {Int. J.
  Thermophys.}\ }\textbf {\bibinfo {volume} {35}},\ \bibinfo {pages} {395}
  (\bibinfo {year} {2014})}\BibitemShut {NoStop}%
\bibitem [{\citenamefont {Gaiser}\ \emph {et~al.}(2015)\citenamefont {Gaiser},
  \citenamefont {Zandt},\ and\ \citenamefont {Fellmuth}}]{gaiser15}%
  \BibitemOpen
  \bibfield  {author} {\bibinfo {author} {\bibfnamefont {C.}~\bibnamefont
  {Gaiser}}, \bibinfo {author} {\bibfnamefont {T.}~\bibnamefont {Zandt}}, \
  and\ \bibinfo {author} {\bibfnamefont {B.}~\bibnamefont {Fellmuth}},\
  }\href@noop {} {\bibfield  {journal} {\bibinfo  {journal} {Metrologia}\
  }\textbf {\bibinfo {volume} {52}},\ \bibinfo {pages} {S217} (\bibinfo {year}
  {2015})}\BibitemShut {NoStop}%
\bibitem [{\citenamefont {Guenz}\ \emph {et~al.}(2017)\citenamefont {Guenz},
  \citenamefont {Gaiser},\ and\ \citenamefont {Richter}}]{guenz17}%
  \BibitemOpen
  \bibfield  {author} {\bibinfo {author} {\bibfnamefont {C.}~\bibnamefont
  {Guenz}}, \bibinfo {author} {\bibfnamefont {C.}~\bibnamefont {Gaiser}}, \
  and\ \bibinfo {author} {\bibfnamefont {M.}~\bibnamefont {Richter}},\
  }\href@noop {} {\bibfield  {journal} {\bibinfo  {journal} {Meas. Sci.
  Technol.}\ }\textbf {\bibinfo {volume} {28}},\ \bibinfo {pages} {027002}
  (\bibinfo {year} {2017})}\BibitemShut {NoStop}%
\bibitem [{\citenamefont {Gaiser}\ and\ \citenamefont
  {Fellmuth}(2019)}]{gaiser19}%
  \BibitemOpen
  \bibfield  {author} {\bibinfo {author} {\bibfnamefont {C.}~\bibnamefont
  {Gaiser}}\ and\ \bibinfo {author} {\bibfnamefont {B.}~\bibnamefont
  {Fellmuth}},\ }\href@noop {} {\bibfield  {journal} {\bibinfo  {journal} {J.
  Chem. Phys.}\ }\textbf {\bibinfo {volume} {150}},\ \bibinfo {pages} {134303}
  (\bibinfo {year} {2019})}\BibitemShut {NoStop}%
\bibitem [{\citenamefont {Johnson}\ and\ \citenamefont
  {Cheng}(1996)}]{johnson96}%
  \BibitemOpen
  \bibfield  {author} {\bibinfo {author} {\bibfnamefont {W.~R.}\ \bibnamefont
  {Johnson}}\ and\ \bibinfo {author} {\bibfnamefont {K.~T.}\ \bibnamefont
  {Cheng}},\ }\href@noop {} {\bibfield  {journal} {\bibinfo  {journal} {Phys.
  Rev. A}\ }\textbf {\bibinfo {volume} {53}},\ \bibinfo {pages} {1375}
  (\bibinfo {year} {1996})}\BibitemShut {NoStop}%
\bibitem [{\citenamefont {Bhatia}\ and\ \citenamefont
  {Drachman}(1998)}]{bhatia98}%
  \BibitemOpen
  \bibfield  {author} {\bibinfo {author} {\bibfnamefont {A.~K.}\ \bibnamefont
  {Bhatia}}\ and\ \bibinfo {author} {\bibfnamefont {R.~J.}\ \bibnamefont
  {Drachman}},\ }\href@noop {} {\bibfield  {journal} {\bibinfo  {journal}
  {Phys. Rev. A}\ }\textbf {\bibinfo {volume} {58}},\ \bibinfo {pages} {4470}
  (\bibinfo {year} {1998})}\BibitemShut {NoStop}%
\bibitem [{\citenamefont {Pachucki}\ and\ \citenamefont
  {Sapirstein}(2000)}]{pachucki00}%
  \BibitemOpen
  \bibfield  {author} {\bibinfo {author} {\bibfnamefont {K.}~\bibnamefont
  {Pachucki}}\ and\ \bibinfo {author} {\bibfnamefont {J.}~\bibnamefont
  {Sapirstein}},\ }\href@noop {} {\bibfield  {journal} {\bibinfo  {journal}
  {Phys. Rev. A}\ }\textbf {\bibinfo {volume} {63}},\ \bibinfo {pages} {012504}
  (\bibinfo {year} {2000})}\BibitemShut {NoStop}%
\bibitem [{\citenamefont {Cencek}\ \emph {et~al.}(2001)\citenamefont {Cencek},
  \citenamefont {Szalewicz},\ and\ \citenamefont {Jeziorski}}]{cencek01}%
  \BibitemOpen
  \bibfield  {author} {\bibinfo {author} {\bibfnamefont {W.}~\bibnamefont
  {Cencek}}, \bibinfo {author} {\bibfnamefont {K.}~\bibnamefont {Szalewicz}}, \
  and\ \bibinfo {author} {\bibfnamefont {B.}~\bibnamefont {Jeziorski}},\
  }\href@noop {} {\bibfield  {journal} {\bibinfo  {journal} {Phys. Rev. Lett.}\
  }\textbf {\bibinfo {volume} {86}},\ \bibinfo {pages} {5675} (\bibinfo {year}
  {2001})}\BibitemShut {NoStop}%
\bibitem [{\citenamefont {\L{}ach}\ \emph {et~al.}(2004)\citenamefont
  {\L{}ach}, \citenamefont {Jeziorski},\ and\ \citenamefont
  {Szalewicz}}]{lach04}%
  \BibitemOpen
  \bibfield  {author} {\bibinfo {author} {\bibfnamefont {G.}~\bibnamefont
  {\L{}ach}}, \bibinfo {author} {\bibfnamefont {B.}~\bibnamefont {Jeziorski}},
  \ and\ \bibinfo {author} {\bibfnamefont {K.}~\bibnamefont {Szalewicz}},\
  }\href@noop {} {\bibfield  {journal} {\bibinfo  {journal} {Phys. Rev. Lett.}\
  }\textbf {\bibinfo {volume} {92}},\ \bibinfo {pages} {233001} (\bibinfo
  {year} {2004})}\BibitemShut {NoStop}%
\bibitem [{\citenamefont {Puchalski}\ \emph {et~al.}(2011)\citenamefont
  {Puchalski}, \citenamefont {Jentschura},\ and\ \citenamefont
  {Mohr}}]{puchalski11}%
  \BibitemOpen
  \bibfield  {author} {\bibinfo {author} {\bibfnamefont {M.}~\bibnamefont
  {Puchalski}}, \bibinfo {author} {\bibfnamefont {U.~D.}\ \bibnamefont
  {Jentschura}}, \ and\ \bibinfo {author} {\bibfnamefont {P.~J.}\ \bibnamefont
  {Mohr}},\ }\href@noop {} {\bibfield  {journal} {\bibinfo  {journal} {Phys.
  Rev. A}\ }\textbf {\bibinfo {volume} {83}},\ \bibinfo {pages} {042508}
  (\bibinfo {year} {2011})}\BibitemShut {NoStop}%
\bibitem [{\citenamefont {Piszczatowski}\ \emph {et~al.}(2015)\citenamefont
  {Piszczatowski}, \citenamefont {Puchalski}, \citenamefont {Komasa},
  \citenamefont {Jeziorski},\ and\ \citenamefont {Szalewicz}}]{piszczu15}%
  \BibitemOpen
  \bibfield  {author} {\bibinfo {author} {\bibfnamefont {K.}~\bibnamefont
  {Piszczatowski}}, \bibinfo {author} {\bibfnamefont {M.}~\bibnamefont
  {Puchalski}}, \bibinfo {author} {\bibfnamefont {J.}~\bibnamefont {Komasa}},
  \bibinfo {author} {\bibfnamefont {B.}~\bibnamefont {Jeziorski}}, \ and\
  \bibinfo {author} {\bibfnamefont {K.}~\bibnamefont {Szalewicz}},\ }\href@noop
  {} {\bibfield  {journal} {\bibinfo  {journal} {Phys. Rev. Lett.}\ }\textbf
  {\bibinfo {volume} {114}},\ \bibinfo {pages} {173004} (\bibinfo {year}
  {2015})}\BibitemShut {NoStop}%
\bibitem [{\citenamefont {Puchalski}\ \emph {et~al.}(2016)\citenamefont
  {Puchalski}, \citenamefont {Piszczatowski}, \citenamefont {Komasa},
  \citenamefont {Jeziorski},\ and\ \citenamefont {Szalewicz}}]{puchalski16}%
  \BibitemOpen
  \bibfield  {author} {\bibinfo {author} {\bibfnamefont {M.}~\bibnamefont
  {Puchalski}}, \bibinfo {author} {\bibfnamefont {K.}~\bibnamefont
  {Piszczatowski}}, \bibinfo {author} {\bibfnamefont {J.}~\bibnamefont
  {Komasa}}, \bibinfo {author} {\bibfnamefont {B.}~\bibnamefont {Jeziorski}}, \
  and\ \bibinfo {author} {\bibfnamefont {K.}~\bibnamefont {Szalewicz}},\
  }\href@noop {} {\bibfield  {journal} {\bibinfo  {journal} {Phys. Rev. A}\
  }\textbf {\bibinfo {volume} {93}},\ \bibinfo {pages} {032515} (\bibinfo
  {year} {2016})}\BibitemShut {NoStop}%
\bibitem [{\citenamefont {Puchalski}\ \emph {et~al.}(2020)\citenamefont
  {Puchalski}, \citenamefont {Szalewicz}, \citenamefont {Lesiuk},\ and\
  \citenamefont {Jeziorski}}]{puchalski20}%
  \BibitemOpen
  \bibfield  {author} {\bibinfo {author} {\bibfnamefont {M.}~\bibnamefont
  {Puchalski}}, \bibinfo {author} {\bibfnamefont {K.}~\bibnamefont
  {Szalewicz}}, \bibinfo {author} {\bibfnamefont {M.}~\bibnamefont {Lesiuk}}, \
  and\ \bibinfo {author} {\bibfnamefont {B.}~\bibnamefont {Jeziorski}},\
  }\href@noop {} {\bibfield  {journal} {\bibinfo  {journal} {Phys. Rev. A}\
  }\textbf {\bibinfo {volume} {101}},\ \bibinfo {pages} {022505} (\bibinfo
  {year} {2020})}\BibitemShut {NoStop}%
\bibitem [{\citenamefont {Lesiuk}\ \emph {et~al.}(2020)\citenamefont {Lesiuk},
  \citenamefont {Przybytek},\ and\ \citenamefont {Jeziorski}}]{lesiuk20}%
  \BibitemOpen
  \bibfield  {author} {\bibinfo {author} {\bibfnamefont {M.}~\bibnamefont
  {Lesiuk}}, \bibinfo {author} {\bibfnamefont {M.}~\bibnamefont {Przybytek}}, \
  and\ \bibinfo {author} {\bibfnamefont {B.}~\bibnamefont {Jeziorski}},\
  }\href@noop {} {\bibfield  {journal} {\bibinfo  {journal} {Phys. Rev. A}\
  }\textbf {\bibinfo {volume} {102}},\ \bibinfo {pages} {052816} (\bibinfo
  {year} {2020})}\BibitemShut {NoStop}%
\bibitem [{\citenamefont {Hellmann}(2022)}]{hellmann22}%
  \BibitemOpen
  \bibfield  {author} {\bibinfo {author} {\bibfnamefont {R.}~\bibnamefont
  {Hellmann}},\ }\href@noop {} {\bibfield  {journal} {\bibinfo  {journal}
  {Phys. Rev. A}\ }\textbf {\bibinfo {volume} {105}},\ \bibinfo {pages}
  {022809} (\bibinfo {year} {2022})}\BibitemShut {NoStop}%
\bibitem [{\citenamefont {Gaiser}\ and\ \citenamefont
  {Fellmuth}(2018)}]{gaiser18}%
  \BibitemOpen
  \bibfield  {author} {\bibinfo {author} {\bibfnamefont {C.}~\bibnamefont
  {Gaiser}}\ and\ \bibinfo {author} {\bibfnamefont {B.}~\bibnamefont
  {Fellmuth}},\ }\href@noop {} {\bibfield  {journal} {\bibinfo  {journal}
  {Phys. Rev. Lett.}\ }\textbf {\bibinfo {volume} {120}},\ \bibinfo {pages}
  {123203} (\bibinfo {year} {2018})}\BibitemShut {NoStop}%
\bibitem [{\citenamefont {Rourke}(2021)}]{rourke21}%
  \BibitemOpen
  \bibfield  {author} {\bibinfo {author} {\bibfnamefont {P.~M.}\ \bibnamefont
  {Rourke}},\ }\href@noop {} {\bibfield  {journal} {\bibinfo  {journal} {J.
  Phys. Chem. Ref. Data}\ }\textbf {\bibinfo {volume} {50}},\ \bibinfo {pages}
  {033104} (\bibinfo {year} {2021})}\BibitemShut {NoStop}%
\bibitem [{\citenamefont {Lupinetti}\ and\ \citenamefont
  {Thakkar}(2005)}]{lupinetti05}%
  \BibitemOpen
  \bibfield  {author} {\bibinfo {author} {\bibfnamefont {C.}~\bibnamefont
  {Lupinetti}}\ and\ \bibinfo {author} {\bibfnamefont {A.~J.}\ \bibnamefont
  {Thakkar}},\ }\href@noop {} {\bibfield  {journal} {\bibinfo  {journal} {J.
  Chem. Phys.}\ }\textbf {\bibinfo {volume} {122}},\ \bibinfo {pages} {044301}
  (\bibinfo {year} {2005})}\BibitemShut {NoStop}%
\bibitem [{\citenamefont {Paw{\l}owski}\ \emph {et~al.}(2005)\citenamefont
  {Paw{\l}owski}, \citenamefont {J{\o}rgensen},\ and\ \citenamefont
  {H{\"a}ttig}}]{pawlowski05}%
  \BibitemOpen
  \bibfield  {author} {\bibinfo {author} {\bibfnamefont {F.}~\bibnamefont
  {Paw{\l}owski}}, \bibinfo {author} {\bibfnamefont {P.}~\bibnamefont
  {J{\o}rgensen}}, \ and\ \bibinfo {author} {\bibfnamefont {C.}~\bibnamefont
  {H{\"a}ttig}},\ }in\ \href@noop {} {\emph {\bibinfo {booktitle} {Advances in
  Quantum Chemistry}}},\ Vol.~\bibinfo {volume} {48}\ (\bibinfo  {publisher}
  {Elsevier},\ \bibinfo {year} {2005})\ pp.\ \bibinfo {pages}
  {9--21}\BibitemShut {NoStop}%
\bibitem [{\citenamefont {Bartlett}\ and\ \citenamefont
  {Musia\l{}}(2007)}]{bartlett07}%
  \BibitemOpen
  \bibfield  {author} {\bibinfo {author} {\bibfnamefont {R.~J.}\ \bibnamefont
  {Bartlett}}\ and\ \bibinfo {author} {\bibfnamefont {M.}~\bibnamefont
  {Musia\l{}}},\ }\href@noop {} {\bibfield  {journal} {\bibinfo  {journal}
  {Rev. Mod. Phys.}\ }\textbf {\bibinfo {volume} {79}},\ \bibinfo {pages} {291}
  (\bibinfo {year} {2007})}\BibitemShut {NoStop}%
\bibitem [{\citenamefont {Crawford}\ and\ \citenamefont
  {Schaefer~III}(2007)}]{crawford07}%
  \BibitemOpen
  \bibfield  {author} {\bibinfo {author} {\bibfnamefont {T.~D.}\ \bibnamefont
  {Crawford}}\ and\ \bibinfo {author} {\bibfnamefont {H.~F.}\ \bibnamefont
  {Schaefer~III}},\ }\enquote {\bibinfo {title} {An introduction to coupled
  cluster theory for computational chemists},}\ in\ \href@noop {} {\emph
  {\bibinfo {booktitle} {Rev. Comp. Chem.}}}\ (\bibinfo  {publisher} {John
  Wiley \& Sons, Ltd},\ \bibinfo {year} {2007})\ pp.\ \bibinfo {pages}
  {33--136}\BibitemShut {NoStop}%
\bibitem [{\citenamefont {Tiesinga}\ \emph {et~al.}(2022)\citenamefont
  {Tiesinga}, \citenamefont {Mohr}, \citenamefont {Newell},\ and\ \citenamefont
  {Taylor}}]{codata18}%
  \BibitemOpen
  \bibfield  {author} {\bibinfo {author} {\bibfnamefont {E.}~\bibnamefont
  {Tiesinga}}, \bibinfo {author} {\bibfnamefont {P.~J.}\ \bibnamefont {Mohr}},
  \bibinfo {author} {\bibfnamefont {D.~B.}\ \bibnamefont {Newell}}, \ and\
  \bibinfo {author} {\bibfnamefont {B.~N.}\ \bibnamefont {Taylor}},\
  }\href@noop {} {\emph {\bibinfo {title} {The 2018 CODATA Recommended Values
  of the Fundamental Physical Constants}}} (\bibinfo {year} {2018, accessed
  December 9, 2022}),\ \bibinfo {note} {available online at
  http://physics.nist.gov/constants}\BibitemShut {NoStop}%
\bibitem [{\citenamefont {Dunning}(1989)}]{dunning89}%
  \BibitemOpen
  \bibfield  {author} {\bibinfo {author} {\bibfnamefont {T.~H.}\ \bibnamefont
  {Dunning}},\ }\href@noop {} {\bibfield  {journal} {\bibinfo  {journal} {J.
  Chem. Phys.}\ }\textbf {\bibinfo {volume} {90}},\ \bibinfo {pages} {1007}
  (\bibinfo {year} {1989})}\BibitemShut {NoStop}%
\bibitem [{\citenamefont {Woon}\ and\ \citenamefont {Dunning}(1993)}]{woon93}%
  \BibitemOpen
  \bibfield  {author} {\bibinfo {author} {\bibfnamefont {D.~E.}\ \bibnamefont
  {Woon}}\ and\ \bibinfo {author} {\bibfnamefont {T.~H.}\ \bibnamefont
  {Dunning}},\ }\href@noop {} {\bibfield  {journal} {\bibinfo  {journal} {J.
  Chem. Phys.}\ }\textbf {\bibinfo {volume} {98}},\ \bibinfo {pages} {1358}
  (\bibinfo {year} {1993})}\BibitemShut {NoStop}%
\bibitem [{\citenamefont {Van~Mourik}\ and\ \citenamefont
  {Dunning~Jr.}(2000)}]{mourik00}%
  \BibitemOpen
  \bibfield  {author} {\bibinfo {author} {\bibfnamefont {T.}~\bibnamefont
  {Van~Mourik}}\ and\ \bibinfo {author} {\bibfnamefont {T.~H.}\ \bibnamefont
  {Dunning~Jr.}},\ }\href@noop {} {\bibfield  {journal} {\bibinfo  {journal}
  {Int. J. Quantum Chem.}\ }\textbf {\bibinfo {volume} {76}},\ \bibinfo {pages}
  {205} (\bibinfo {year} {2000})}\BibitemShut {NoStop}%
\bibitem [{\citenamefont {Dunning}\ \emph {et~al.}(2001)\citenamefont
  {Dunning}, \citenamefont {Peterson},\ and\ \citenamefont
  {Wilson}}]{dunning01}%
  \BibitemOpen
  \bibfield  {author} {\bibinfo {author} {\bibfnamefont {T.~H.}\ \bibnamefont
  {Dunning}}, \bibinfo {author} {\bibfnamefont {K.~A.}\ \bibnamefont
  {Peterson}}, \ and\ \bibinfo {author} {\bibfnamefont {A.~K.}\ \bibnamefont
  {Wilson}},\ }\href@noop {} {\bibfield  {journal} {\bibinfo  {journal} {J.
  Chem. Phys.}\ }\textbf {\bibinfo {volume} {114}},\ \bibinfo {pages} {9244}
  (\bibinfo {year} {2001})}\BibitemShut {NoStop}%
\bibitem [{\citenamefont {Peterson}\ and\ \citenamefont
  {Dunning}(2002)}]{peterson02}%
  \BibitemOpen
  \bibfield  {author} {\bibinfo {author} {\bibfnamefont {K.~A.}\ \bibnamefont
  {Peterson}}\ and\ \bibinfo {author} {\bibfnamefont {T.~H.}\ \bibnamefont
  {Dunning}},\ }\href@noop {} {\bibfield  {journal} {\bibinfo  {journal} {J.
  Chem. Phys.}\ }\textbf {\bibinfo {volume} {117}},\ \bibinfo {pages} {10548}
  (\bibinfo {year} {2002})}\BibitemShut {NoStop}%
\bibitem [{\citenamefont {Cinal}(2020)}]{cinal20}%
  \BibitemOpen
  \bibfield  {author} {\bibinfo {author} {\bibfnamefont {M.}~\bibnamefont
  {Cinal}},\ }\href@noop {} {\bibfield  {journal} {\bibinfo  {journal} {J.
  Math. Chem.}\ }\textbf {\bibinfo {volume} {58}},\ \bibinfo {pages} {1571}
  (\bibinfo {year} {2020})}\BibitemShut {NoStop}%
\bibitem [{\citenamefont {Aidas}\ \emph {et~al.}(2014)\citenamefont {Aidas},
  \citenamefont {Angeli}, \citenamefont {Bak}, \citenamefont {Bakken},
  \citenamefont {Bast}, \citenamefont {Boman}, \citenamefont {Christiansen},
  \citenamefont {Cimiraglia}, \citenamefont {Coriani}, \citenamefont {Dahle},
  \citenamefont {Dalskov}, \citenamefont {Ekstr\"{o}m}, \citenamefont
  {Enevoldsen}, \citenamefont {Eriksen}, \citenamefont {Ettenhuber},
  \citenamefont {Fern\'{a}ndez}, \citenamefont {Ferrighi}, \citenamefont
  {Fliegl}, \citenamefont {Frediani}, \citenamefont {Hald}, \citenamefont
  {Halkier}, \citenamefont {H\"{a}ttig}, \citenamefont {Heiberg}, \citenamefont
  {Helgaker}, \citenamefont {Hennum}, \citenamefont {Hettema}, \citenamefont
  {Hjerten\ae{}s}, \citenamefont {H\o{}st}, \citenamefont {H\o{}yvik},
  \citenamefont {Iozzi}, \citenamefont {Jans\'{i}k}, \citenamefont {Jensen},
  \citenamefont {Jonsson}, \citenamefont {J\o{}rgensen}, \citenamefont
  {Kauczor}, \citenamefont {Kirpekar}, \citenamefont {Kj\ae{}rgaard},
  \citenamefont {Klopper}, \citenamefont {Knecht}, \citenamefont {Kobayashi},
  \citenamefont {Koch}, \citenamefont {Kongsted}, \citenamefont {Krapp},
  \citenamefont {Kristensen}, \citenamefont {Ligabue}, \citenamefont
  {Lutn\ae{}s}, \citenamefont {Melo}, \citenamefont {Mikkelsen}, \citenamefont
  {Myhre}, \citenamefont {Neiss}, \citenamefont {Nielsen}, \citenamefont
  {Norman}, \citenamefont {Olsen}, \citenamefont {Olsen}, \citenamefont
  {Osted}, \citenamefont {Packer}, \citenamefont {Pawlowski}, \citenamefont
  {Pedersen}, \citenamefont {Provasi}, \citenamefont {Reine}, \citenamefont
  {Rinkevicius}, \citenamefont {Ruden}, \citenamefont {Ruud}, \citenamefont
  {Rybkin}, \citenamefont {Sa\l{}ek}, \citenamefont {Samson}, \citenamefont
  {de~Mer\'{a}s}, \citenamefont {Saue}, \citenamefont {Sauer}, \citenamefont
  {Schimmelpfennig}, \citenamefont {Sneskov}, \citenamefont {Steindal},
  \citenamefont {Sylvester-Hvid}, \citenamefont {Taylor}, \citenamefont
  {Teale}, \citenamefont {Tellgren}, \citenamefont {Tew}, \citenamefont
  {Thorvaldsen}, \citenamefont {Th\o{}gersen}, \citenamefont {Vahtras},
  \citenamefont {Watson}, \citenamefont {Wilson}, \citenamefont {Ziolkowski},\
  and\ \citenamefont {\AA{}gren}}]{daltonpaper}%
  \BibitemOpen
  \bibfield  {author} {\bibinfo {author} {\bibfnamefont {K.}~\bibnamefont
  {Aidas}}, \bibinfo {author} {\bibfnamefont {C.}~\bibnamefont {Angeli}},
  \bibinfo {author} {\bibfnamefont {K.~L.}\ \bibnamefont {Bak}}, \bibinfo
  {author} {\bibfnamefont {V.}~\bibnamefont {Bakken}}, \bibinfo {author}
  {\bibfnamefont {R.}~\bibnamefont {Bast}}, \bibinfo {author} {\bibfnamefont
  {L.}~\bibnamefont {Boman}}, \bibinfo {author} {\bibfnamefont
  {O.}~\bibnamefont {Christiansen}}, \bibinfo {author} {\bibfnamefont
  {R.}~\bibnamefont {Cimiraglia}}, \bibinfo {author} {\bibfnamefont
  {S.}~\bibnamefont {Coriani}}, \bibinfo {author} {\bibfnamefont
  {P.}~\bibnamefont {Dahle}}, \bibinfo {author} {\bibfnamefont {E.~K.}\
  \bibnamefont {Dalskov}}, \bibinfo {author} {\bibfnamefont {U.}~\bibnamefont
  {Ekstr\"{o}m}}, \bibinfo {author} {\bibfnamefont {T.}~\bibnamefont
  {Enevoldsen}}, \bibinfo {author} {\bibfnamefont {J.~J.}\ \bibnamefont
  {Eriksen}}, \bibinfo {author} {\bibfnamefont {P.}~\bibnamefont {Ettenhuber}},
  \bibinfo {author} {\bibfnamefont {B.}~\bibnamefont {Fern\'{a}ndez}}, \bibinfo
  {author} {\bibfnamefont {L.}~\bibnamefont {Ferrighi}}, \bibinfo {author}
  {\bibfnamefont {H.}~\bibnamefont {Fliegl}}, \bibinfo {author} {\bibfnamefont
  {L.}~\bibnamefont {Frediani}}, \bibinfo {author} {\bibfnamefont
  {K.}~\bibnamefont {Hald}}, \bibinfo {author} {\bibfnamefont {A.}~\bibnamefont
  {Halkier}}, \bibinfo {author} {\bibfnamefont {C.}~\bibnamefont {H\"{a}ttig}},
  \bibinfo {author} {\bibfnamefont {H.}~\bibnamefont {Heiberg}}, \bibinfo
  {author} {\bibfnamefont {T.}~\bibnamefont {Helgaker}}, \bibinfo {author}
  {\bibfnamefont {A.~C.}\ \bibnamefont {Hennum}}, \bibinfo {author}
  {\bibfnamefont {H.}~\bibnamefont {Hettema}}, \bibinfo {author} {\bibfnamefont
  {E.}~\bibnamefont {Hjerten\ae{}s}}, \bibinfo {author} {\bibfnamefont
  {S.}~\bibnamefont {H\o{}st}}, \bibinfo {author} {\bibfnamefont {I.-M.}\
  \bibnamefont {H\o{}yvik}}, \bibinfo {author} {\bibfnamefont {M.~F.}\
  \bibnamefont {Iozzi}}, \bibinfo {author} {\bibfnamefont {B.}~\bibnamefont
  {Jans\'{i}k}}, \bibinfo {author} {\bibfnamefont {H.~J.~{\relax Aa}.}\
  \bibnamefont {Jensen}}, \bibinfo {author} {\bibfnamefont {D.}~\bibnamefont
  {Jonsson}}, \bibinfo {author} {\bibfnamefont {P.}~\bibnamefont
  {J\o{}rgensen}}, \bibinfo {author} {\bibfnamefont {J.}~\bibnamefont
  {Kauczor}}, \bibinfo {author} {\bibfnamefont {S.}~\bibnamefont {Kirpekar}},
  \bibinfo {author} {\bibfnamefont {T.}~\bibnamefont {Kj\ae{}rgaard}}, \bibinfo
  {author} {\bibfnamefont {W.}~\bibnamefont {Klopper}}, \bibinfo {author}
  {\bibfnamefont {S.}~\bibnamefont {Knecht}}, \bibinfo {author} {\bibfnamefont
  {R.}~\bibnamefont {Kobayashi}}, \bibinfo {author} {\bibfnamefont
  {H.}~\bibnamefont {Koch}}, \bibinfo {author} {\bibfnamefont {J.}~\bibnamefont
  {Kongsted}}, \bibinfo {author} {\bibfnamefont {A.}~\bibnamefont {Krapp}},
  \bibinfo {author} {\bibfnamefont {K.}~\bibnamefont {Kristensen}}, \bibinfo
  {author} {\bibfnamefont {A.}~\bibnamefont {Ligabue}}, \bibinfo {author}
  {\bibfnamefont {O.~B.}\ \bibnamefont {Lutn\ae{}s}}, \bibinfo {author}
  {\bibfnamefont {J.~I.}\ \bibnamefont {Melo}}, \bibinfo {author}
  {\bibfnamefont {K.~V.}\ \bibnamefont {Mikkelsen}}, \bibinfo {author}
  {\bibfnamefont {R.~H.}\ \bibnamefont {Myhre}}, \bibinfo {author}
  {\bibfnamefont {C.}~\bibnamefont {Neiss}}, \bibinfo {author} {\bibfnamefont
  {C.~B.}\ \bibnamefont {Nielsen}}, \bibinfo {author} {\bibfnamefont
  {P.}~\bibnamefont {Norman}}, \bibinfo {author} {\bibfnamefont
  {J.}~\bibnamefont {Olsen}}, \bibinfo {author} {\bibfnamefont {J.~M.~H.}\
  \bibnamefont {Olsen}}, \bibinfo {author} {\bibfnamefont {A.}~\bibnamefont
  {Osted}}, \bibinfo {author} {\bibfnamefont {M.~J.}\ \bibnamefont {Packer}},
  \bibinfo {author} {\bibfnamefont {F.}~\bibnamefont {Pawlowski}}, \bibinfo
  {author} {\bibfnamefont {T.~B.}\ \bibnamefont {Pedersen}}, \bibinfo {author}
  {\bibfnamefont {P.~F.}\ \bibnamefont {Provasi}}, \bibinfo {author}
  {\bibfnamefont {S.}~\bibnamefont {Reine}}, \bibinfo {author} {\bibfnamefont
  {Z.}~\bibnamefont {Rinkevicius}}, \bibinfo {author} {\bibfnamefont {T.~A.}\
  \bibnamefont {Ruden}}, \bibinfo {author} {\bibfnamefont {K.}~\bibnamefont
  {Ruud}}, \bibinfo {author} {\bibfnamefont {V.~V.}\ \bibnamefont {Rybkin}},
  \bibinfo {author} {\bibfnamefont {P.}~\bibnamefont {Sa\l{}ek}}, \bibinfo
  {author} {\bibfnamefont {C.~C.~M.}\ \bibnamefont {Samson}}, \bibinfo {author}
  {\bibfnamefont {A.~S.}\ \bibnamefont {de~Mer\'{a}s}}, \bibinfo {author}
  {\bibfnamefont {T.}~\bibnamefont {Saue}}, \bibinfo {author} {\bibfnamefont
  {S.~P.~A.}\ \bibnamefont {Sauer}}, \bibinfo {author} {\bibfnamefont
  {B.}~\bibnamefont {Schimmelpfennig}}, \bibinfo {author} {\bibfnamefont
  {K.}~\bibnamefont {Sneskov}}, \bibinfo {author} {\bibfnamefont {A.~H.}\
  \bibnamefont {Steindal}}, \bibinfo {author} {\bibfnamefont {K.~O.}\
  \bibnamefont {Sylvester-Hvid}}, \bibinfo {author} {\bibfnamefont {P.~R.}\
  \bibnamefont {Taylor}}, \bibinfo {author} {\bibfnamefont {A.~M.}\
  \bibnamefont {Teale}}, \bibinfo {author} {\bibfnamefont {E.~I.}\ \bibnamefont
  {Tellgren}}, \bibinfo {author} {\bibfnamefont {D.~P.}\ \bibnamefont {Tew}},
  \bibinfo {author} {\bibfnamefont {A.~J.}\ \bibnamefont {Thorvaldsen}},
  \bibinfo {author} {\bibfnamefont {L.}~\bibnamefont {Th\o{}gersen}}, \bibinfo
  {author} {\bibfnamefont {O.}~\bibnamefont {Vahtras}}, \bibinfo {author}
  {\bibfnamefont {M.~A.}\ \bibnamefont {Watson}}, \bibinfo {author}
  {\bibfnamefont {D.~J.~D.}\ \bibnamefont {Wilson}}, \bibinfo {author}
  {\bibfnamefont {M.}~\bibnamefont {Ziolkowski}}, \ and\ \bibinfo {author}
  {\bibfnamefont {H.}~\bibnamefont {\AA{}gren}},\ }\href@noop {} {\bibfield
  {journal} {\bibinfo  {journal} {WIREs Comput.~Mol.~Sci.}\ }\textbf {\bibinfo
  {volume} {4}},\ \bibinfo {pages} {269} (\bibinfo {year} {2014})}\BibitemShut
  {NoStop}%
\bibitem [{sup()}]{supporting}%
  \BibitemOpen
  \href@noop {} {}\bibinfo {note} {{See Supplemental Material at [URL will be
  inserted by publisher] for composition and exponents of the optimized
  Gaussian basis sets.}}\BibitemShut {Stop}%
\bibitem [{\citenamefont {Minnhagen}(1973)}]{minnhagen73}%
  \BibitemOpen
  \bibfield  {author} {\bibinfo {author} {\bibfnamefont {L.}~\bibnamefont
  {Minnhagen}},\ }\href@noop {} {\bibfield  {journal} {\bibinfo  {journal} {J.
  Opt. Soc. Am. A}\ }\textbf {\bibinfo {volume} {63}},\ \bibinfo {pages} {1185}
  (\bibinfo {year} {1973})}\BibitemShut {NoStop}%
\bibitem [{\citenamefont {Purvis}\ and\ \citenamefont
  {Bartlett}(1982)}]{purvis82}%
  \BibitemOpen
  \bibfield  {author} {\bibinfo {author} {\bibfnamefont {G.~D.}\ \bibnamefont
  {Purvis}}\ and\ \bibinfo {author} {\bibfnamefont {R.~J.}\ \bibnamefont
  {Bartlett}},\ }\href@noop {} {\bibfield  {journal} {\bibinfo  {journal} {J.
  Chem. Phys.}\ }\textbf {\bibinfo {volume} {76}},\ \bibinfo {pages} {1910}
  (\bibinfo {year} {1982})}\BibitemShut {NoStop}%
\bibitem [{\citenamefont {Scuseria}\ \emph {et~al.}(1987)\citenamefont
  {Scuseria}, \citenamefont {Scheiner}, \citenamefont {Lee}, \citenamefont
  {Rice},\ and\ \citenamefont {Schaefer}}]{scuseria87}%
  \BibitemOpen
  \bibfield  {author} {\bibinfo {author} {\bibfnamefont {G.~E.}\ \bibnamefont
  {Scuseria}}, \bibinfo {author} {\bibfnamefont {A.~C.}\ \bibnamefont
  {Scheiner}}, \bibinfo {author} {\bibfnamefont {T.~J.}\ \bibnamefont {Lee}},
  \bibinfo {author} {\bibfnamefont {J.~E.}\ \bibnamefont {Rice}}, \ and\
  \bibinfo {author} {\bibfnamefont {H.~F.}\ \bibnamefont {Schaefer}},\
  }\href@noop {} {\bibfield  {journal} {\bibinfo  {journal} {J. Chem. Phys.}\
  }\textbf {\bibinfo {volume} {86}},\ \bibinfo {pages} {2881} (\bibinfo {year}
  {1987})}\BibitemShut {NoStop}%
\bibitem [{\citenamefont {Raghavachari}\ \emph {et~al.}(1989)\citenamefont
  {Raghavachari}, \citenamefont {Trucks}, \citenamefont {Pople},\ and\
  \citenamefont {Head-Gordon}}]{ragha89}%
  \BibitemOpen
  \bibfield  {author} {\bibinfo {author} {\bibfnamefont {K.}~\bibnamefont
  {Raghavachari}}, \bibinfo {author} {\bibfnamefont {G.~W.}\ \bibnamefont
  {Trucks}}, \bibinfo {author} {\bibfnamefont {J.~A.}\ \bibnamefont {Pople}}, \
  and\ \bibinfo {author} {\bibfnamefont {M.}~\bibnamefont {Head-Gordon}},\
  }\href@noop {} {\bibfield  {journal} {\bibinfo  {journal} {Chem. Phys.
  Lett.}\ }\textbf {\bibinfo {volume} {157}},\ \bibinfo {pages} {479} (\bibinfo
  {year} {1989})}\BibitemShut {NoStop}%
\bibitem [{\citenamefont {Koch}\ \emph {et~al.}(1997)\citenamefont {Koch},
  \citenamefont {Christiansen}, \citenamefont {J\o{}rgensen}, \citenamefont
  {Sanchez~de Mer{\'a}s},\ and\ \citenamefont {Helgaker}}]{koch97}%
  \BibitemOpen
  \bibfield  {author} {\bibinfo {author} {\bibfnamefont {H.}~\bibnamefont
  {Koch}}, \bibinfo {author} {\bibfnamefont {O.}~\bibnamefont {Christiansen}},
  \bibinfo {author} {\bibfnamefont {P.}~\bibnamefont {J\o{}rgensen}}, \bibinfo
  {author} {\bibfnamefont {A.~M.}\ \bibnamefont {Sanchez~de Mer{\'a}s}}, \ and\
  \bibinfo {author} {\bibfnamefont {T.}~\bibnamefont {Helgaker}},\ }\href@noop
  {} {\bibfield  {journal} {\bibinfo  {journal} {J. Chem. Phys.}\ }\textbf
  {\bibinfo {volume} {106}},\ \bibinfo {pages} {1808} (\bibinfo {year}
  {1997})}\BibitemShut {NoStop}%
\bibitem [{\citenamefont {Noga}\ and\ \citenamefont {Bartlett}(1987)}]{noga87}%
  \BibitemOpen
  \bibfield  {author} {\bibinfo {author} {\bibfnamefont {J.}~\bibnamefont
  {Noga}}\ and\ \bibinfo {author} {\bibfnamefont {R.~J.}\ \bibnamefont
  {Bartlett}},\ }\href@noop {} {\bibfield  {journal} {\bibinfo  {journal} {J.
  Chem. Phys.}\ }\textbf {\bibinfo {volume} {86}},\ \bibinfo {pages} {7041}
  (\bibinfo {year} {1987})}\BibitemShut {NoStop}%
\bibitem [{\citenamefont {Scuseria}\ and\ \citenamefont
  {Schaefer}(1988)}]{scuseria88}%
  \BibitemOpen
  \bibfield  {author} {\bibinfo {author} {\bibfnamefont {G.~E.}\ \bibnamefont
  {Scuseria}}\ and\ \bibinfo {author} {\bibfnamefont {H.~F.}\ \bibnamefont
  {Schaefer}},\ }\href@noop {} {\bibfield  {journal} {\bibinfo  {journal}
  {Chem. Phys. Lett.}\ }\textbf {\bibinfo {volume} {152}},\ \bibinfo {pages}
  {382 } (\bibinfo {year} {1988})}\BibitemShut {NoStop}%
\bibitem [{\citenamefont {Stanton}\ \emph {et~al.}()\citenamefont {Stanton},
  \citenamefont {Gauss}, \citenamefont {Cheng}, \citenamefont {Harding},
  \citenamefont {Matthews},\ and\ \citenamefont {Szalay}}]{cfour}%
  \BibitemOpen
  \bibfield  {author} {\bibinfo {author} {\bibfnamefont {J.~F.}\ \bibnamefont
  {Stanton}}, \bibinfo {author} {\bibfnamefont {J.}~\bibnamefont {Gauss}},
  \bibinfo {author} {\bibfnamefont {L.}~\bibnamefont {Cheng}}, \bibinfo
  {author} {\bibfnamefont {M.~E.}\ \bibnamefont {Harding}}, \bibinfo {author}
  {\bibfnamefont {D.~A.}\ \bibnamefont {Matthews}}, \ and\ \bibinfo {author}
  {\bibfnamefont {P.~G.}\ \bibnamefont {Szalay}},\ }\href@noop {} {\enquote
  {\bibinfo {title} {{CFOUR, Coupled-Cluster techniques for Computational
  Chemistry, a quantum-chemical program package}},}\ }\bibinfo {note} {{W}ith
  contributions from {A}.{A}. {A}uer, {R}.{J}. {B}artlett, {U}. {B}enedikt,
  {C}. {B}erger, {D}.{E}. {B}ernholdt, {Y}.{J}. {B}omble, {O}. {C}hristiansen,
  {F}. Engel, {R}. Faber, {M}. {H}eckert, {O}. {H}eun, {M}. Hilgenberg, {C}.
  {H}uber, {T}.-{C}. {J}agau, {D}. {J}onsson, {J}. {J}us{\'e}lius, {T}. Kirsch,
  {K}. {K}lein, {W}.{J}. {L}auderdale, {F}. {L}ipparini, {T}. {M}etzroth,
  {L}.{A}. {M}{\"u}ck, {D}.{P}. {O}'{N}eill, {D}.{R}. {P}rice, {E}. {P}rochnow,
  {C}. {P}uzzarini, {K}. {R}uud, {F}. {S}chiffmann, {W}. {S}chwalbach, {C}.
  {S}immons, {S}. {S}topkowicz, {A}. {T}ajti, {J}. {V}{\'a}zquez, {F}. {W}ang,
  {J}.{D}. {W}atts and the integral packages {MOLECULE} ({J}. {A}lml{\"o}f and
  {P}.{R}. {T}aylor), {PROPS} ({P}.{R}. {T}aylor), {ABACUS} ({T}. {H}elgaker,
  {H}.{J}. {A}a. {J}ensen, {P}. {J}{\o}rgensen, and {J}. {O}lsen), and {ECP}
  routines by {A}. {V}. {M}itin and {C}. van {W}{\"u}llen. {F}or the current
  version, see http://www.cfour.de.}\BibitemShut {Stop}%
\bibitem [{\citenamefont {K\'{a}llay}\ \emph {et~al.}(2020)\citenamefont
  {K\'{a}llay}, \citenamefont {Nagy}, \citenamefont {Mester}, \citenamefont
  {Rolik}, \citenamefont {Samu}, \citenamefont {Csontos}, \citenamefont
  {Cs\'{o}ka}, \citenamefont {Szab\'{o}}, \citenamefont {Gyevi-Nagy},
  \citenamefont {H\'{e}gely}, \citenamefont {Ladj\'{a}nszki}, \citenamefont
  {Szegedy}, \citenamefont {Lad\'{o}czki}, \citenamefont {Petrov},
  \citenamefont {Farkas}, \citenamefont {Mezei},\ and\ \citenamefont
  {Ganyecz}}]{kallay20}%
  \BibitemOpen
  \bibfield  {author} {\bibinfo {author} {\bibfnamefont {M.}~\bibnamefont
  {K\'{a}llay}}, \bibinfo {author} {\bibfnamefont {P.~R.}\ \bibnamefont
  {Nagy}}, \bibinfo {author} {\bibfnamefont {D.}~\bibnamefont {Mester}},
  \bibinfo {author} {\bibfnamefont {Z.}~\bibnamefont {Rolik}}, \bibinfo
  {author} {\bibfnamefont {G.}~\bibnamefont {Samu}}, \bibinfo {author}
  {\bibfnamefont {J.}~\bibnamefont {Csontos}}, \bibinfo {author} {\bibfnamefont
  {J.}~\bibnamefont {Cs\'{o}ka}}, \bibinfo {author} {\bibfnamefont {P.~B.}\
  \bibnamefont {Szab\'{o}}}, \bibinfo {author} {\bibfnamefont {L.}~\bibnamefont
  {Gyevi-Nagy}}, \bibinfo {author} {\bibfnamefont {B.}~\bibnamefont
  {H\'{e}gely}}, \bibinfo {author} {\bibfnamefont {I.}~\bibnamefont
  {Ladj\'{a}nszki}}, \bibinfo {author} {\bibfnamefont {L.}~\bibnamefont
  {Szegedy}}, \bibinfo {author} {\bibfnamefont {B.}~\bibnamefont
  {Lad\'{o}czki}}, \bibinfo {author} {\bibfnamefont {K.}~\bibnamefont
  {Petrov}}, \bibinfo {author} {\bibfnamefont {M.}~\bibnamefont {Farkas}},
  \bibinfo {author} {\bibfnamefont {P.~D.}\ \bibnamefont {Mezei}}, \ and\
  \bibinfo {author} {\bibfnamefont {A.}~\bibnamefont {Ganyecz}},\ }\href@noop
  {} {\bibfield  {journal} {\bibinfo  {journal} {J. Chem. Phys.}\ }\textbf
  {\bibinfo {volume} {152}},\ \bibinfo {pages} {074107} (\bibinfo {year}
  {2020})}\BibitemShut {NoStop}%
\bibitem [{\citenamefont {Kucharski}\ and\ \citenamefont
  {Bartlett}(1991)}]{kucharski91}%
  \BibitemOpen
  \bibfield  {author} {\bibinfo {author} {\bibfnamefont {S.~A.}\ \bibnamefont
  {Kucharski}}\ and\ \bibinfo {author} {\bibfnamefont {R.~J.}\ \bibnamefont
  {Bartlett}},\ }\href@noop {} {\bibfield  {journal} {\bibinfo  {journal}
  {Theor. Chim. Acta}\ }\textbf {\bibinfo {volume} {80}},\ \bibinfo {pages}
  {387} (\bibinfo {year} {1991})}\BibitemShut {NoStop}%
\bibitem [{\citenamefont {Oliphant}\ and\ \citenamefont
  {Adamowicz}(1991)}]{oliphant91}%
  \BibitemOpen
  \bibfield  {author} {\bibinfo {author} {\bibfnamefont {N.}~\bibnamefont
  {Oliphant}}\ and\ \bibinfo {author} {\bibfnamefont {L.}~\bibnamefont
  {Adamowicz}},\ }\href@noop {} {\bibfield  {journal} {\bibinfo  {journal} {J.
  Chem. Phys.}\ }\textbf {\bibinfo {volume} {95}},\ \bibinfo {pages} {6645}
  (\bibinfo {year} {1991})}\BibitemShut {NoStop}%
\bibitem [{\citenamefont {Kucharski}\ and\ \citenamefont
  {Bartlett}(1992)}]{kucharski92}%
  \BibitemOpen
  \bibfield  {author} {\bibinfo {author} {\bibfnamefont {S.~A.}\ \bibnamefont
  {Kucharski}}\ and\ \bibinfo {author} {\bibfnamefont {R.~J.}\ \bibnamefont
  {Bartlett}},\ }\href@noop {} {\bibfield  {journal} {\bibinfo  {journal} {J.
  Chem. Phys.}\ }\textbf {\bibinfo {volume} {97}},\ \bibinfo {pages} {4282}
  (\bibinfo {year} {1992})}\BibitemShut {NoStop}%
\bibitem [{\citenamefont {Kucharski}\ and\ \citenamefont
  {Musiał}(2010)}]{kucharski10}%
  \BibitemOpen
  \bibfield  {author} {\bibinfo {author} {\bibfnamefont {S.~A.}\ \bibnamefont
  {Kucharski}}\ and\ \bibinfo {author} {\bibfnamefont {M.}~\bibnamefont
  {Musiał}},\ }\href@noop {} {\bibfield  {journal} {\bibinfo  {journal} {Mol.
  Phys.}\ }\textbf {\bibinfo {volume} {108}},\ \bibinfo {pages} {2975}
  (\bibinfo {year} {2010})}\BibitemShut {NoStop}%
\bibitem [{\citenamefont {Musia{\l}}\ \emph {et~al.}(2000)\citenamefont
  {Musia{\l}}, \citenamefont {Kucharski},\ and\ \citenamefont
  {Bartlett}}]{musial00}%
  \BibitemOpen
  \bibfield  {author} {\bibinfo {author} {\bibfnamefont {M.}~\bibnamefont
  {Musia{\l}}}, \bibinfo {author} {\bibfnamefont {S.~A.}\ \bibnamefont
  {Kucharski}}, \ and\ \bibinfo {author} {\bibfnamefont {R.~J.}\ \bibnamefont
  {Bartlett}},\ }\href@noop {} {\bibfield  {journal} {\bibinfo  {journal}
  {Chem. Phys. Lett.}\ }\textbf {\bibinfo {volume} {320}},\ \bibinfo {pages}
  {542} (\bibinfo {year} {2000})}\BibitemShut {NoStop}%
\bibitem [{\citenamefont {Musia{\l}}\ \emph {et~al.}(2002)\citenamefont
  {Musia{\l}}, \citenamefont {Kucharski},\ and\ \citenamefont
  {Bartlett}}]{musial02}%
  \BibitemOpen
  \bibfield  {author} {\bibinfo {author} {\bibfnamefont {M.}~\bibnamefont
  {Musia{\l}}}, \bibinfo {author} {\bibfnamefont {S.}~\bibnamefont
  {Kucharski}}, \ and\ \bibinfo {author} {\bibfnamefont {R.}~\bibnamefont
  {Bartlett}},\ }\href@noop {} {\bibfield  {journal} {\bibinfo  {journal} {J.
  Chem. Phys.}\ }\textbf {\bibinfo {volume} {116}},\ \bibinfo {pages} {4382}
  (\bibinfo {year} {2002})}\BibitemShut {NoStop}%
\bibitem [{\citenamefont {K{\'a}llay}\ and\ \citenamefont
  {Surj{\'a}n}(2001)}]{kallay01}%
  \BibitemOpen
  \bibfield  {author} {\bibinfo {author} {\bibfnamefont {M.}~\bibnamefont
  {K{\'a}llay}}\ and\ \bibinfo {author} {\bibfnamefont {P.~R.}\ \bibnamefont
  {Surj{\'a}n}},\ }\href@noop {} {\bibfield  {journal} {\bibinfo  {journal} {J.
  Chem. Phys.}\ }\textbf {\bibinfo {volume} {115}},\ \bibinfo {pages} {2945}
  (\bibinfo {year} {2001})}\BibitemShut {NoStop}%
\bibitem [{\citenamefont {Olsen}(2000)}]{olsen00}%
  \BibitemOpen
  \bibfield  {author} {\bibinfo {author} {\bibfnamefont {J.}~\bibnamefont
  {Olsen}},\ }\href@noop {} {\bibfield  {journal} {\bibinfo  {journal} {J.
  Chem. Phys.}\ }\textbf {\bibinfo {volume} {113}},\ \bibinfo {pages} {7140}
  (\bibinfo {year} {2000})}\BibitemShut {NoStop}%
\bibitem [{\citenamefont {Hirata}(2003)}]{hirata03}%
  \BibitemOpen
  \bibfield  {author} {\bibinfo {author} {\bibfnamefont {S.}~\bibnamefont
  {Hirata}},\ }\href@noop {} {\bibfield  {journal} {\bibinfo  {journal} {J.
  Phys. Chem. A}\ }\textbf {\bibinfo {volume} {107}},\ \bibinfo {pages} {9887}
  (\bibinfo {year} {2003})}\BibitemShut {NoStop}%
\bibitem [{\citenamefont {Lesiuk}\ and\ \citenamefont
  {Jeziorski}(2019)}]{lesiuk19}%
  \BibitemOpen
  \bibfield  {author} {\bibinfo {author} {\bibfnamefont {M.}~\bibnamefont
  {Lesiuk}}\ and\ \bibinfo {author} {\bibfnamefont {B.}~\bibnamefont
  {Jeziorski}},\ }\href@noop {} {\bibfield  {journal} {\bibinfo  {journal} {J.
  Chem. Theory Comput.}\ }\textbf {\bibinfo {volume} {15}},\ \bibinfo {pages}
  {5398} (\bibinfo {year} {2019})}\BibitemShut {NoStop}%
\bibitem [{\citenamefont {Bomble}\ \emph {et~al.}(2005)\citenamefont {Bomble},
  \citenamefont {Stanton}, \citenamefont {Kállay},\ and\ \citenamefont
  {Gauss}}]{bomble05}%
  \BibitemOpen
  \bibfield  {author} {\bibinfo {author} {\bibfnamefont {Y.~J.}\ \bibnamefont
  {Bomble}}, \bibinfo {author} {\bibfnamefont {J.~F.}\ \bibnamefont {Stanton}},
  \bibinfo {author} {\bibfnamefont {M.}~\bibnamefont {Kállay}}, \ and\
  \bibinfo {author} {\bibfnamefont {J.}~\bibnamefont {Gauss}},\ }\href@noop {}
  {\bibfield  {journal} {\bibinfo  {journal} {J. Chem. Phys.}\ }\textbf
  {\bibinfo {volume} {123}},\ \bibinfo {pages} {054101} (\bibinfo {year}
  {2005})}\BibitemShut {NoStop}%
\bibitem [{\citenamefont {K{\'a}llay}\ and\ \citenamefont
  {Gauss}(2005)}]{kallay05}%
  \BibitemOpen
  \bibfield  {author} {\bibinfo {author} {\bibfnamefont {M.}~\bibnamefont
  {K{\'a}llay}}\ and\ \bibinfo {author} {\bibfnamefont {J.}~\bibnamefont
  {Gauss}},\ }\href@noop {} {\bibfield  {journal} {\bibinfo  {journal} {J.
  Chem. Phys.}\ }\textbf {\bibinfo {volume} {123}},\ \bibinfo {pages} {214105}
  (\bibinfo {year} {2005})}\BibitemShut {NoStop}%
\bibitem [{\citenamefont {Kucharski}\ and\ \citenamefont
  {Bartlett}(1989)}]{kucharski89}%
  \BibitemOpen
  \bibfield  {author} {\bibinfo {author} {\bibfnamefont {S.~A.}\ \bibnamefont
  {Kucharski}}\ and\ \bibinfo {author} {\bibfnamefont {R.~J.}\ \bibnamefont
  {Bartlett}},\ }\href {\doibase https://doi.org/10.1016/0009-2614(89)87388-9}
  {\bibfield  {journal} {\bibinfo  {journal} {Chem. Phys. Lett.}\ }\textbf
  {\bibinfo {volume} {158}},\ \bibinfo {pages} {550} (\bibinfo {year}
  {1989})}\BibitemShut {NoStop}%
\bibitem [{\citenamefont {Bethe}\ and\ \citenamefont
  {Salpeter}(1975)}]{bethe75}%
  \BibitemOpen
  \bibfield  {author} {\bibinfo {author} {\bibfnamefont {H.~A.}\ \bibnamefont
  {Bethe}}\ and\ \bibinfo {author} {\bibfnamefont {E.~E.}\ \bibnamefont
  {Salpeter}},\ }\href@noop {} {\emph {\bibinfo {title} {Quantum Mechanics of
  One- and Two- Electron Systems}}}\ (\bibinfo  {publisher} {Springer:
  Berlin},\ \bibinfo {year} {1975})\BibitemShut {NoStop}%
\bibitem [{\citenamefont {Cowan}\ and\ \citenamefont
  {Griffin}(1976)}]{cowan76}%
  \BibitemOpen
  \bibfield  {author} {\bibinfo {author} {\bibfnamefont {R.~D.}\ \bibnamefont
  {Cowan}}\ and\ \bibinfo {author} {\bibfnamefont {D.~C.}\ \bibnamefont
  {Griffin}},\ }\href@noop {} {\bibfield  {journal} {\bibinfo  {journal} {J.
  Opt. Soc. Am. A}\ }\textbf {\bibinfo {volume} {66}},\ \bibinfo {pages} {1010}
  (\bibinfo {year} {1976})}\BibitemShut {NoStop}%
\bibitem [{\citenamefont {Coriani}\ \emph {et~al.}(2004)\citenamefont
  {Coriani}, \citenamefont {Helgaker}, \citenamefont {J\o{}rgensen},\ and\
  \citenamefont {Klopper}}]{coriani04}%
  \BibitemOpen
  \bibfield  {author} {\bibinfo {author} {\bibfnamefont {S.}~\bibnamefont
  {Coriani}}, \bibinfo {author} {\bibfnamefont {T.}~\bibnamefont {Helgaker}},
  \bibinfo {author} {\bibfnamefont {P.}~\bibnamefont {J\o{}rgensen}}, \ and\
  \bibinfo {author} {\bibfnamefont {W.}~\bibnamefont {Klopper}},\ }\href@noop
  {} {\bibfield  {journal} {\bibinfo  {journal} {J. Chem. Phys.}\ }\textbf
  {\bibinfo {volume} {121}},\ \bibinfo {pages} {6591} (\bibinfo {year}
  {2004})}\BibitemShut {NoStop}%
\bibitem [{\citenamefont {Douglas}\ and\ \citenamefont
  {Kroll}(1974)}]{douglas74}%
  \BibitemOpen
  \bibfield  {author} {\bibinfo {author} {\bibfnamefont {M.}~\bibnamefont
  {Douglas}}\ and\ \bibinfo {author} {\bibfnamefont {N.~M.}\ \bibnamefont
  {Kroll}},\ }\href@noop {} {\bibfield  {journal} {\bibinfo  {journal} {Ann.
  Phys.}\ }\textbf {\bibinfo {volume} {82}},\ \bibinfo {pages} {89} (\bibinfo
  {year} {1974})}\BibitemShut {NoStop}%
\bibitem [{\citenamefont {Hess}(1985)}]{hess85}%
  \BibitemOpen
  \bibfield  {author} {\bibinfo {author} {\bibfnamefont {B.~A.}\ \bibnamefont
  {Hess}},\ }\href@noop {} {\bibfield  {journal} {\bibinfo  {journal} {Phys.
  Rev. A}\ }\textbf {\bibinfo {volume} {32}},\ \bibinfo {pages} {756} (\bibinfo
  {year} {1985})}\BibitemShut {NoStop}%
\bibitem [{\citenamefont {Reiher}(2006)}]{reiher06}%
  \BibitemOpen
  \bibfield  {author} {\bibinfo {author} {\bibfnamefont {M.}~\bibnamefont
  {Reiher}},\ }\href@noop {} {\bibfield  {journal} {\bibinfo  {journal} {Theor.
  Chem. Acc.}\ }\textbf {\bibinfo {volume} {116}},\ \bibinfo {pages} {241}
  (\bibinfo {year} {2006})}\BibitemShut {NoStop}%
\bibitem [{\citenamefont {Kutzelnigg}(2008)}]{kutz08}%
  \BibitemOpen
  \bibfield  {author} {\bibinfo {author} {\bibfnamefont {W.}~\bibnamefont
  {Kutzelnigg}},\ }\href@noop {} {\bibfield  {journal} {\bibinfo  {journal}
  {Int. J. Quantum Chem.}\ }\textbf {\bibinfo {volume} {108}},\ \bibinfo
  {pages} {2280} (\bibinfo {year} {2008})}\BibitemShut {NoStop}%
\bibitem [{\citenamefont {Middendorf}\ \emph {et~al.}(2012)\citenamefont
  {Middendorf}, \citenamefont {H\"{o}fener}, \citenamefont {Klopper},\ and\
  \citenamefont {Helgaker}}]{middendorf12}%
  \BibitemOpen
  \bibfield  {author} {\bibinfo {author} {\bibfnamefont {N.}~\bibnamefont
  {Middendorf}}, \bibinfo {author} {\bibfnamefont {S.}~\bibnamefont
  {H\"{o}fener}}, \bibinfo {author} {\bibfnamefont {W.}~\bibnamefont
  {Klopper}}, \ and\ \bibinfo {author} {\bibfnamefont {T.}~\bibnamefont
  {Helgaker}},\ }\href@noop {} {\bibfield  {journal} {\bibinfo  {journal}
  {Chem. Phys.}\ }\textbf {\bibinfo {volume} {401}},\ \bibinfo {pages} {146 }
  (\bibinfo {year} {2012})}\BibitemShut {NoStop}%
\bibitem [{\citenamefont {Bischoff}\ \emph {et~al.}(2010)\citenamefont
  {Bischoff}, \citenamefont {Valeev}, \citenamefont {Klopper},\ and\
  \citenamefont {Janssen}}]{bischoff10}%
  \BibitemOpen
  \bibfield  {author} {\bibinfo {author} {\bibfnamefont {F.~A.}\ \bibnamefont
  {Bischoff}}, \bibinfo {author} {\bibfnamefont {E.~F.}\ \bibnamefont
  {Valeev}}, \bibinfo {author} {\bibfnamefont {W.}~\bibnamefont {Klopper}}, \
  and\ \bibinfo {author} {\bibfnamefont {C.~L.}\ \bibnamefont {Janssen}},\
  }\href@noop {} {\bibfield  {journal} {\bibinfo  {journal} {J. Chem. Phys.}\
  }\textbf {\bibinfo {volume} {132}},\ \bibinfo {pages} {214104} (\bibinfo
  {year} {2010})}\BibitemShut {NoStop}%
\bibitem [{\citenamefont {Ottschofski}\ and\ \citenamefont
  {Kutzelnigg}(1997)}]{otto97}%
  \BibitemOpen
  \bibfield  {author} {\bibinfo {author} {\bibfnamefont {E.}~\bibnamefont
  {Ottschofski}}\ and\ \bibinfo {author} {\bibfnamefont {W.}~\bibnamefont
  {Kutzelnigg}},\ }\href@noop {} {\bibfield  {journal} {\bibinfo  {journal} {J.
  Chem. Phys.}\ }\textbf {\bibinfo {volume} {106}},\ \bibinfo {pages} {6634}
  (\bibinfo {year} {1997})}\BibitemShut {NoStop}%
\bibitem [{\citenamefont {Przybytek}\ \emph {et~al.}(2010)\citenamefont
  {Przybytek}, \citenamefont {Cencek}, \citenamefont {Komasa}, \citenamefont
  {\L{}ach}, \citenamefont {Jeziorski},\ and\ \citenamefont
  {Szalewicz}}]{przybytek10}%
  \BibitemOpen
  \bibfield  {author} {\bibinfo {author} {\bibfnamefont {M.}~\bibnamefont
  {Przybytek}}, \bibinfo {author} {\bibfnamefont {W.}~\bibnamefont {Cencek}},
  \bibinfo {author} {\bibfnamefont {J.}~\bibnamefont {Komasa}}, \bibinfo
  {author} {\bibfnamefont {G.}~\bibnamefont {\L{}ach}}, \bibinfo {author}
  {\bibfnamefont {B.}~\bibnamefont {Jeziorski}}, \ and\ \bibinfo {author}
  {\bibfnamefont {K.}~\bibnamefont {Szalewicz}},\ }\href@noop {} {\bibfield
  {journal} {\bibinfo  {journal} {Phys. Rev. Lett.}\ }\textbf {\bibinfo
  {volume} {104}},\ \bibinfo {pages} {183003} (\bibinfo {year}
  {2010})}\BibitemShut {NoStop}%
\bibitem [{\citenamefont {Przybytek}\ \emph {et~al.}(2017)\citenamefont
  {Przybytek}, \citenamefont {Cencek}, \citenamefont {Jeziorski},\ and\
  \citenamefont {Szalewicz}}]{przybytek17}%
  \BibitemOpen
  \bibfield  {author} {\bibinfo {author} {\bibfnamefont {M.}~\bibnamefont
  {Przybytek}}, \bibinfo {author} {\bibfnamefont {W.}~\bibnamefont {Cencek}},
  \bibinfo {author} {\bibfnamefont {B.}~\bibnamefont {Jeziorski}}, \ and\
  \bibinfo {author} {\bibfnamefont {K.}~\bibnamefont {Szalewicz}},\ }\href@noop
  {} {\bibfield  {journal} {\bibinfo  {journal} {Phys. Rev. Lett.}\ }\textbf
  {\bibinfo {volume} {119}},\ \bibinfo {pages} {123401} (\bibinfo {year}
  {2017})}\BibitemShut {NoStop}%
\bibitem [{\citenamefont {Cencek}\ \emph {et~al.}(2012)\citenamefont {Cencek},
  \citenamefont {Przybytek}, \citenamefont {Komasa}, \citenamefont {Mehl},
  \citenamefont {Jeziorski},\ and\ \citenamefont {Szalewicz}}]{cencek12}%
  \BibitemOpen
  \bibfield  {author} {\bibinfo {author} {\bibfnamefont {W.}~\bibnamefont
  {Cencek}}, \bibinfo {author} {\bibfnamefont {M.}~\bibnamefont {Przybytek}},
  \bibinfo {author} {\bibfnamefont {J.}~\bibnamefont {Komasa}}, \bibinfo
  {author} {\bibfnamefont {J.~B.}\ \bibnamefont {Mehl}}, \bibinfo {author}
  {\bibfnamefont {B.}~\bibnamefont {Jeziorski}}, \ and\ \bibinfo {author}
  {\bibfnamefont {K.}~\bibnamefont {Szalewicz}},\ }\href@noop {} {\bibfield
  {journal} {\bibinfo  {journal} {J. Chem. Phys.}\ }\textbf {\bibinfo {volume}
  {136}},\ \bibinfo {pages} {224303} (\bibinfo {year} {2012})}\BibitemShut
  {NoStop}%
\bibitem [{\citenamefont {Pachucki}(2006)}]{pachucki06}%
  \BibitemOpen
  \bibfield  {author} {\bibinfo {author} {\bibfnamefont {K.}~\bibnamefont
  {Pachucki}},\ }\href@noop {} {\bibfield  {journal} {\bibinfo  {journal}
  {Phys. Rev. A}\ }\textbf {\bibinfo {volume} {74}},\ \bibinfo {pages} {022512}
  (\bibinfo {year} {2006})}\BibitemShut {NoStop}%
\bibitem [{\citenamefont {Saue}\ \emph {et~al.}(2020)\citenamefont {Saue},
  \citenamefont {Bast}, \citenamefont {Gomes}, \citenamefont {Jensen},
  \citenamefont {Visscher}, \citenamefont {Aucar}, \citenamefont {Di~Remigio},
  \citenamefont {Dyall}, \citenamefont {Eliav}, \citenamefont {Fasshauer} \emph
  {et~al.}}]{saue20}%
  \BibitemOpen
  \bibfield  {author} {\bibinfo {author} {\bibfnamefont {T.}~\bibnamefont
  {Saue}}, \bibinfo {author} {\bibfnamefont {R.}~\bibnamefont {Bast}}, \bibinfo
  {author} {\bibfnamefont {A.~S.~P.}\ \bibnamefont {Gomes}}, \bibinfo {author}
  {\bibfnamefont {H.~J.~A.}\ \bibnamefont {Jensen}}, \bibinfo {author}
  {\bibfnamefont {L.}~\bibnamefont {Visscher}}, \bibinfo {author}
  {\bibfnamefont {I.~A.}\ \bibnamefont {Aucar}}, \bibinfo {author}
  {\bibfnamefont {R.}~\bibnamefont {Di~Remigio}}, \bibinfo {author}
  {\bibfnamefont {K.~G.}\ \bibnamefont {Dyall}}, \bibinfo {author}
  {\bibfnamefont {E.}~\bibnamefont {Eliav}}, \bibinfo {author} {\bibfnamefont
  {E.}~\bibnamefont {Fasshauer}},  \emph {et~al.},\ }\href@noop {} {\bibfield
  {journal} {\bibinfo  {journal} {J. Chem. Phys.}\ }\textbf {\bibinfo {volume}
  {152}},\ \bibinfo {pages} {204104} (\bibinfo {year} {2020})}\BibitemShut
  {NoStop}%
\bibitem [{dir()}]{dirac23}%
  \BibitemOpen
  \href@noop {} {}\bibinfo {note} {{DIRAC}, a relativistic ab initio electronic
  structure program, Release {DIRAC23} (2023), written by R.~Bast,
  A.~S.~P.~Gomes, T.~Saue and L.~Visscher and H.~J.~{\relax Aa}.~Jensen, with
  contributions from I.~A.~Aucar, V.~Bakken, C.~Chibueze, J.~Creutzberg,
  K.~G.~Dyall, S.~Dubillard, U.~Ekstr{\"o}m, E.~Eliav, T.~Enevoldsen,
  E.~Fa{\ss}hauer, T.~Fleig, O.~Fossgaard, L.~Halbert, E.~D.~Hedeg{\aa}rd,
  T.~Helgaker, B.~Helmich--Paris, J.~Henriksson, M.~van~Horn, M.~Ilia{\v{s}},
  Ch.~R.~Jacob, S.~Knecht, S.~Komorovsk{\'y}, O.~Kullie, J.~K.~L{\ae}rdahl,
  C.~V.~Larsen, Y.~S.~Lee, N.~H.~List, H.~S.~Nataraj, M.~K.~Nayak, P.~Norman,
  A.~Nyvang, G.~Olejniczak, J.~Olsen, J.~M.~H.~Olsen, A.~Papadopoulos,
  Y.~C.~Park, J.~K.~Pedersen, M.~Pernpointner, J.~V.~Pototschnig,
  R.~di~Remigio, M.~Repisky, K.~Ruud, P.~Sa{\l}ek, B.~Schimmelpfennig,
  B.~Senjean, A.~Shee, J.~Sikkema, A.~Sunaga, A.~J.~Thorvaldsen, J.~Thyssen,
  J.~van~Stralen, M.~L.~Vidal, S.~Villaume, O.~Visser, T.~Winther, S.~Yamamoto
  and X.~Yuan (available at https://doi.org/10.5281/zenodo.7670749, see also
  https://www.diracprogram.org)}\BibitemShut {NoStop}%
\bibitem [{\citenamefont {Dyall}(1994)}]{dyall94}%
  \BibitemOpen
  \bibfield  {author} {\bibinfo {author} {\bibfnamefont {K.~G.}\ \bibnamefont
  {Dyall}},\ }\href@noop {} {\bibfield  {journal} {\bibinfo  {journal} {J.
  Chem. Phys.}\ }\textbf {\bibinfo {volume} {100}},\ \bibinfo {pages} {2118}
  (\bibinfo {year} {1994})}\BibitemShut {NoStop}%
\bibitem [{\citenamefont {Caswell}\ and\ \citenamefont
  {Lepage}(1986)}]{caswell86}%
  \BibitemOpen
  \bibfield  {author} {\bibinfo {author} {\bibfnamefont {W.}~\bibnamefont
  {Caswell}}\ and\ \bibinfo {author} {\bibfnamefont {G.}~\bibnamefont
  {Lepage}},\ }\href@noop {} {\bibfield  {journal} {\bibinfo  {journal} {Phys.
  Lett. B}\ }\textbf {\bibinfo {volume} {167}},\ \bibinfo {pages} {437 }
  (\bibinfo {year} {1986})}\BibitemShut {NoStop}%
\bibitem [{\citenamefont {Pachucki}(1993)}]{pachucki93}%
  \BibitemOpen
  \bibfield  {author} {\bibinfo {author} {\bibfnamefont {K.}~\bibnamefont
  {Pachucki}},\ }\href@noop {} {\bibfield  {journal} {\bibinfo  {journal} {Ann.
  Phys.}\ }\textbf {\bibinfo {volume} {226}},\ \bibinfo {pages} {1} (\bibinfo
  {year} {1993})}\BibitemShut {NoStop}%
\bibitem [{\citenamefont {Pachucki}(1998)}]{pachucki98}%
  \BibitemOpen
  \bibfield  {author} {\bibinfo {author} {\bibfnamefont {K.}~\bibnamefont
  {Pachucki}},\ }\href@noop {} {\bibfield  {journal} {\bibinfo  {journal} {J.
  Phys. B}\ }\textbf {\bibinfo {volume} {31}},\ \bibinfo {pages} {5123}
  (\bibinfo {year} {1998})}\BibitemShut {NoStop}%
\bibitem [{\citenamefont {Schwartz}(1961)}]{schwartz61}%
  \BibitemOpen
  \bibfield  {author} {\bibinfo {author} {\bibfnamefont {C.}~\bibnamefont
  {Schwartz}},\ }\href@noop {} {\bibfield  {journal} {\bibinfo  {journal}
  {Phys. Rev.}\ }\textbf {\bibinfo {volume} {123}},\ \bibinfo {pages} {1700}
  (\bibinfo {year} {1961})}\BibitemShut {NoStop}%
\bibitem [{\citenamefont {Araki}(1957)}]{araki57}%
  \BibitemOpen
  \bibfield  {author} {\bibinfo {author} {\bibfnamefont {H.}~\bibnamefont
  {Araki}},\ }\href@noop {} {\bibfield  {journal} {\bibinfo  {journal} {Prog.
  Theor. Phys.}\ }\textbf {\bibinfo {volume} {17}},\ \bibinfo {pages} {619}
  (\bibinfo {year} {1957})}\BibitemShut {NoStop}%
\bibitem [{\citenamefont {Sucher}(1958)}]{sucher58}%
  \BibitemOpen
  \bibfield  {author} {\bibinfo {author} {\bibfnamefont {J.}~\bibnamefont
  {Sucher}},\ }\href@noop {} {\bibfield  {journal} {\bibinfo  {journal} {Phys.
  Rev.}\ }\textbf {\bibinfo {volume} {109}},\ \bibinfo {pages} {1010} (\bibinfo
  {year} {1958})}\BibitemShut {NoStop}%
\bibitem [{\citenamefont {Balcerzak}\ \emph {et~al.}(2017)\citenamefont
  {Balcerzak}, \citenamefont {Lesiuk},\ and\ \citenamefont
  {Moszynski}}]{balcerzak17}%
  \BibitemOpen
  \bibfield  {author} {\bibinfo {author} {\bibfnamefont {J.~G.}\ \bibnamefont
  {Balcerzak}}, \bibinfo {author} {\bibfnamefont {M.}~\bibnamefont {Lesiuk}}, \
  and\ \bibinfo {author} {\bibfnamefont {R.}~\bibnamefont {Moszynski}},\
  }\href@noop {} {\bibfield  {journal} {\bibinfo  {journal} {Phys. Rev. A}\
  }\textbf {\bibinfo {volume} {96}},\ \bibinfo {pages} {052510} (\bibinfo
  {year} {2017})}\BibitemShut {NoStop}%
\bibitem [{\citenamefont {Lesiuk}\ \emph {et~al.}(2019)\citenamefont {Lesiuk},
  \citenamefont {Przybytek}, \citenamefont {Balcerzak}, \citenamefont
  {Musia{\l}},\ and\ \citenamefont {Moszynski}}]{lesiuk19a}%
  \BibitemOpen
  \bibfield  {author} {\bibinfo {author} {\bibfnamefont {M.}~\bibnamefont
  {Lesiuk}}, \bibinfo {author} {\bibfnamefont {M.}~\bibnamefont {Przybytek}},
  \bibinfo {author} {\bibfnamefont {J.~G.}\ \bibnamefont {Balcerzak}}, \bibinfo
  {author} {\bibfnamefont {M.}~\bibnamefont {Musia{\l}}}, \ and\ \bibinfo
  {author} {\bibfnamefont {R.}~\bibnamefont {Moszynski}},\ }\href@noop {}
  {\bibfield  {journal} {\bibinfo  {journal} {J. Chem. Theory Comput.}\
  }\textbf {\bibinfo {volume} {15}},\ \bibinfo {pages} {2470} (\bibinfo {year}
  {2019})}\BibitemShut {NoStop}%
\bibitem [{\citenamefont {Jaquet}\ and\ \citenamefont
  {Lesiuk}(2020)}]{jaquet20}%
  \BibitemOpen
  \bibfield  {author} {\bibinfo {author} {\bibfnamefont {R.}~\bibnamefont
  {Jaquet}}\ and\ \bibinfo {author} {\bibfnamefont {M.}~\bibnamefont
  {Lesiuk}},\ }\href@noop {} {\bibfield  {journal} {\bibinfo  {journal} {J.
  Chem. Phys.}\ }\textbf {\bibinfo {volume} {152}},\ \bibinfo {pages} {104109}
  (\bibinfo {year} {2020})}\BibitemShut {NoStop}%
\bibitem [{\citenamefont {Czachorowski}\ \emph {et~al.}(2020)\citenamefont
  {Czachorowski}, \citenamefont {Przybytek}, \citenamefont {Lesiuk},
  \citenamefont {Puchalski},\ and\ \citenamefont {Jeziorski}}]{czachor20}%
  \BibitemOpen
  \bibfield  {author} {\bibinfo {author} {\bibfnamefont {P.}~\bibnamefont
  {Czachorowski}}, \bibinfo {author} {\bibfnamefont {M.}~\bibnamefont
  {Przybytek}}, \bibinfo {author} {\bibfnamefont {M.}~\bibnamefont {Lesiuk}},
  \bibinfo {author} {\bibfnamefont {M.}~\bibnamefont {Puchalski}}, \ and\
  \bibinfo {author} {\bibfnamefont {B.}~\bibnamefont {Jeziorski}},\ }\href@noop
  {} {\bibfield  {journal} {\bibinfo  {journal} {Phys. Rev. A}\ }\textbf
  {\bibinfo {volume} {102}},\ \bibinfo {pages} {042810} (\bibinfo {year}
  {2020})}\BibitemShut {NoStop}%
\bibitem [{\citenamefont {Eides}\ \emph {et~al.}(2001)\citenamefont {Eides},
  \citenamefont {Grotch},\ and\ \citenamefont {Shelyuto}}]{eides01}%
  \BibitemOpen
  \bibfield  {author} {\bibinfo {author} {\bibfnamefont {M.~I.}\ \bibnamefont
  {Eides}}, \bibinfo {author} {\bibfnamefont {H.}~\bibnamefont {Grotch}}, \
  and\ \bibinfo {author} {\bibfnamefont {V.~A.}\ \bibnamefont {Shelyuto}},\
  }\href@noop {} {\bibfield  {journal} {\bibinfo  {journal} {Phys. Rep.}\
  }\textbf {\bibinfo {volume} {342}},\ \bibinfo {pages} {63 } (\bibinfo {year}
  {2001})}\BibitemShut {NoStop}%
\bibitem [{\citenamefont {Puchalski}\ \emph {et~al.}(2010)\citenamefont
  {Puchalski}, \citenamefont {Kedziera},\ and\ \citenamefont
  {Pachucki}}]{puchalski10}%
  \BibitemOpen
  \bibfield  {author} {\bibinfo {author} {\bibfnamefont {M.}~\bibnamefont
  {Puchalski}}, \bibinfo {author} {\bibfnamefont {D.}~\bibnamefont {Kedziera}},
  \ and\ \bibinfo {author} {\bibfnamefont {K.}~\bibnamefont {Pachucki}},\
  }\href@noop {} {\bibfield  {journal} {\bibinfo  {journal} {Phys. Rev. A}\
  }\textbf {\bibinfo {volume} {82}},\ \bibinfo {pages} {062509} (\bibinfo
  {year} {2010})}\BibitemShut {NoStop}%
\bibitem [{\citenamefont {Vries}\ \emph {et~al.}(1987)\citenamefont {Vries},
  \citenamefont {Jager},\ and\ \citenamefont {Vries}}]{devries87}%
  \BibitemOpen
  \bibfield  {author} {\bibinfo {author} {\bibfnamefont {H.~D.}\ \bibnamefont
  {Vries}}, \bibinfo {author} {\bibfnamefont {C.~D.}\ \bibnamefont {Jager}}, \
  and\ \bibinfo {author} {\bibfnamefont {C.~D.}\ \bibnamefont {Vries}},\
  }\href@noop {} {\bibfield  {journal} {\bibinfo  {journal} {At. Data Nucl.
  Data Tables}\ }\textbf {\bibinfo {volume} {36}},\ \bibinfo {pages} {495 }
  (\bibinfo {year} {1987})}\BibitemShut {NoStop}%
\bibitem [{\citenamefont {Born}\ \emph {et~al.}(1955)\citenamefont {Born},
  \citenamefont {Huang},\ and\ \citenamefont {Lax}}]{born55}%
  \BibitemOpen
  \bibfield  {author} {\bibinfo {author} {\bibfnamefont {M.}~\bibnamefont
  {Born}}, \bibinfo {author} {\bibfnamefont {K.}~\bibnamefont {Huang}}, \ and\
  \bibinfo {author} {\bibfnamefont {M.}~\bibnamefont {Lax}},\ }\href@noop {}
  {\bibfield  {journal} {\bibinfo  {journal} {Am. J. Phys.}\ }\textbf {\bibinfo
  {volume} {23}},\ \bibinfo {pages} {474} (\bibinfo {year} {1955})}\BibitemShut
  {NoStop}%
\bibitem [{\citenamefont {Gauss}\ \emph {et~al.}(2006)\citenamefont {Gauss},
  \citenamefont {Tajti}, \citenamefont {K{\'a}llay}, \citenamefont {Stanton},\
  and\ \citenamefont {Szalay}}]{gauss06}%
  \BibitemOpen
  \bibfield  {author} {\bibinfo {author} {\bibfnamefont {J.}~\bibnamefont
  {Gauss}}, \bibinfo {author} {\bibfnamefont {A.}~\bibnamefont {Tajti}},
  \bibinfo {author} {\bibfnamefont {M.}~\bibnamefont {K{\'a}llay}}, \bibinfo
  {author} {\bibfnamefont {J.~F.}\ \bibnamefont {Stanton}}, \ and\ \bibinfo
  {author} {\bibfnamefont {P.~G.}\ \bibnamefont {Szalay}},\ }\href@noop {}
  {\bibfield  {journal} {\bibinfo  {journal} {J. Chem. Phys.}\ }\textbf
  {\bibinfo {volume} {125}},\ \bibinfo {pages} {144111} (\bibinfo {year}
  {2006})}\BibitemShut {NoStop}%
\bibitem [{\citenamefont {Tajti}\ \emph {et~al.}(2007)\citenamefont {Tajti},
  \citenamefont {Szalay},\ and\ \citenamefont {Gauss}}]{tajti07}%
  \BibitemOpen
  \bibfield  {author} {\bibinfo {author} {\bibfnamefont {A.}~\bibnamefont
  {Tajti}}, \bibinfo {author} {\bibfnamefont {P.~G.}\ \bibnamefont {Szalay}}, \
  and\ \bibinfo {author} {\bibfnamefont {J.}~\bibnamefont {Gauss}},\
  }\href@noop {} {\bibfield  {journal} {\bibinfo  {journal} {J. Chem. Phys.}\
  }\textbf {\bibinfo {volume} {127}},\ \bibinfo {pages} {014102} (\bibinfo
  {year} {2007})}\BibitemShut {NoStop}%
\bibitem [{\citenamefont {Langevin}(1905)}]{langevin1905}%
  \BibitemOpen
  \bibfield  {author} {\bibinfo {author} {\bibfnamefont {P.}~\bibnamefont
  {Langevin}},\ }\href@noop {} {\bibfield  {journal} {\bibinfo  {journal} {J.
  Phys. Theor. Appl.}\ }\textbf {\bibinfo {volume} {4}},\ \bibinfo {pages}
  {678} (\bibinfo {year} {1905})}\BibitemShut {NoStop}%
\bibitem [{\citenamefont {Saito}(2009)}]{saito09}%
  \BibitemOpen
  \bibfield  {author} {\bibinfo {author} {\bibfnamefont {S.~L.}\ \bibnamefont
  {Saito}},\ }\href@noop {} {\bibfield  {journal} {\bibinfo  {journal} {At.
  Data Nucl. Data Tables}\ }\textbf {\bibinfo {volume} {95}},\ \bibinfo {pages}
  {836} (\bibinfo {year} {2009})}\BibitemShut {NoStop}%
\bibitem [{\citenamefont {Kumar}\ and\ \citenamefont
  {Thakkar}(2010)}]{kumar10}%
  \BibitemOpen
  \bibfield  {author} {\bibinfo {author} {\bibfnamefont {A.}~\bibnamefont
  {Kumar}}\ and\ \bibinfo {author} {\bibfnamefont {A.~J.}\ \bibnamefont
  {Thakkar}},\ }\href@noop {} {\bibfield  {journal} {\bibinfo  {journal} {J.
  Chem. Phys.}\ }\textbf {\bibinfo {volume} {132}},\ \bibinfo {pages} {074301}
  (\bibinfo {year} {2010})}\BibitemShut {NoStop}%
\bibitem [{\citenamefont {Orcutt}\ and\ \citenamefont {Cole}(1967)}]{orcutt67}%
  \BibitemOpen
  \bibfield  {author} {\bibinfo {author} {\bibfnamefont {R.}~\bibnamefont
  {Orcutt}}\ and\ \bibinfo {author} {\bibfnamefont {R.}~\bibnamefont {Cole}},\
  }\href@noop {} {\bibfield  {journal} {\bibinfo  {journal} {J. Chem. Phys.}\
  }\textbf {\bibinfo {volume} {46}},\ \bibinfo {pages} {697} (\bibinfo {year}
  {1967})}\BibitemShut {NoStop}%
\bibitem [{\citenamefont {Buckley}\ \emph {et~al.}(2000)\citenamefont
  {Buckley}, \citenamefont {Hamelin},\ and\ \citenamefont
  {Moldover}}]{buckley00}%
  \BibitemOpen
  \bibfield  {author} {\bibinfo {author} {\bibfnamefont {T.~J.}\ \bibnamefont
  {Buckley}}, \bibinfo {author} {\bibfnamefont {J.}~\bibnamefont {Hamelin}}, \
  and\ \bibinfo {author} {\bibfnamefont {M.~R.}\ \bibnamefont {Moldover}},\
  }\href@noop {} {\bibfield  {journal} {\bibinfo  {journal} {Rev. Sci.
  Instrum.}\ }\textbf {\bibinfo {volume} {71}},\ \bibinfo {pages} {2914}
  (\bibinfo {year} {2000})}\BibitemShut {NoStop}%
\bibitem [{\citenamefont {Egan}\ \emph {et~al.}(2019)\citenamefont {Egan},
  \citenamefont {Stone}, \citenamefont {Scherschligt},\ and\ \citenamefont
  {Harvey}}]{egan19}%
  \BibitemOpen
  \bibfield  {author} {\bibinfo {author} {\bibfnamefont {P.~F.}\ \bibnamefont
  {Egan}}, \bibinfo {author} {\bibfnamefont {J.~A.}\ \bibnamefont {Stone}},
  \bibinfo {author} {\bibfnamefont {J.~K.}\ \bibnamefont {Scherschligt}}, \
  and\ \bibinfo {author} {\bibfnamefont {A.~H.}\ \bibnamefont {Harvey}},\
  }\href@noop {} {\bibfield  {journal} {\bibinfo  {journal} {J. Vac. Sci.
  Technol.}\ }\textbf {\bibinfo {volume} {37}},\ \bibinfo {pages} {031603}
  (\bibinfo {year} {2019})}\BibitemShut {NoStop}%
\bibitem [{\citenamefont {Havens}(1933)}]{havens33}%
  \BibitemOpen
  \bibfield  {author} {\bibinfo {author} {\bibfnamefont {G.~G.}\ \bibnamefont
  {Havens}},\ }\href@noop {} {\bibfield  {journal} {\bibinfo  {journal} {Phys.
  Rev.}\ }\textbf {\bibinfo {volume} {43}},\ \bibinfo {pages} {992} (\bibinfo
  {year} {1933})}\BibitemShut {NoStop}%
\bibitem [{\citenamefont {Mann}(1936)}]{mann36}%
  \BibitemOpen
  \bibfield  {author} {\bibinfo {author} {\bibfnamefont {K.~E.}\ \bibnamefont
  {Mann}},\ }\href@noop {} {\bibfield  {journal} {\bibinfo  {journal} {Z.
  Phys.}\ }\textbf {\bibinfo {volume} {98}},\ \bibinfo {pages} {548} (\bibinfo
  {year} {1936})}\BibitemShut {NoStop}%
\bibitem [{\citenamefont {Abonnenc}(1939)}]{abonnenc39}%
  \BibitemOpen
  \bibfield  {author} {\bibinfo {author} {\bibfnamefont {L.}~\bibnamefont
  {Abonnenc}},\ }\href@noop {} {\bibfield  {journal} {\bibinfo  {journal} {CR
  Acad. Sci}\ }\textbf {\bibinfo {volume} {208}},\ \bibinfo {pages} {986}
  (\bibinfo {year} {1939})}\BibitemShut {NoStop}%
\bibitem [{\citenamefont {Barter}\ \emph {et~al.}(1960)\citenamefont {Barter},
  \citenamefont {Meisenheimer},\ and\ \citenamefont {Stevenson}}]{barter60}%
  \BibitemOpen
  \bibfield  {author} {\bibinfo {author} {\bibfnamefont {C.}~\bibnamefont
  {Barter}}, \bibinfo {author} {\bibfnamefont {R.}~\bibnamefont
  {Meisenheimer}}, \ and\ \bibinfo {author} {\bibfnamefont {D.}~\bibnamefont
  {Stevenson}},\ }\href@noop {} {\bibfield  {journal} {\bibinfo  {journal} {J.
  Phys. Chem.}\ }\textbf {\bibinfo {volume} {64}},\ \bibinfo {pages} {1312}
  (\bibinfo {year} {1960})}\BibitemShut {NoStop}%
\bibitem [{\citenamefont {Yoshizawa}\ and\ \citenamefont
  {Hada}(2009)}]{yoshizawa09}%
  \BibitemOpen
  \bibfield  {author} {\bibinfo {author} {\bibfnamefont {T.}~\bibnamefont
  {Yoshizawa}}\ and\ \bibinfo {author} {\bibfnamefont {M.}~\bibnamefont
  {Hada}},\ }\href@noop {} {\bibfield  {journal} {\bibinfo  {journal} {J. Comp.
  Chem.}\ }\textbf {\bibinfo {volume} {30}},\ \bibinfo {pages} {2550} (\bibinfo
  {year} {2009})}\BibitemShut {NoStop}%
\bibitem [{\citenamefont {Ruud}\ \emph {et~al.}(1994)\citenamefont {Ruud},
  \citenamefont {Skaane}, \citenamefont {Helgaker}, \citenamefont {Bak},\ and\
  \citenamefont {Joergensen}}]{ruud94}%
  \BibitemOpen
  \bibfield  {author} {\bibinfo {author} {\bibfnamefont {K.}~\bibnamefont
  {Ruud}}, \bibinfo {author} {\bibfnamefont {H.}~\bibnamefont {Skaane}},
  \bibinfo {author} {\bibfnamefont {T.}~\bibnamefont {Helgaker}}, \bibinfo
  {author} {\bibfnamefont {K.~L.}\ \bibnamefont {Bak}}, \ and\ \bibinfo
  {author} {\bibfnamefont {P.}~\bibnamefont {Joergensen}},\ }\href@noop {}
  {\bibfield  {journal} {\bibinfo  {journal} {J. Am. Chem. Soc.}\ }\textbf
  {\bibinfo {volume} {116}},\ \bibinfo {pages} {10135} (\bibinfo {year}
  {1994})}\BibitemShut {NoStop}%
\bibitem [{\citenamefont {Jaszu{\'n}ski}\ \emph {et~al.}(1995)\citenamefont
  {Jaszu{\'n}ski}, \citenamefont {J{\o}rgensen},\ and\ \citenamefont
  {Rizzo}}]{jaszunski95}%
  \BibitemOpen
  \bibfield  {author} {\bibinfo {author} {\bibfnamefont {M.}~\bibnamefont
  {Jaszu{\'n}ski}}, \bibinfo {author} {\bibfnamefont {P.}~\bibnamefont
  {J{\o}rgensen}}, \ and\ \bibinfo {author} {\bibfnamefont {A.}~\bibnamefont
  {Rizzo}},\ }\href@noop {} {\bibfield  {journal} {\bibinfo  {journal} {Theor.
  Chim. Acta}\ }\textbf {\bibinfo {volume} {90}},\ \bibinfo {pages} {291}
  (\bibinfo {year} {1995})}\BibitemShut {NoStop}%
\bibitem [{\citenamefont {Reinsch}\ and\ \citenamefont
  {Meyer}(1976)}]{reinsch76}%
  \BibitemOpen
  \bibfield  {author} {\bibinfo {author} {\bibfnamefont {E.-A.}\ \bibnamefont
  {Reinsch}}\ and\ \bibinfo {author} {\bibfnamefont {W.}~\bibnamefont
  {Meyer}},\ }\href@noop {} {\bibfield  {journal} {\bibinfo  {journal} {Phys.
  Rev. A}\ }\textbf {\bibinfo {volume} {14}},\ \bibinfo {pages} {915} (\bibinfo
  {year} {1976})}\BibitemShut {NoStop}%
\bibitem [{\citenamefont {Levy}\ and\ \citenamefont {Perdew}(1985)}]{levy85}%
  \BibitemOpen
  \bibfield  {author} {\bibinfo {author} {\bibfnamefont {M.}~\bibnamefont
  {Levy}}\ and\ \bibinfo {author} {\bibfnamefont {J.~P.}\ \bibnamefont
  {Perdew}},\ }\href@noop {} {\bibfield  {journal} {\bibinfo  {journal} {Phys.
  Rev. A}\ }\textbf {\bibinfo {volume} {32}},\ \bibinfo {pages} {2010}
  (\bibinfo {year} {1985})}\BibitemShut {NoStop}%
\bibitem [{\citenamefont {Desclaux}(1973)}]{desclaux73}%
  \BibitemOpen
  \bibfield  {author} {\bibinfo {author} {\bibfnamefont {J.}~\bibnamefont
  {Desclaux}},\ }\href@noop {} {\bibfield  {journal} {\bibinfo  {journal} {At.
  Data Nucl. Data Tables}\ }\textbf {\bibinfo {volume} {12}},\ \bibinfo {pages}
  {311} (\bibinfo {year} {1973})}\BibitemShut {NoStop}%
\end{thebibliography}%

\end{document}